\documentclass[a4paper,11pt]{article}
\pdfoutput=1 

\usepackage{jheppub} 
\usepackage[dvipsnames]{xcolor}
\usepackage{caption}
\usepackage{subcaption}
\usepackage{bold-extra}
\usepackage{booktabs}
\usepackage{float}
\usepackage{hyperref}
\usepackage[T1]{fontenc} 
\usepackage{lmodern}
\usepackage{tgtermes} 
\makeatletter
\renewcommand{\textsc}[1]{%
  {\fontencoding{T1}%
   \fontfamily{qtm}
   \fontseries{\f@series}
   \selectfont
   \scshape
   #1%
  }%
}
\makeatother
 
\newcommand{\jewel}{\textbf{\textsc{Jewel}}}
\newcommand{\bjewel}{\textbf{\textsc{Jewel + Pythia}}}

\title{Apples to Apples in Jet Quenching: robustness of Machine Learning classification of quenched jets to Underlying Event contamination}

\author[a,b]{J. A. Gonçalves}
\author[a,b]{J. G. Milhano}

\affiliation[a]{Laboratório de Instrumentação e Física Experimental de Partículas (LIP), Avenida
Professor Gama Pinto 2, 1649-003 Lisboa, Portugal}
\affiliation[b]{Instituto Superior Técnico, Universidade de Lisboa, Avenida Rovisco Pais 1, 1609-001,
Lisboa, Portugal}

\emailAdd{jgponcalves@lip.pt}
\emailAdd{gmilhano@lip.pt}

\abstract{Progress in the theoretical understanding of parton branching dynamics within an expanding Quark Gluon Plasma relies on detailed and fair comparisons with experimental data for reconstructed jets. 
Such comparisons are only meaningful when the computed jet, be it analytically or via event generation, accounts for the complexity of jets reconstructed in the challenging environment of heavy-ion collisions. 
Jet reconstruction in heavy-ion collisions involves a necessarily imperfect subtraction of the large and fluctuating underlying event: reconstructed jets always include underlying event contamination. 
To identify true jet quenching effects, modifications due to the interaction of the branching partonic system with the Quark Gluon Plasma, we establish a baseline that accounts for possible background contamination on unmodified jets. In practical terms, jet quenching effects are only those not present in jets produced in proton-proton collisions that have been embedded in a realistic heavy-ion background and where subtraction has been carried out analogously to that in the heavy ion case.
With this setup, we assess the sensitivity to underlying event of commonly discussed jet quenching observables and its impact on the robustness of Machine Learning studies, aimed at classifying jets according to their degree of modification by the Quark Gluon Plasma, that rely on those observables. We find the discrimination power of a simple Boosted Decision Tree to be robust in the realistic scenario where both medium response and underlying event are present, giving support to portability to the experimental context.}

\begin{document} 
\maketitle
\flushbottom

\section{Introduction}
\label{sec:intro}

In the past few years, the exponential growth of Machine Learning (ML) and Deep Learning (DL) applications in high energy physics (see \cite{Feickert:2021ajf,Larkoski:2017jix} for comprehensive reviews and \cite{hepmllivingreview} for an up-to-date repository of relevant works) has not left the physics of heavy-ion collisions unscathed.
Among the many applications in that context, see \cite{Zhou:2023pti} for a review, the modification of jets by interaction with Quark Gluon Plasma (QGP), commonly referred to as jet quenching, has received particular attention \cite{Chien:2018rgm,Apolinario:2021olp,Lai:2021ckt,Du:2020pmp,Du:2021pqa,Yang:2022yfr,Lee:2022kdn,Liu:2022hzd,CrispimRomao:2023ssj,Qureshi:2024ceh,Du:2023qst}.

Jets are collimated bunches of hadrons resulting from the branching and subsequent hadronization of energetic partons. 
In heavy-ion collisions, branching occurs within the QGP. 
Jets reconstructed in heavy-ion collisions differ from their proton-proton counterparts (where no QGP is present) both due to changes in the branching pattern and to the parton-induced excitations of the QGP, the medium response (MR), that fall within the catchment area of the jet (see \cite{Mehtar-Tani:2013pia,Qin:2015srf,Connors:2017ptx,Apolinario:2022vzg} for reviews). 
These modifications occur throughout the branching process and, as such, are sensitive to a wide range of spatial and momentum scales. This, together with the excellent theoretical control of jet physics in the absence of QGP, makes jets a versatile probe of QGP properties.
 
However, the extraction of QGP properties from modifications of jet observables becomes less straightforward when underlying event (UE) contributions to jet observables are accounted for \cite{Apolinario:2012si}. In heavy-ion collisions, UE is dominated by the hadronization of QGP that did not interact with the traversing partons. 
In the busy environment of a central heavy-ion collision, UE contributions within the catchment area of a $R=0.4$ jet have a total average transverse momentum of $\sim 100$\, GeV with fluctuations of the order of 15\%. 
All experimental measurements of jets in heavy-ion collisions involve the subtraction of UE prior to jet reconstruction.
While sophisticated UE subtraction procedures \cite{Connors:2017ptx} have been devised, subtraction can never be perfect on a jet-by-jet basis. Jets reconstructed in heavy-ion collisions will always include some contribution from UE which can mimic effects due to medium response (the modification of QGP arising from jet development within it). 

The identification of \textit{bona fide} jet quenching effects, measured modifications of jet properties due to interaction of the developing parton cascade with QGP, requires the understanding of the extent to which UE contamination affects jet observables.
The effect of UE is of particular relevance in the context of ML/DL approaches that aim to distinguish jets that have been modified by interaction with QGP from those that have not.

Some of the studies conducted so far using ML/DL \cite{Apolinario:2021olp,CrispimRomao:2023ssj} to identify quenched jets on the basis of their reconstructed properties, rely on setups where both MR and UE are ignored. 
Such studies identify jet modifications with respect to a proton-proton baseline that arise exclusively from changes in the fragmentation pattern and resulting from the transport of soft particles away from the jet reconstruction radius. They consistently find that a subset of jets in PbPb collisions to be distinguishable from their pp counterparts.
Studies that have included effects due to MR \cite{Du:2020pmp,Du:2021pqa,Liu:2022hzd,Qureshi:2024ceh} find an almost perfect distinction between jets produced in pp and PbPb collisions. While the presence of MR in PbPb collisions, together with its obvious absence in the pp case, is an important distinguishing feature, its interplay with UE contamination, necessarily present in any realistic scenario, has not been addressed in full.

In this work we argue that the identification of true quenching effects can be achieved by comparison of realistic PbPb samples, including both MR and UE, with a baseline pp sample where PbPb-like UE contamination effects are fully accounted for. This apples-to-apples comparison singles out measured modifications that resulted from interaction with QGP.

The paper is organized as follows. In Sec.~\ref{sec:samples} we describe the various event samples used throughout the study. In Sec.~\ref{sec:anlysis} we lay out the analysis performed over the jet events, including detailed information on how the samples where generated. In Sec.~\ref{sec:observablesensitivity} we discuss the sensitivity of specific observables to UE contamination. 
Sec.~\ref{sec:mlrobustness} addresses the robustness of an example set of ML analyses to UE effects, and Sec.~\ref{sec:conclusions} summarizes the conclusions of our work. A series of appendices provides further detailed information.

\section{Simulated Data}
\label{sec:samples}

This study is based on hadron level jet samples generated, both for proton-proton (vacuum) and Pb-Pb (medium) collisions, with \jewel\ 2.3.0 \cite{Zapp:2012ak,Zapp:2013vla} at $\sqrt{s_{NN}}=5.02$ TeV. The hard matrix element was generated with a lower transverse momentum cut-off of $50$ GeV, and the generation spectrum was re-weighted as $p_T^5$, so as to oversample the large $p_T$ region with the resulting event weights used throughout the study. The proton PDFs are given by CT14nlo \cite{Dulat:2015mca}, and those for an isospin averaged nucleon in Pb by EPPS16nlo\_CT14nlo\_Pb208 \cite{Eskola:2016oht}, both provided through LHAPDF6 \cite{Buckley:2014ana}.

For the PbPb case, the simple parameterized QGP medium described in detail in \cite{Zapp:2013zya} is used with standard parameters $T_i = 590$ MeV (the maximal temperature at the center of a collision with zero impact parameter), $\tau_i = 0.4$ (the time at which $T_i$ is reached), and $T_c = 170$ MeV (the temperature at which interaction with medium stops). All PbPb samples were generated with the QGP medium extending over $|\eta| <4$ and for $0-10\%$ centrality, both with and without medium response. 
When including medium response, the \jewel\ specific subtraction procedure introduced in \cite{Milhano:2022kzx}, and briefly discussed in App.~\ref{app:jewsub}, was used. We stress that this is an essential step to generate a physically meaningful medium response in \jewel. Without the \jewel-specific subtraction, jets will include unphysical contributions that cannot be effectively subtracted by use of generic background subtraction methods.

Additional pp and PbPb samples were prepared by embedding events into a realistic PbPb underlying event as described in App. \ref{app:UE} and performing an underlying event (UE) subtraction prior to any reconstruction as detailed in App. \ref{app:uesub}.

Overall, six samples are considered throughout:
\begin{itemize}
    \item \bjewel{}\ pp \textbf{[pp]}:\\
        Generated proton-proton. The standard baseline in the absence of QGP effects.
    \item \bjewel{}\ pp with PbPb UE \textbf{[pp + UE]}:\\
        Generated proton-proton baseline embedded in a PbPb UE. UE subtraction is carried out prior to jet reconstruction. This sample, the preferred baseline in our study, accounts for a physical scenario where jets in PbPb are unmodified by interaction with QGP, and observed modifications result exclusively from imperfect UE subtraction.
    \item \bjewel{}\ PbPb without medium response \textbf{[PbPb]}:\\
        Generated PbPb sample without medium response. All jet modifications result from modifications to the parton shower effected by interaction with QGP.
    \item \bjewel{}\ PbPb with medium response \textbf{[PbPb + MR]}:\\
        Generated PbPb sample with medium response (including the necessary \jewel\ specific subtraction). This sample includes effects from parton shower modification and medium response.
    \item \bjewel{}\ PbPb with UE \textbf{[PbPb + UE]}:\\
        Generated PbPb sample embedded in PbPb UE, with UE subtraction carried out before jet reconstruction. This sample includes effects due to modification of the parton shower and from imperfect UE subtraction.
    \item \bjewel{}\ PbPb with medium response and UE \textbf{[PbPb + MR + UE]}:\\
        Generated PbPb sample with medium response (including the necessary \jewel\ specific subtraction), embedded in PbPb UE, with UE subtraction carried out before jet reconstruction. This sample, the closest proxy to experimental data, includes effects from parton shower modification, medium response, and from imperfect UE subtraction.
\end{itemize}

\section{Analysis}
\label{sec:anlysis}

All particles within $\lvert\eta\rvert < 4$ and with $p_T > 100$\,MeV are considered for jet reconstruction. Jets, with radius parameter $R=0.4$, are reconstructed with the anti-$k_T$ algorithm \cite{Cacciari:2008gp} as implemented in FastJet \cite{Cacciari:2011ma}. Unless otherwise specified, jets are only included if $\lvert\eta_{\mathrm{jet}}\rvert < 3$, and for $p_T > 100$\,GeV.
For di-jet systems, the cuts are $p_T > 120$\,GeV for the leading jet, $p_T > 50$\,GeV for the sub-leading jet, and their azimuthal separation at least $5\pi/6$.

\subsection{UE subtraction}
\label{subsec:UEsub}

UE subtraction is performed using the ICS algorithm \cite{Berta:2019hnj} with recommended parameters for $R=0.4$ anti-$k_T$ jets (see App. \ref{app:uesub} for further details). Subtraction is applied event-wide with jet reconstruction performed only after subtraction. 

In order to assess the impact of contamination from imperfectly subtracted UE we have matched, event-by-event, the highest $p_T$ jet in the sample without UE (\textbf{pp} in the proton-proton case and \textbf{PbPb+MR} in the PbPb case) to the highest $p_T$ jet in the corresponding UE embedded and subtracted sample (respectively \textbf{pp+UE} and \textbf{PbPb+MR+UE}) for which the rapidity-azimuth distance between jet axes $\Delta R = \sqrt{(\Delta\phi)^2 + (\Delta\eta)^2}$ is less than 0.4. We note that this matching jet is not necessarily the highest $p_T$ jet in the event as UE contamination can swap the $p_T$ ordering in balanced dijet pairs. The closeness requirement $\Delta R \leq 0.4$ is essential for the same jet, without and with UE contamination, to be correctly matched.
In the rare cases where no matching jet is found, the event is discarded. 

For each pair of matched jets we compute (with $i=$ \textbf{pp}, \textbf{PbPb+MR} for proton-proton and PbPb, respectively): $\frac{\delta p_T}{p_T^{[i]}} = \frac{p_T^{[i+\mathbf{UE}]} - p_T^{[i]}}{p_T^{[i]}}$, the relative transverse momentum change effected by UE embedding and subtraction; $\frac{\delta m}{p_T^{[i]}} = \frac{m^{[i+\mathbf{UE}]} - m^{[i]}}{p_T^{[i]}}$, the change in mass normalized to the jet $p_T$ before UE embedding and subtraction; $\delta \eta = \eta^{[i+\mathbf{UE}]} - \eta^{[i]}$, the difference in pseudorapidity of the jet axes; and $\delta \varphi = |\varphi^{[i+\mathbf{UE}]} - \varphi^{[i]}|$, the absolute difference of azimuthal orientation of jet axes. 

\begin{figure}[!htbp]
     \centering
     \begin{subfigure}[b]{0.49\textwidth}
         \centering
         \includegraphics[width=\textwidth, angle=0]{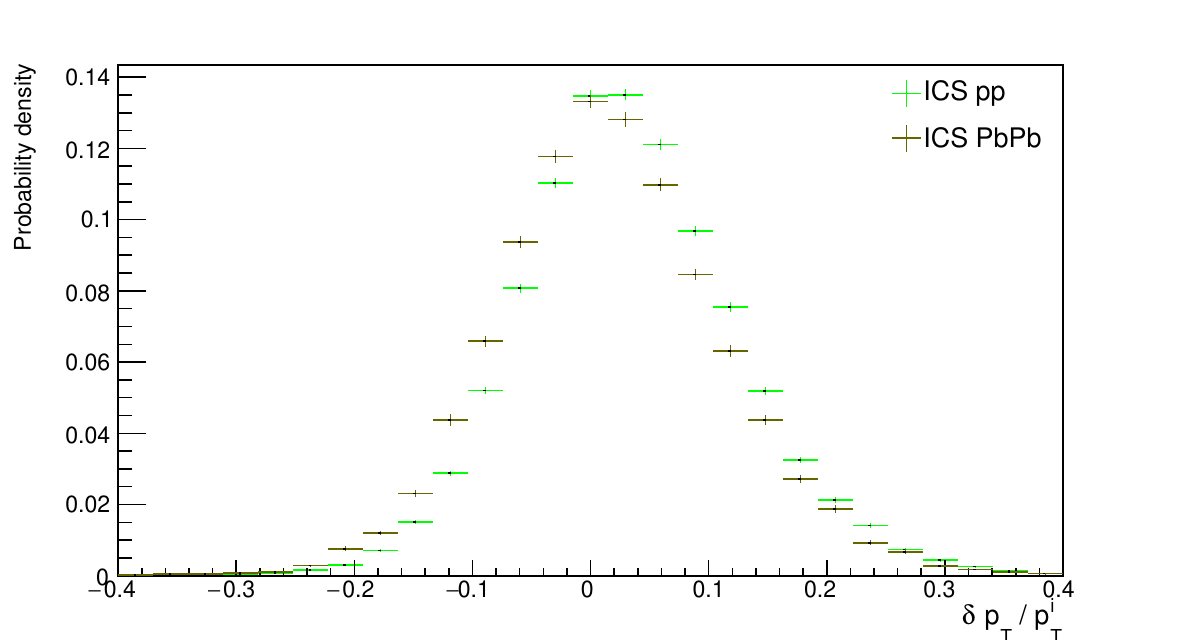}
         \caption{}
         \label{sfig:deltapt}
     \end{subfigure}
     \hfill
     \begin{subfigure}[b]{0.49\textwidth}
         \centering
         \includegraphics[width=\textwidth, angle=0]{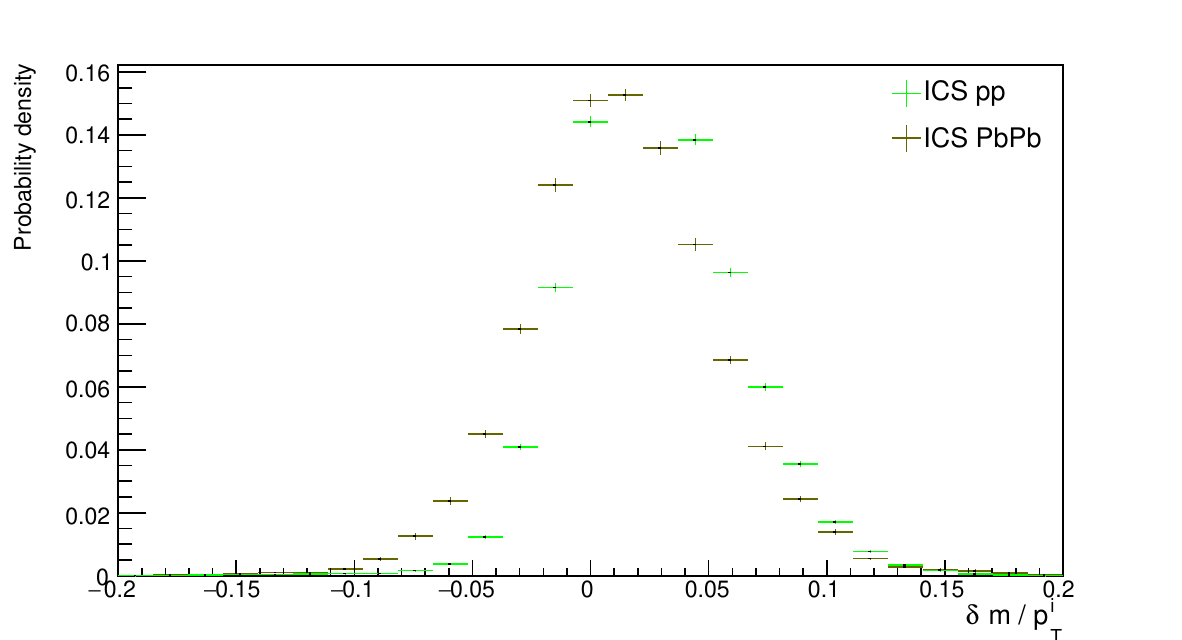}
          \caption{}
         \label{sfig:tdeltam}
     \end{subfigure}
     \\
     \begin{subfigure}[b]{0.49\textwidth}
         \centering
         \includegraphics[width=\textwidth, angle=0]{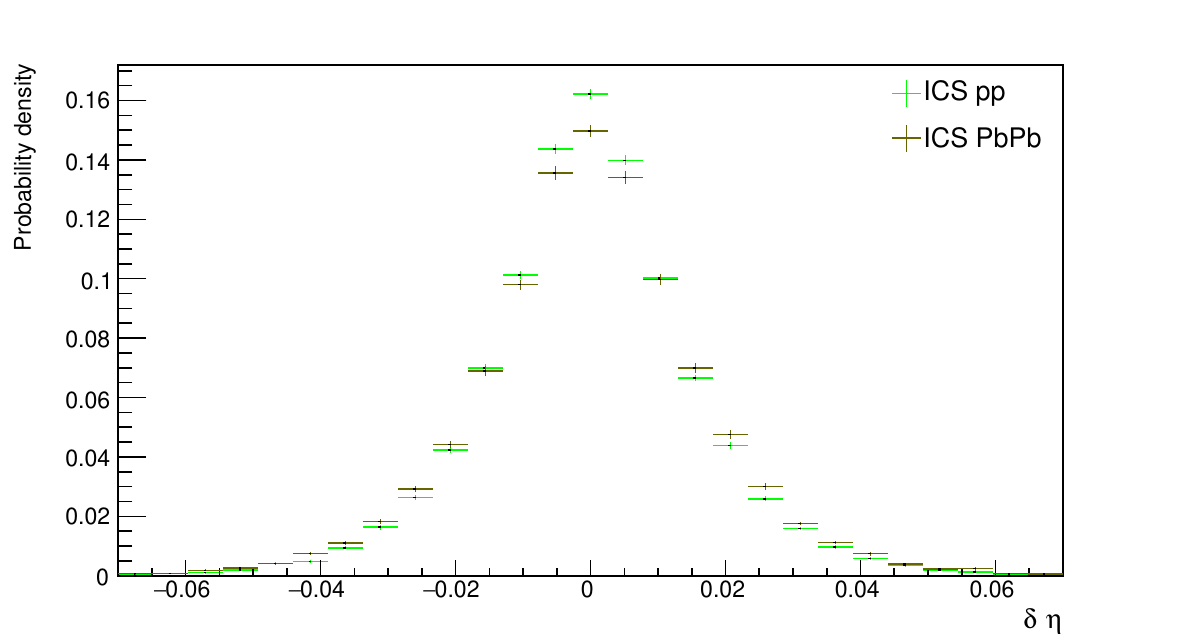}
         \caption{}
         \label{sfig:deltaeta}
     \end{subfigure}
     \hfill
     \begin{subfigure}[b]{0.49\textwidth}
         \centering
         \includegraphics[width=\textwidth, angle=0]{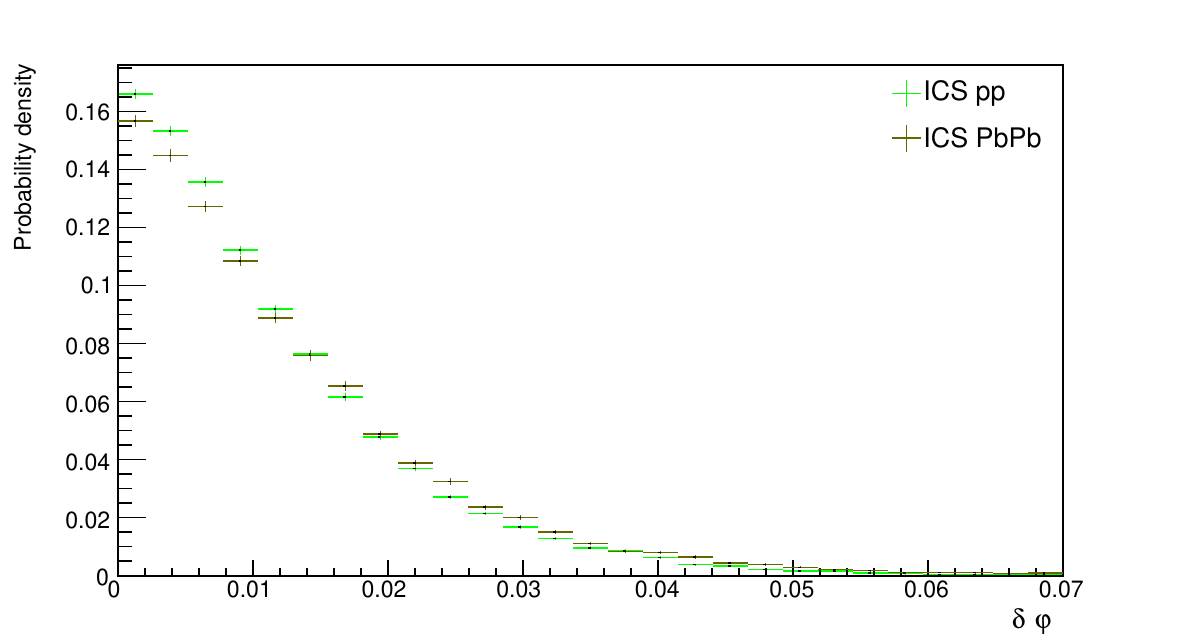}
         \caption{}
         \label{sfig:deltaphi}
     \end{subfigure}
        \caption{Residual UE contamination, after embedding and subtraction, on: (a) relative jet transverse momentum \(\frac{\delta p_T}{p_T^{[i]}} = \frac{p_T^{[i+\mathbf{UE}]} - p_T^{[i]}}{p_T^{[i]}}\); (b) jet mass normalized to the jet initial momentum \(\frac{\delta m}{p_T^{[i]}} = \frac{m^{[i+\mathbf{UE}]} - m^{[i]}}{p_T^{[i]}}\); (c) pseudorapidity \(\delta \eta = \eta^{[i+\mathbf{UE}]} - \eta^{[i]}\); and (d) azimuthal angle \(\delta \varphi = |\varphi^{[i+\mathbf{UE}]} - \varphi^{[i]}|\).}
        \label{fig:subtval}
\end{figure}

The results are shown in Fig.~\ref{fig:subtval} where we can first note that the embedding and ICS subtraction we performed yield comparable results for both pp and PbPb cases. 
The presence of medium modification (both those resulting from modifications of the shower and those due to medium response) does not alter the performance of our background subtraction procedure.

Jet $p_T$, Fig.~\ref{sfig:deltapt}, is smeared around an average positive shift (3.4\% for pp and 1.8\% for PbPb) with a width (one standard deviation) of 9\% in both cases. 
The shift in jet mass, normalized by the jet $p_T$, is shown in Fig.~\ref{sfig:tdeltam}. Embedding and subtraction result in an average increase in normalized mass (0.03 and 0.01 in pp and PbPb respectively) with a spread of about 0.04 in both cases. 
Pseudorapidity (Fig.~\ref{sfig:deltaeta}) and azimuthal orientation (Fig.~\ref{sfig:deltaphi}) of the jet axis 
are very mildly affected by imperfect UE subtraction. In both cases, the distributions are peaked at values very close to zero and are narrow with one standard deviation widths of 0.02 for $\delta\eta$ and 0.01 for $\delta\phi$.

All in all, the embedding and subtraction procedure we implemented smears both jet $p_T$ and mass, and introduces a systematic positive shift of their average values.

\section{Sensitivity of jet observables to UE contamination}
\label{sec:observablesensitivity}

In principle, residual UE contamination affects all jet observables. Of particular interest to us here is the extent to which such contamination can mimic effects usually understood as jet quenching, that is, emerging from the interaction of partons in a developing shower with the QGP they traverse. The interpretation of modifications of jet properties as jet quenching can only be made once UE contamination has been accounted for.
While UE contamination can be reduced by either further optimization of the scheme described in Sec.~\ref{subsec:UEsub} and App.~\ref{app:uesub}, or by using arguably more sophisticated strategies (for a review of background subtraction methods see \cite{Connors:2017ptx} and references therein), reconstructed jets in heavy-ion collisions will always include some UE contamination.

Our assessment of the sensitivity of observables to UE contamination relies on the comparison of the effect of UE contamination in pp and PbPb. When the effect of UE contamination is very similar in both the pp and PbPb samples, it cancels out in the ratio \textbf{PbPb + MR + UE}/\textbf{pp + UE} making it (nearly) identical to the the ratio \textbf{PbPb + MR}/\textbf{pp}. Observables for which these two ratios are nearly identical over the entire considered range of the observable, are classified as robust. Significant differences between the ratios imply that UE contamination leads to modifications that can be confounded with quenching effects and we classify observables where that occurs as sensitive.

Below we show examples\footnote{results for further observables are shown in App.~\ref{app:jetobs}.} of observables whose modification is robust and observables where the sensitivity to UE remnants cannot be separated from putative quenching effects. 
Experimental data is shown, simply for illustrative purposes, whenever available in HEPData for compatible kinematical cuts.

\subsection{Observables that are robust in the presence of UE contamination}
\label{subsec:robustobservables}

\subsubsection{Inclusive jet \texorpdfstring{\(R_{AA}\)}{}}

\begin{figure}[!htbp]
     \centering
     \begin{subfigure}[b]{0.49\textwidth}
         \centering
         \includegraphics[width=\textwidth, angle=0]{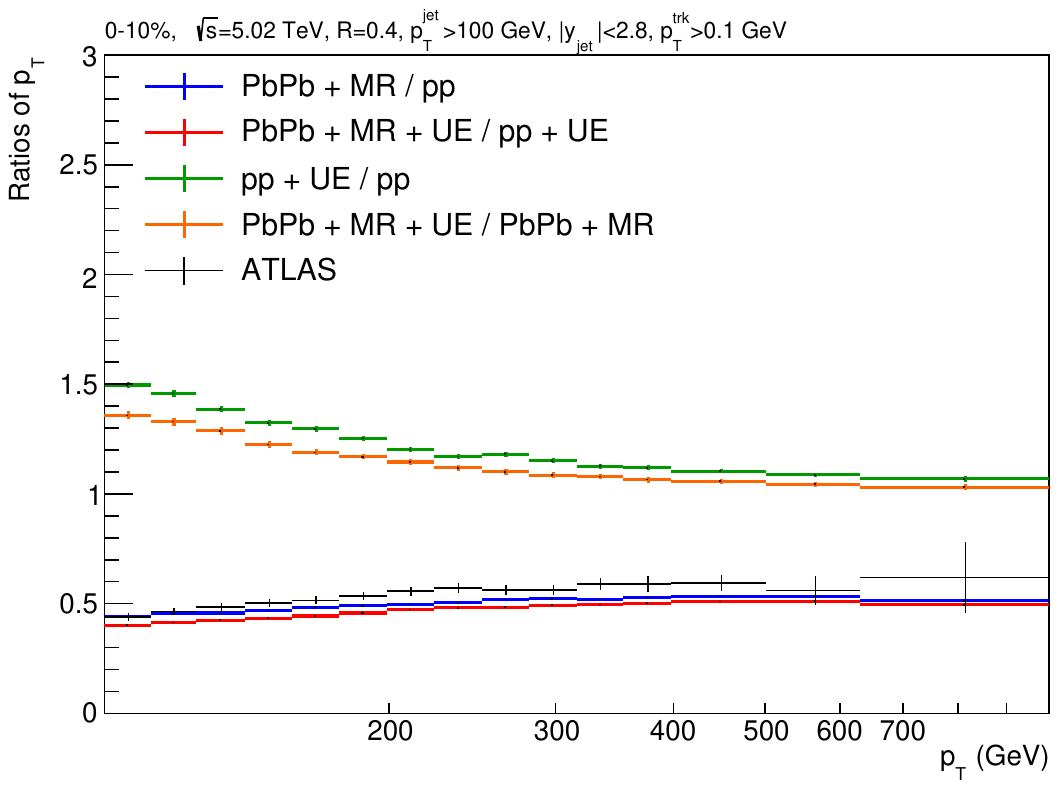}
         \caption{}
         \label{sfig:pt}
     \end{subfigure}
     \hfill
     \begin{subfigure}[b]{0.49\textwidth}
         \centering
         \includegraphics[width=\textwidth, angle=0]{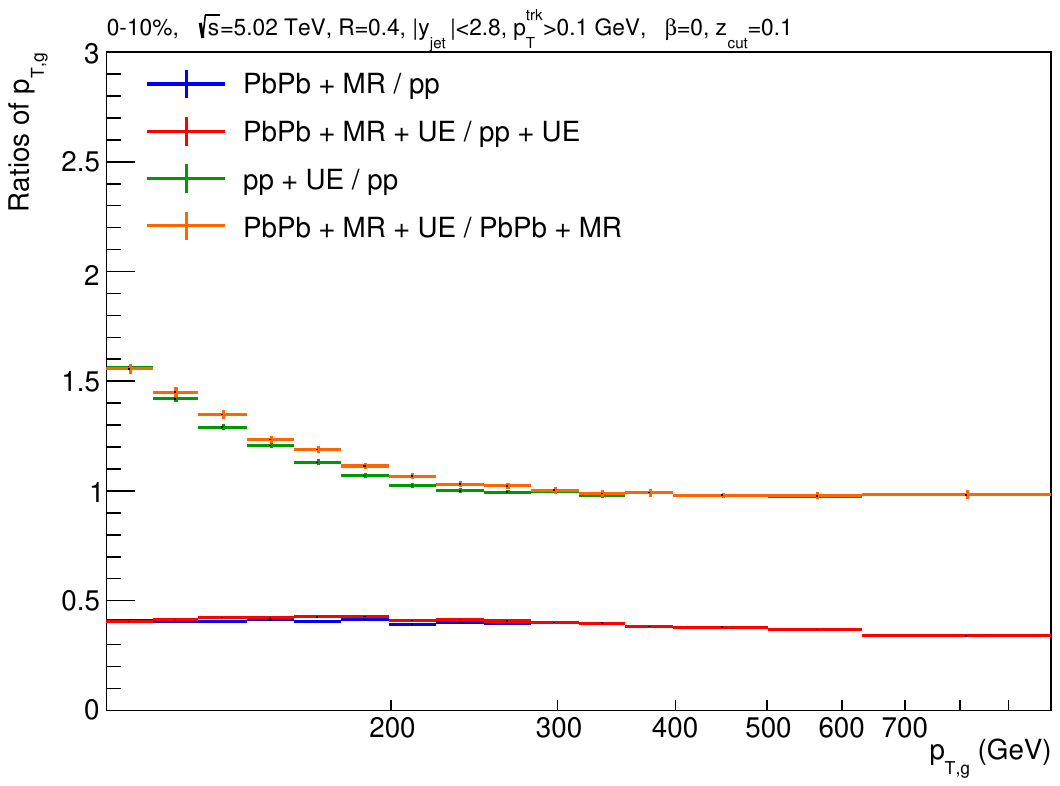}
          \caption{}
         \label{sfig:ptg}
     \end{subfigure}
     \caption{Inclusive jet \(R_{AA}\) without (blue) and with (red) the inclusion of UE contamination. The \textbf{pp + UE} to \textbf{pp} (green) and \textbf{PbPb + MR + UE} to \textbf{PbPb + MR} (orange) ratios show the effect of the contamination separately on each sample. (a) for ungroomed jets,  with experimental data \cite{ATLAS:2018gwx} shown in black, and (b) for Soft Drop groomed jets.}
        \label{fig:ptratios}
\end{figure}

Fig.~\ref{sfig:pt} shows the nuclear modification factor $R_{AA}$ for inclusive jets computed both with (red) and without (blue) taking UE contamination into account. 
In both cases there is very good agreement with published experimental data \cite{ATLAS:2018gwx}. The effect of UE contamination leads to a very small reduction of $R_{AA}$ across the $p_T$ range we considered. 
This robustness follows from $R_{AA}$ being a ratio and the effect of UE being very similar in PbPb and pp (see Fig.~\ref{fig:subtval}). 
This is despite the effect of UE contamination not being small, particularly for the lower $p_T$ jets, as shown in the green curve for pp and in the orange curve for PbPb, also in Fig.~\ref{sfig:pt}. The small observed reduction of $R_{AA}$ follows from the UE effect being marginally larger for pp than for PbPb.

The effect of UE contamination is further reduced, see Fig.~\ref{sfig:ptg}, if one considers jets groomed with Soft Drop \cite{Dasgupta:2013ihk,Larkoski:2014wba} (with $\beta = 0$ and $z_{cut} = 0.1$). Here, the removal of large angle soft components from the jet has both the effect of extending the range for which UE contamination is negligible ($p_{T,g} > 200$~GeV) and making the residual contamination almost identical in pp and PbPb whenever it is sizable ($p_{T,g} \lesssim 200$~GeV).

\subsubsection{Dijet asymmetry}

A similar picture emerges when we consider the transverse momentum asymmetry in back-to-back dijet pairs quantified by the ratio $x_j$:
\begin{equation}
        x_j = \frac{p_T^{\mathsf{subleading}}}{p_T^{\mathsf{leading}}}\, ,
\end{equation}
where $p_T^{\mathsf{leading}}$ and $p_T^{\mathsf{subleading}}$ are, respectively, the transverse momenta of the leading and sub-leading jets in the dijet pair.

\begin{figure}[!htbp]
     \centering
     \begin{subfigure}[b]{0.49\textwidth}
         \centering
         \includegraphics[width=\textwidth, angle=0]{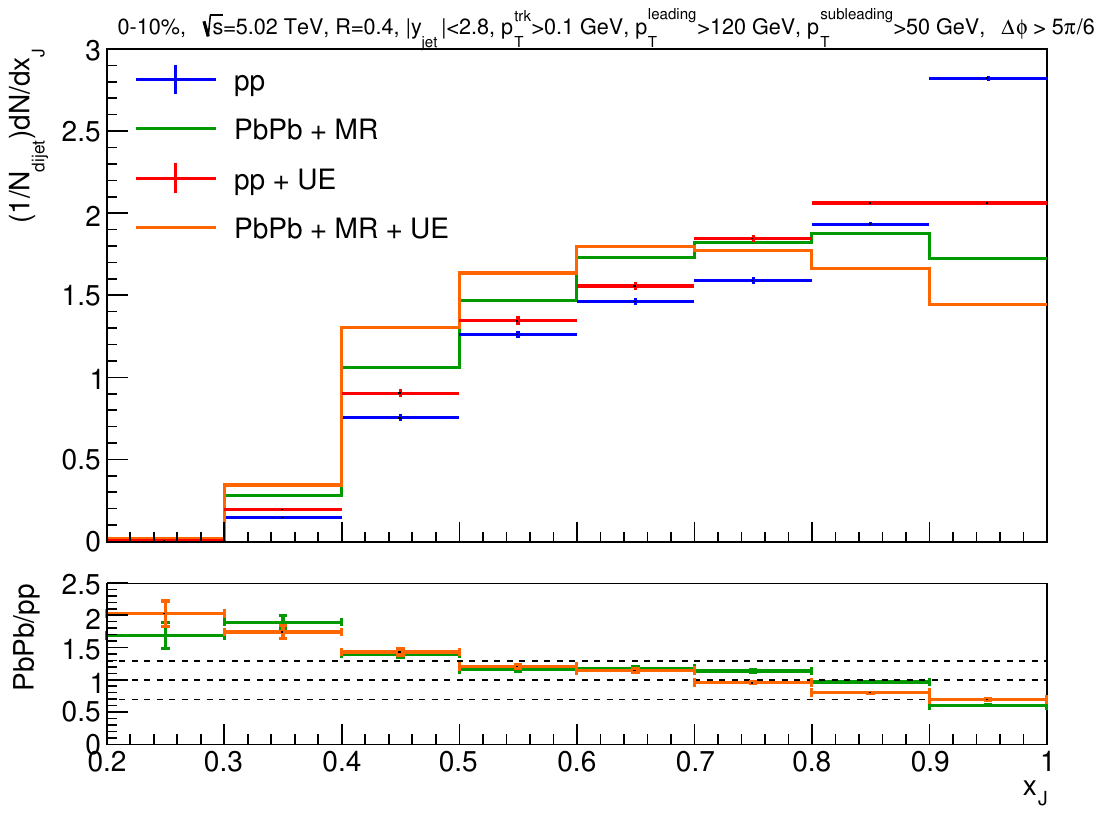}
         \caption{}
         \label{sfig:xj}
     \end{subfigure}
     \hfill
     \begin{subfigure}[b]{0.49\textwidth}
         \centering
         \includegraphics[width=\textwidth, angle=0]{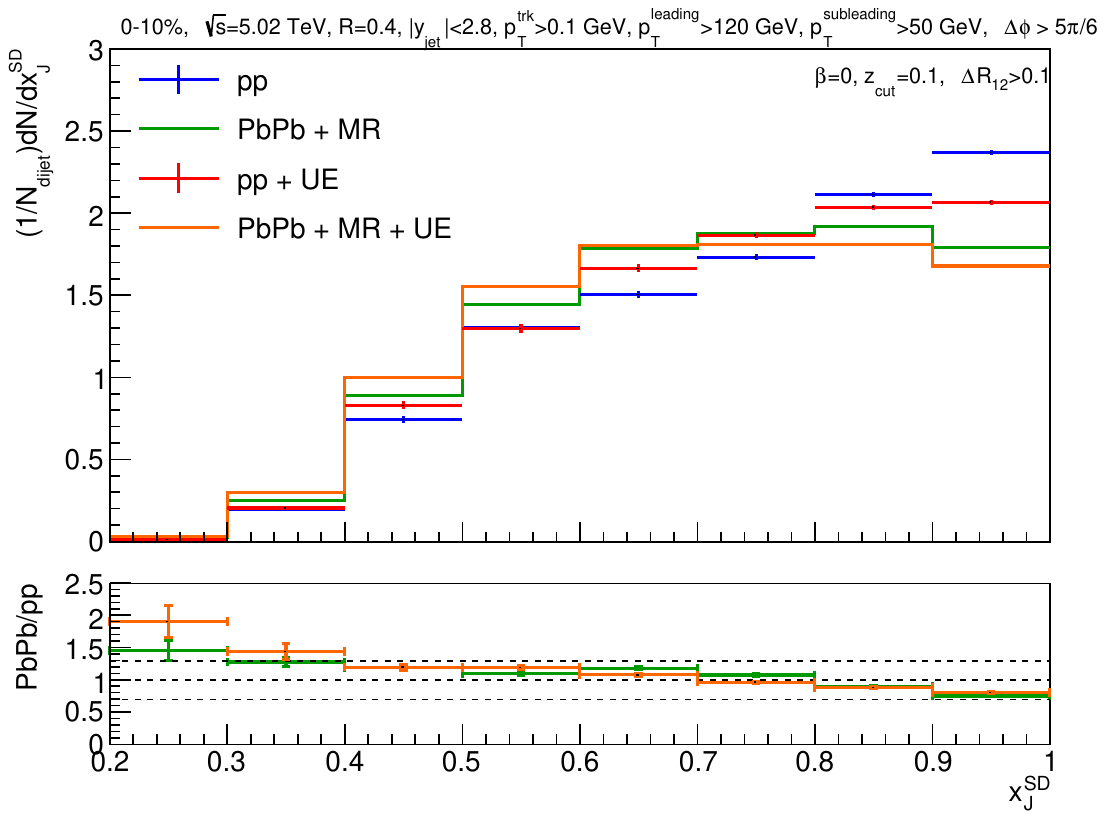}
          \caption{}
         \label{sfig:xjg}
     \end{subfigure}
     \caption{Transverse momentum fraction $x_j$ for \textbf{pp} (blue), \textbf{pp + UE} (red), \textbf{PbPb + MR} (green), and \textbf{PbPb + MR + UE} (orange). Ratios of \textbf{PbPb + MR} to \textbf{pp} without (green) and with (orange) UE are shown in the lower panels. (a) for ungroomed dijet pairs, and (b) for Soft Drop groomed dijet pairs.}
        \label{fig:xjall}
\end{figure}

The main effect of residual UE, see Fig.~\ref{sfig:xj}, is to increase the asymmetry by reducing the occurrence of nearly balanced pairs ($x_j \sim 1$) and thus broadening the distribution. The effect is more pronounced in the pp case where near balanced pairs are more abundant. The ratio of the distributions shown in the lower panel of Fig.~\ref{sfig:xj}, however, suffers only minor modifications. The more sizable differences are seen for very sparsely populated bins ($x_j < 0.3$) where statistical fluctuations are significant.

Computing the same observable for Soft Drop groomed jets, see Fig.~\ref{sfig:xjg}, the effect of UE contamination is reduced. Again, this is consistent with UE remnants being soft and consequently partially removed by grooming.

\subsubsection{Momentum dispersion}

The momentum dispersion $p_T^D$ measures the second moment of the $p_T$ distribution of jet constituents, reflecting the hardness of fragmentation within the jet. It is given by: 
\begin{equation}
    p_T^D = \frac{1}{p_T^{\mathsf{jet}}} \, \sqrt{\sum_k {\big(p_T^{(k)}\big)}^2}\, , 
\end{equation}
where $p_T^{\mathsf{jet}}$ is the total transverse momentum of the jet, $p_T^{(k)}$ is the transverse momenta of constituent $k$, and the sum runs over all jet constituents. 

\begin{figure}[!htbp]
     \centering
     \begin{subfigure}[b]{0.49\textwidth}
         \centering
         \includegraphics[width=\textwidth, angle=0]{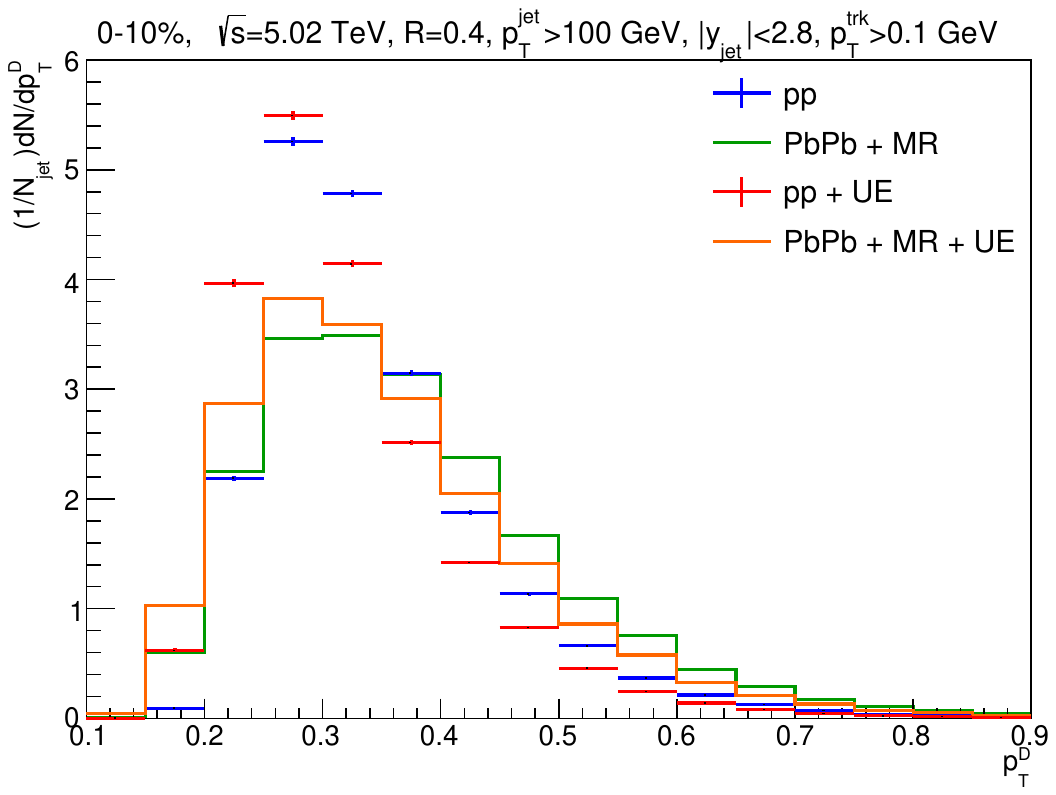}
         \caption{}
         \label{sfig:ptd}
     \end{subfigure}
     \hfill
     \begin{subfigure}[b]{0.49\textwidth}
         \centering
         \includegraphics[width=\textwidth, angle=0]{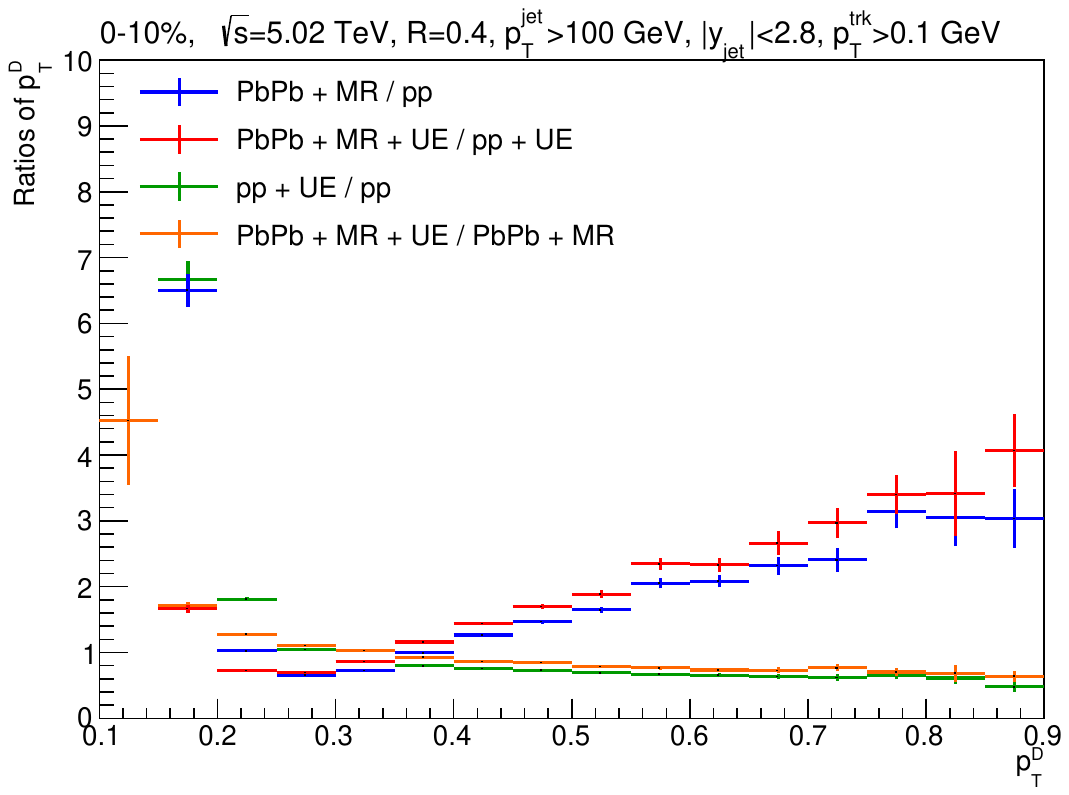}
          \caption{}
         \label{sfig:ptdratio}
     \end{subfigure}
     \caption{Transverse momentum dispersion $p_T^D$. (a) distributions for \textbf{pp} (blue), \textbf{pp + UE} (red), \textbf{PbPb + MR} (green), and \textbf{PbPb + MR + UE} (orange); (b) ratios of \textbf{PbPb + MR} to \textbf{pp} distributions without (blue) and with (red) UE, and the effect of UE contamination for \textbf{pp} (green) and \textbf{PbPb + MR} (orange) separately.}
        \label{fig:momdisp}
\end{figure}

The results, in Fig.~\ref{sfig:ptd}, show a remarkable insensitivity to UE contamination. Importantly, the effect is very similar in pp and PbPb and, consequently, the ratio of the $p_T^D$ distributions (Fig.~\ref{sfig:ptdratio}) is even more robust to UE contamination than the individual distributions.

A common feature for observables we found to be robust to UE contamination, including the jet fragmentation function shown in App.~\ref{app:fragfunc},  is that they do not involve any angular information, depending exclusively on transverse momenta.
The similarity of the modifications due to UE in both pp and PbPb leads to a cancellation that renders the observables, in particular when PbPb to pp ratios are considered, insensitive to the UE that remains after subtraction.

\subsection{Observables that are sensitive to UE contamination}
\label{subsec:sensitiveobservables}

\subsubsection{Jet profile}

The jet profile, the distribution of transverse momentum around the jet axis, is given by:
\begin{equation}
    \rho(\Delta r) = \frac{1}{p_T^{\mathsf{jet}}} \, \sum_{k: \Delta_{kJ} = \Delta r}\, p_T^{(k)} \, ,
\end{equation}
where $p_T^{\mathsf{jet}}$ is the total transverse momentum of the jet, $p_T^{(k)}$ is the transverse momenta of constituent $k$, and the sum runs over all charged jet constituents, with constituents whose distance to the jet axis $\Delta_{kJ}$ is within $\Delta r = [r,r+\delta r]$ contributing to the value of the distribution at $r$. 

\begin{figure}[!htbp]
     \centering
     \begin{subfigure}[b]{0.49\textwidth}
         \centering
         \includegraphics[width=\textwidth]{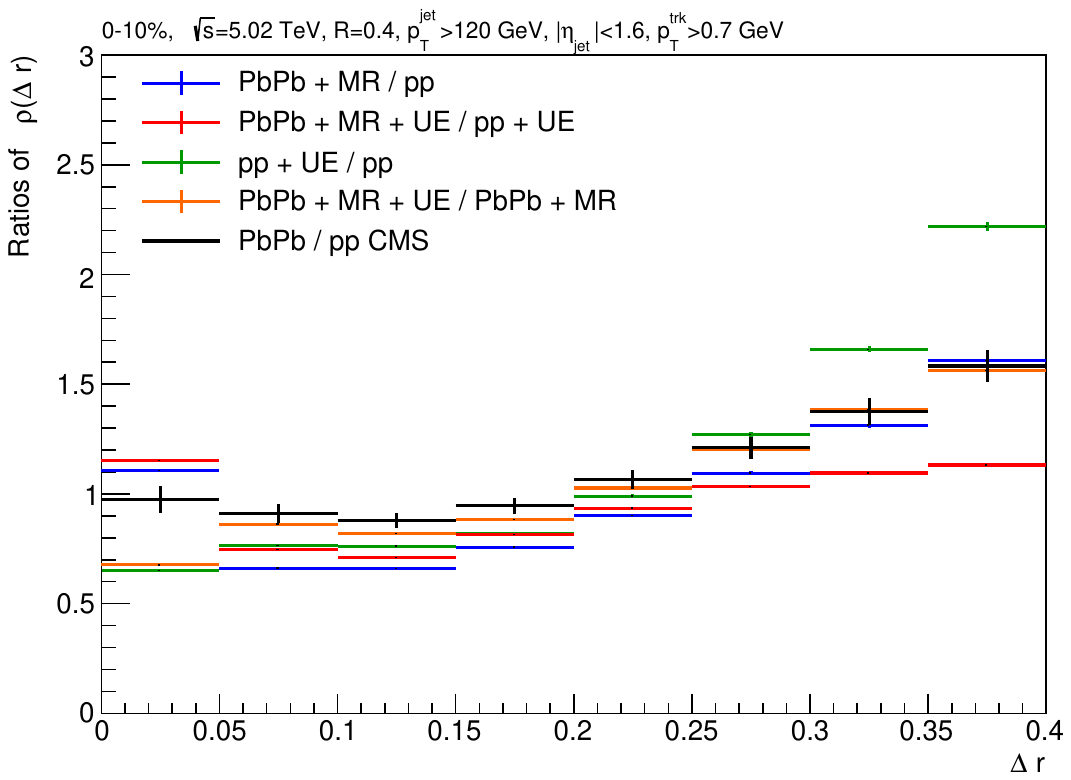}
         \caption{}
         \label{sfig:rho}
     \end{subfigure}
     \\
     \begin{subfigure}[b]{0.49\textwidth}
         \centering
         \includegraphics[width=\textwidth]{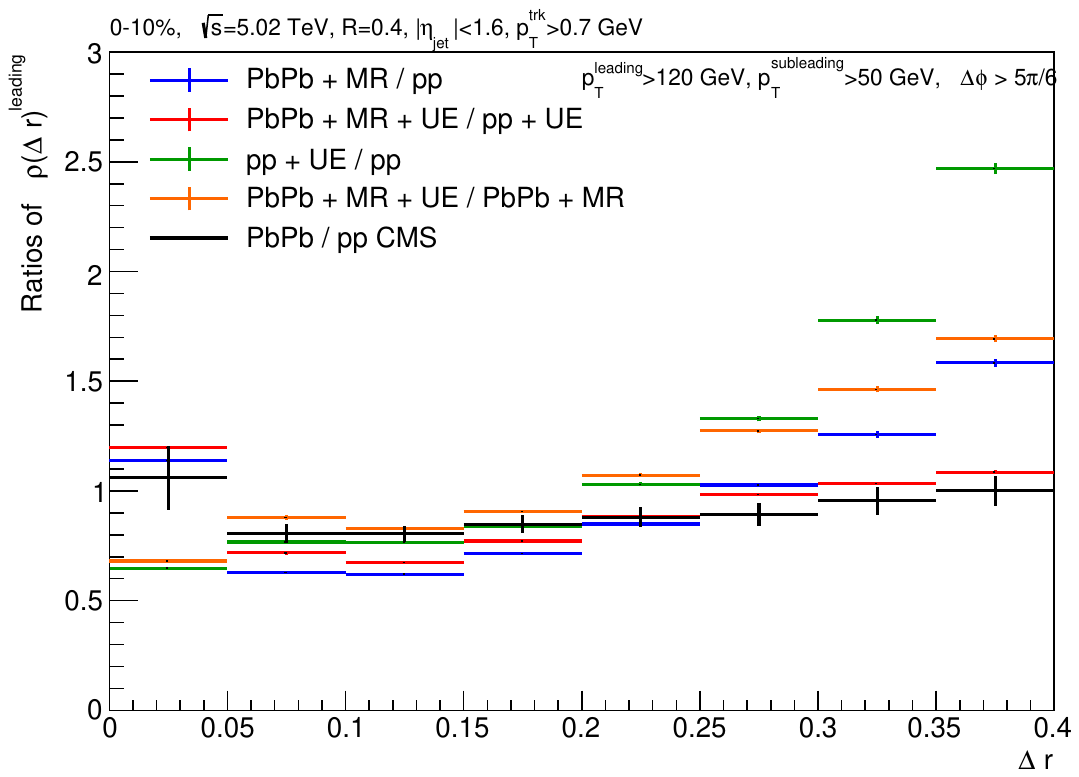}
          \caption{}
         \label{sfig:rhol}
     \end{subfigure}
     \hfill
     \begin{subfigure}[b]{0.49\textwidth}
         \centering
         \includegraphics[width=\textwidth]{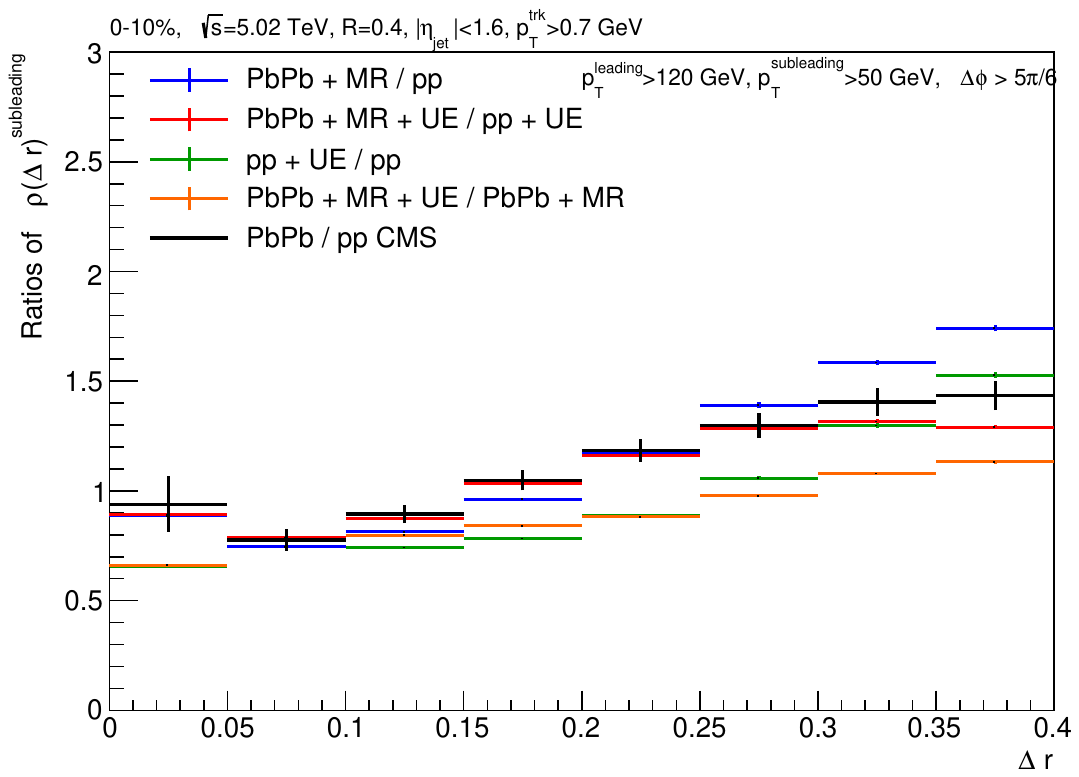}
          \caption{}
         \label{sfig:rhos}
     \end{subfigure}
     \caption{Ratios of \textbf{PbPb + MR} to \textbf{pp} jet profiles, without UE (blue) and with UE (red), and the effect of UE contamination for \textbf{pp} (green) and \textbf{PbPb + MR} (orange) separately. (a) inclusive jets, experimental data (black) from \cite{CMS:2018zze}; (b) leading jet in a dijet pair, experimental data (black) from \cite{CMS:2021nhn}; and (c) subleading jet in a dijet pair, experimental data (black) from \cite{CMS:2021nhn}.}
        \label{fig:jetprofile}
\end{figure}

We start by noting that \jewel\ results for inclusive jets, Fig.~\ref{sfig:rho}, without UE contamination are compatible with experimental data \cite{CMS:2018zze} for large $\Delta r$ and in less agreement for low and intermediate $\Delta r$.
UE contamination slightly improves agreement at low and intermediate $\Delta r$, while severely worsening it at large $\Delta r$. The jet profile shows a significant sensitivity to UE contamination, making the identification of jet modification due to interaction with the QGP ill-defined. 

When leading and subleading jets in dijet pairs are considered separately, Fig.~\ref{sfig:rhol} and Fig.~\ref{sfig:rhos}, the agreement with experimental data \cite{CMS:2021nhn} when UE is not included is much worse overall. While agreement improves with the inclusion of UE contamination, the interpretation of the observed modifications as due to interaction with the QGP remains problematic. 

\subsubsection{Energy-Energy Correlators}
\label{subsubsection:eec}

The Energy-Energy Correlator (EEC) \cite{Basham:1979gh,Basham:1978bw,Sveshnikov:1995vi,Dixon:2019uzg,Lee:2022ige,Chen:2023zzh,Komiske:2022enw} has been the focus of several recent studies aimed at identifying QGP-induced jet modifications \cite{Andres:2022ovj,Andres:2023xwr,Yang:2023dwc,Barata:2023bhh,Barata:2023zqg,Andres:2024ksi,Bossi:2024qho,Andres:2024pyz,Andres:2024hdd,Barata:2024ieg}. For our purposes we compute the normalized EEC through:
\begin{equation}
    EEC(\theta) = \frac{1}{N_{pairs}} \sum_{\mathsf{jets}}\,\sum_{i \neq j}\, \frac{p_T^i p_T^j}{p_{T,\,\mathsf{jet}}^2}\, \delta(\theta - \theta_{ij})\, ,
\end{equation}
where $\theta_{ij} = \sqrt{(y_i - y_j)^2 + (\phi_i - \phi_j)^2}$ is the angular distance between the directions of $p_T^i$ and $p_T^j$ in a jet and $p_{T,\,\mathsf{jet}}$ is the total transverse momentum of the jet.

\begin{figure}[!htbp]
    \centering
    \includegraphics[width=.7\textwidth]{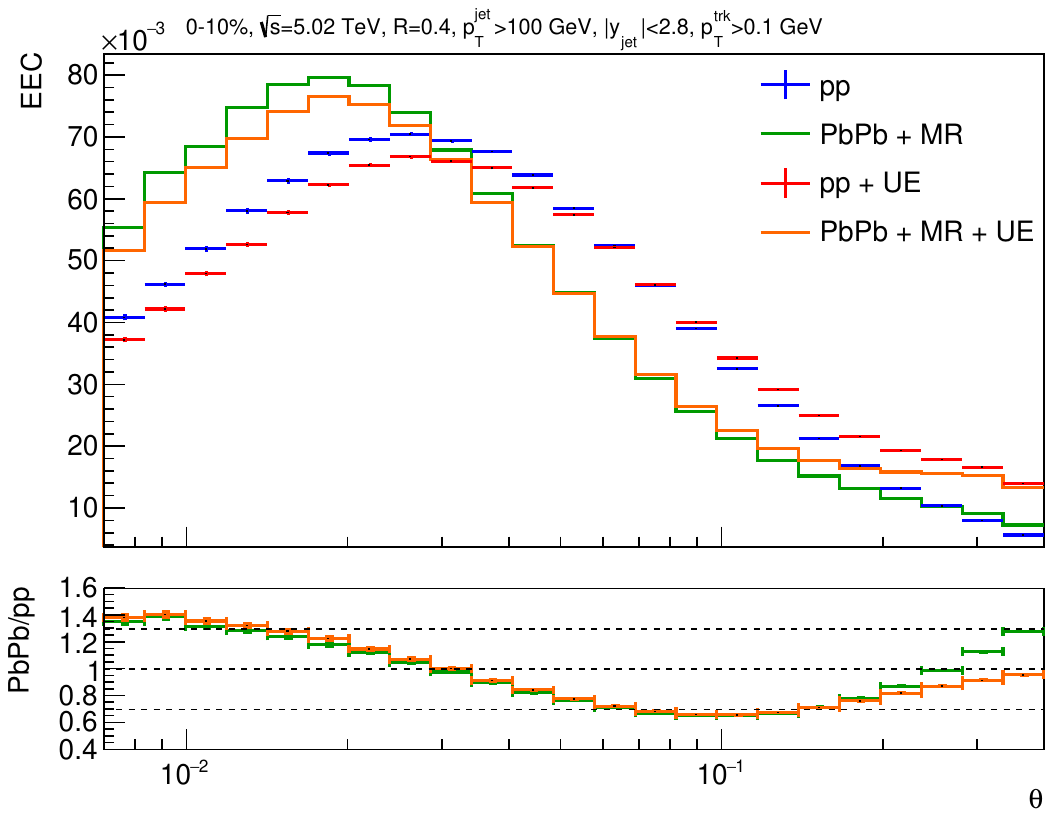}
    \caption{EEC for \textbf{pp} (blue), \textbf{pp + UE} (red), \textbf{PbPb + MR} (green), and \textbf{PbPb + MR + UE} (orange). Ratios of \textbf{PbPb + MR} to \textbf{pp} without (green) and with (orange) UE are shown in the lower panels.} 
    \label{fig:eec}
\end{figure}

The sensitivity of the EEC to UE contamination, shown in Fig.~\ref{fig:eec}, is restricted to the large angle region which in the absence of a QGP is fully determined from perturbative QCD \cite{Komiske:2022enw}. The effect of UE contamination is larger in the pp sample, effectively reducing the enhancement due to medium response. These features of the EEC, being robust across most of its domain and sensitive where all QGP induced effects and UE overlap, make it very suitable for the cross check on the ability to discriminate quenched and unquenched jets carried out in Sec.~\ref{subsec:crosscheck}.

\subsubsection{Lund jet planes}
\label{subsubsec:lund}

Lund jet planes \cite{Dreyer:2018nbf} are constructed as follows.
Each jet is reclustered using the Cambridge-Aachen (C/A) algorithm which, at each clustering step, recombines the pair of pseudo-jets $i$ and $j$ (final state particles are taken as the initial set of pseudo-jets) with the smallest rapidity-azimuth distance $\Delta_{ij} = \sqrt{(y_i-y_j)^2 + (\phi_i -\phi_k)^2}$ into a pseudo-jet with momentum $p = p_i +p_j$. The procedure is repeated until a single pseudo-jet (the reclustered jet) remains.
The primary Lund plane is then generated by undoing, step-by-step, the clustering sequence. Starting with the full jet, 
one clustering step is undone to obtain pseudo-jets $a$ and $b$ with $p_{T,a} > p_{T,b}$. The Lund coordinates $\Delta \equiv \Delta_{ab}$ (the rapidity-azimuth separation of the two pseudo-jets) and $k_t \equiv p_{T,b}\Delta_{ab}$ (the relative transverse momentum of the two pseudo-jets) are recorded. The procedure is repeated for the harder branch $a$ until the products of the declustering are final state particles.
The entire procedure was carried out using the \textsf{LundGenerator} from \textsf{fastjet-contrib}.

\begin{figure}[!htbp]
    \centering
    \begin{subfigure}[b]{0.48\textwidth}
        \centering
        \includegraphics[width=\textwidth]{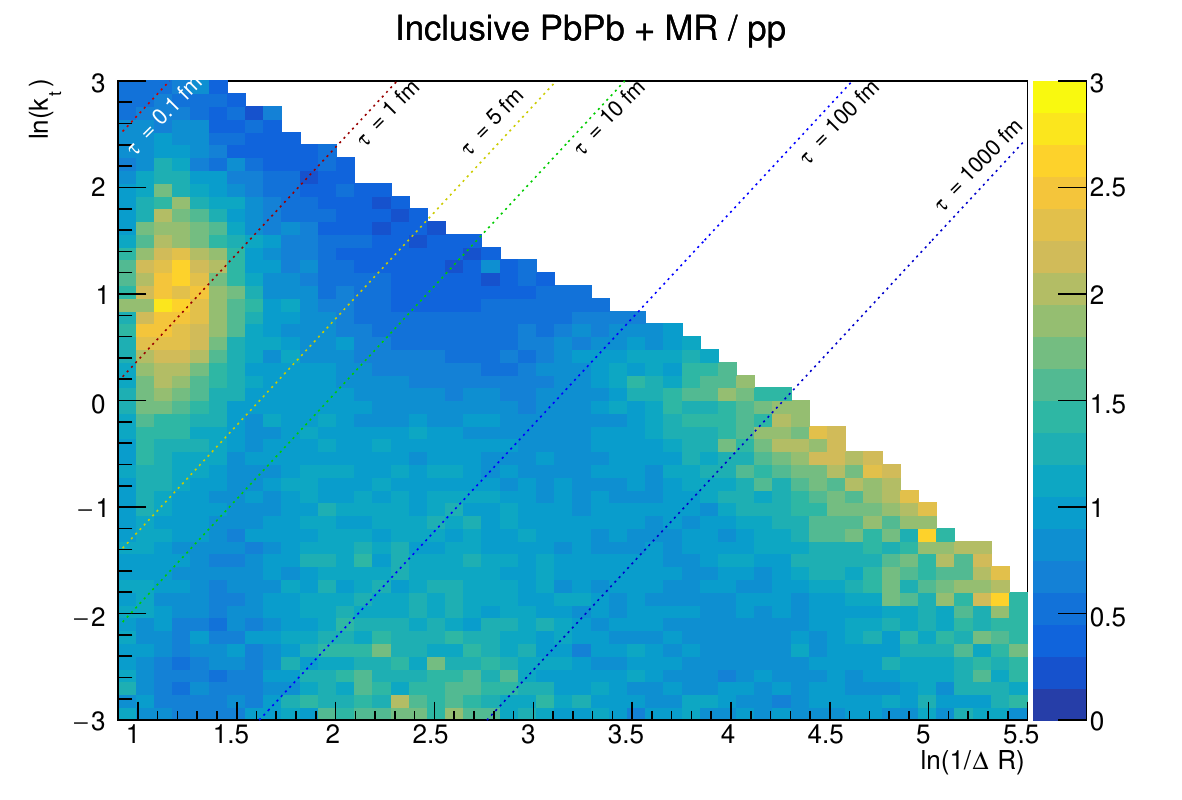}
         \caption{}
        \label{sfig:lund_gen}
    \end{subfigure}
     \hfill
    \begin{subfigure}[b]{0.48\textwidth}
        \centering
        \includegraphics[width=\textwidth]{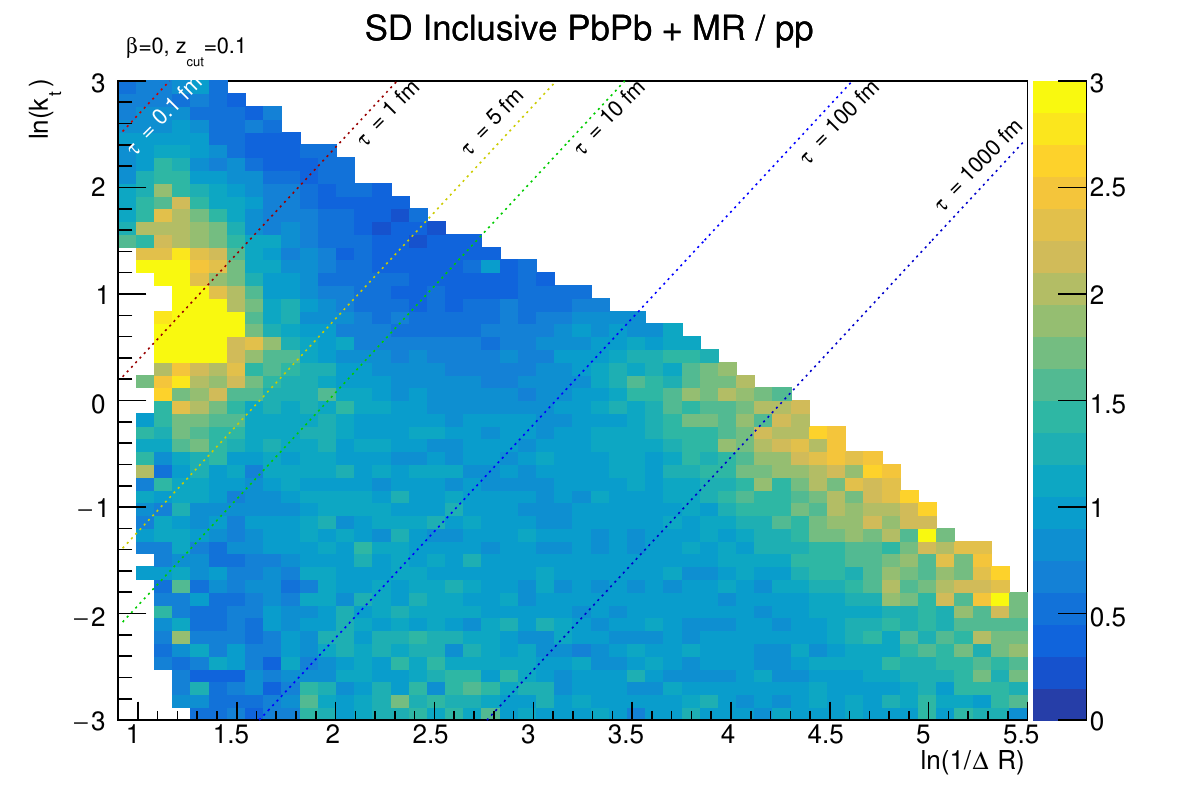}
         \caption{}
        \label{sfig:lund_gen_sd}
    \end{subfigure}
    \\
    \begin{subfigure}[b]{0.48\textwidth}
        \centering
        \includegraphics[width=\textwidth]{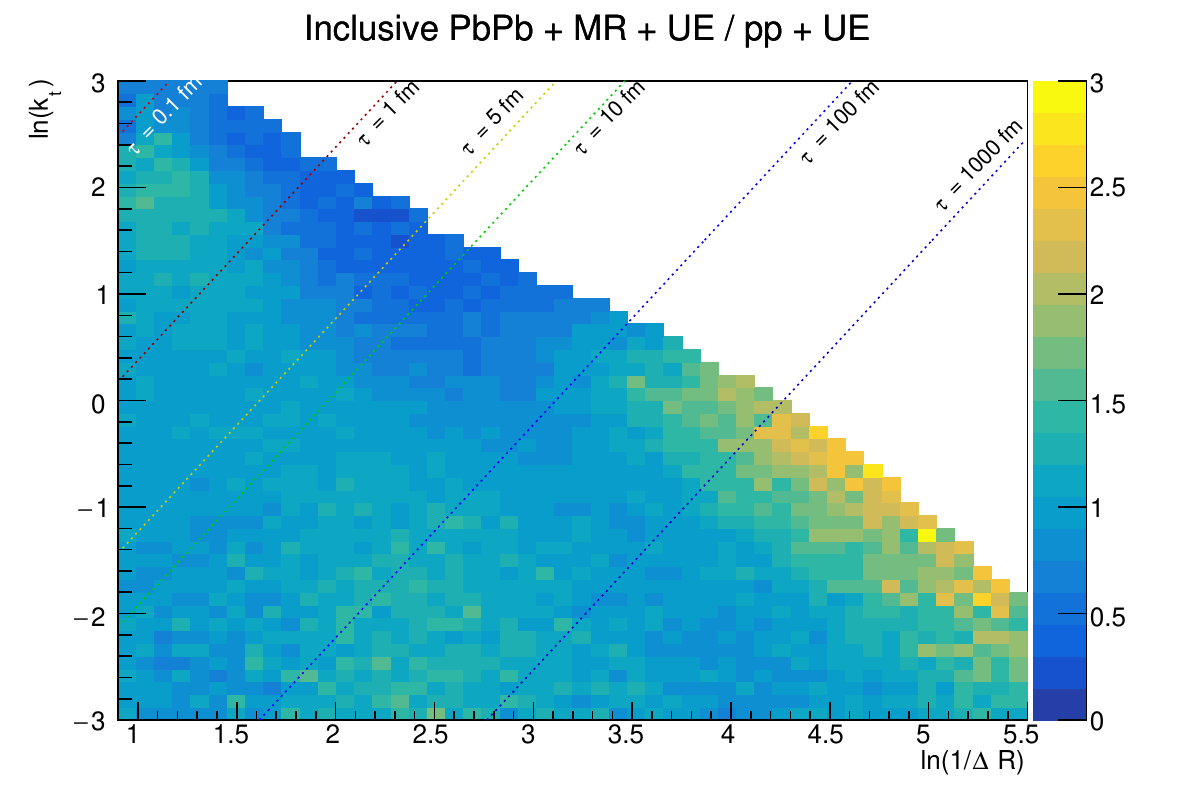}
         \caption{}
        \label{sfig:lund_sub}
    \end{subfigure}
     \hfill
    \begin{subfigure}[b]{0.48\textwidth}
        \centering
        \includegraphics[width=\textwidth]{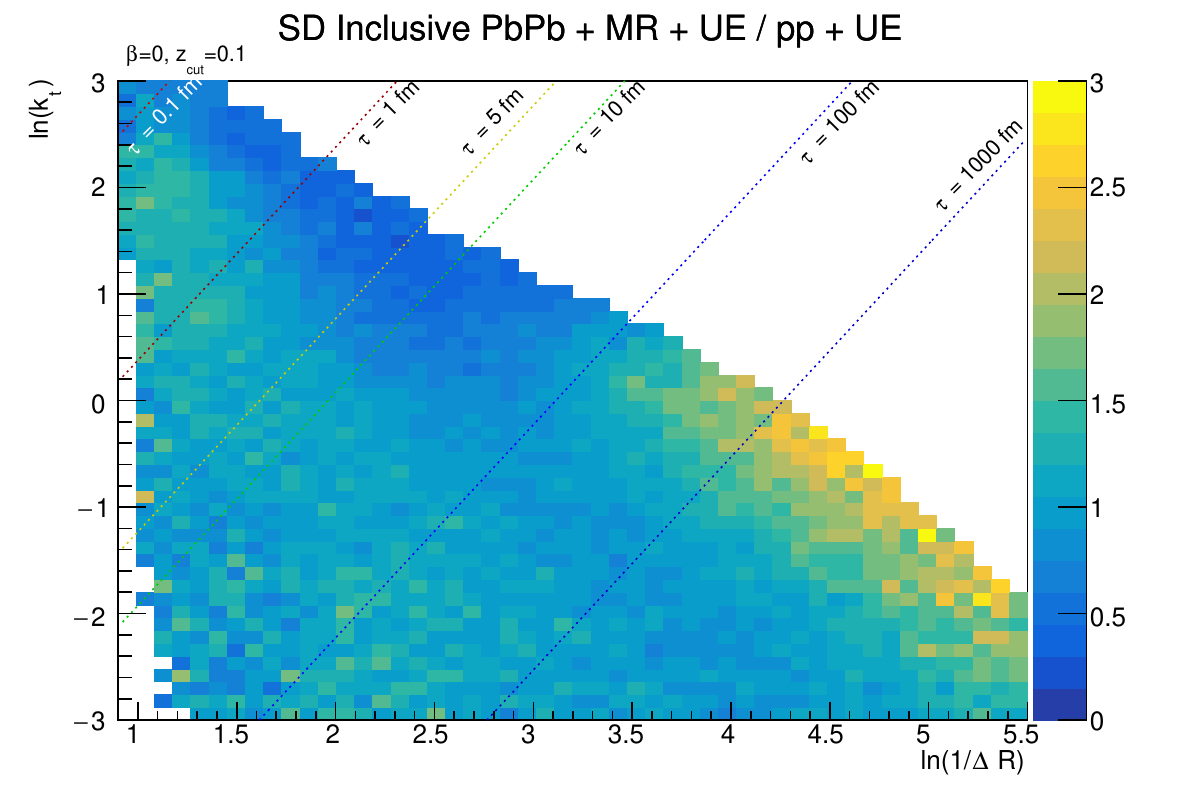}
         \caption{}
        \label{sfig:lund_sub_sd}
    \end{subfigure}
    \caption{Ratios of PbPb to pp primary Jet Lund planes. (a) \textbf{PbPb + MR} to \textbf{pp} for ungroomed jets; (b) \textbf{PbPb + MR} to \textbf{pp} for Soft Drop groomed jets; (c) \textbf{PbPb + MR + UE} to \textbf{pp + UE} for ungroomed jets; and (d) \textbf{PbPb + MR + UE} to \textbf{pp + UE} for Soft Drop groomed jets.}
    \label{fig:lundR}
\end{figure}

Fig.~\ref{fig:lundR} shows ratios between the PbPb and pp primary Lund jet planes for inclusive jets without and with the effect of UE contamination, for both ungroomed and Soft Drop groomed jets.
When UE is not included, Fig.~\ref{sfig:lund_gen}, a prominent enhancement at large angles and $k_T \sim 2.5\,$GeV, is present. This enhancement, which is due to MR, is even more prominent when jets are groomed using Soft Drop (Fig.~\ref{sfig:lund_gen_sd}). This seemingly counterintuitive feature arises from the incomplete removal of MR by the applied SD cut ($\beta = 0\, , z_{\text{cut}=0.1}$). Since the ratio is built from self-normalised Lund Planes (densities), the relative contribution of the MR contribution that was not groomed away increases.

When UE contamination is accounted for, the enhancement is no longer present as shown in Fig.~\ref{sfig:lund_sub} and Fig.~\ref{sfig:lund_sub_sd}. 
This signals the sensitivity of the moderate $k_t$ large angle region of the primary Lund jet plane to UE contamination for both ungroomed and Soft Drop groomed jets. The observable is otherwise robust. 
Analogous results, shown in App.~\ref{app:jetobs}, are obtained when leading and subleading jets in a dijet pair are considered separately.

\section{Robustness of Machine Learning based studies to UE contamination}
\label{sec:mlrobustness}

As outlined in the previous section, the effect of UE contamination in several commonly used jet observables is to mimic effects due to medium response. 
It is then conceivable that ML studies, particularly if using those observables as input to distinguish quenched jets from their unmodified vacuum counterparts, may identify as relevant features spurious effects due to UE contamination.
A commonly used setup for proof-of-principle studies \cite{Apolinario:2021olp,CrispimRomao:2023ssj} is to not include either medium response or underlying event. 
In such cases, all discrimination relies solely on modifications of the parton shower due to interaction with QGP. 
While these provide an extremely valuable insight, that ML architectures can identify the effect of QGP in parton branching. Their portability to analyses of experimental data is, at best, questionable.
Some studies \cite{Du:2020pmp,Du:2021pqa} have also included medium response which significantly improves discrimination power. This result can be easily understood as the presence of medium response in quenched jets, and its obvious absence in pp samples, serves as a powerful distinguishing feature that any \textit{bona fide} ML architecture ought to identify. The portability of such results to experimental analyses is also highly problematic.
Jets in experimental data will always contain some UE contamination which supresses the discrimination power afforded by the presence of medium response.

The procedure we put forward in this paper, of establishing robust quenching effects by comparing PbPb samples that include both medium response and UE contamination with pp samples that also include PbPb UE contamination, forms the basis for ML studies that can eventually be ported to experimental analyses.

To assess how UE contamination affects ML studies we have taken as reference the analyses carried out in \cite{CrispimRomao:2023ssj} where neither medium response nor UE were included. That study considered as input a set of 31 per-jet observables: global properties, angularities \cite{Larkoski:2014pca}, N-Subjettiness \cite{Thaler:2010tr}, jet charges \cite{Krohn:2012fg}, and SoftDrop \cite{Larkoski:2014wba} and dynamical grooming \cite{Mehtar-Tani:2019rrk} specific observables.

A first study identified both linear (using a Principal Component Analysis) and non-linear (using a Deep Auto-Encoder) relations between observables with the aim of finding a minimal set of degrees of freedom that fully accounted for the information content of the entire set of observables. The main findings can be summarized as: (i) subsets of observables are highly mutually correlated and consequently the information content of the full set can be described by a small number of effective degrees of freedom, smaller when non-linear relations are accounted for; and (ii) that most relations between observables are not modified by quenching effects.

A second study assessed the ability of a Boosted Decision Tree (BDT) to discriminate, on the basis of the full set of observables, between quenched and unquenched jets. Further implementation details can be found in \cite{CrispimRomao:2023ssj}.

\subsection{Relations between observables -- PCA and AE}

We start by addressing the effect of UE on the ability to find a reduced number of effective degrees of freedom that account for the information content of the full set of observables.

A Principal Component Analysis (PCA) identifies the main directions of the dataset, that is the linear combinations of the set of 31 observables that best explain the distributions of values measured for the observables, by minimizing the reconstruction error
\begin{equation}
    \min_{\mathbf{V}} \mathbb{E}[\| x - \mathbf{V}\cdot \mathbf{V}^T \cdot x \|^2] \, ,
    \label{eq:pca_rec_error}
\end{equation}
where $x$ is the vector of the observables computed for jet $i$, $\mathbf{V}$ is a rectangular matrix built by stacking together a fixed number of principal components (the main directions) represented as column vectors, and $\mathbb{E}[x]=\sum_i^{N_{jet}} w_i x_i/ \sum_i^{N_{jet}} w_i$ is the weighted expected value taken over the entire dataset accounting for event generation weights. 
When all the principal components are used to build $\mathbf{V}$, one has $\mathbf{V}\cdot \mathbf{V}^T=\mathbf{1}$ and the reconstruction error trivially vanishes.
The extent to which a fixed number of principal components, smaller than the total number of observables, can explain the entire dataset can be quantified by the coefficient of determination $R^2$
\begin{equation}
    R^2(x, \hat x) = 1 - \frac{\mathbb{E} [\| x - \hat x \|^2]}{\mathbb{E}[\| x - \mathbb{E}[x]\|^2]} \, ,
    \label{eq:r2}
\end{equation}
where $x$ are the observables vectors, and $\hat x = \mathbf{V} \cdot \mathbf{V}^T\cdot x$ is the reconstructed $x$ after being rotated into the principal components and back. This metric measures how well the observables are reconstructed after being projected into the principal components and back to the original basis, therefore quantifying how much information was retained.
It equals $0$ when the reconstruction only reproduces the sample average value for each observable, and is $1$ when the value of each observable for each individual jet is reproduced accurately. 

The analysis using a Deep Auto-Encoder ($AE$) goes beyond PCA in that it also captures non-linear relations between observables. This neural network architecture attempts to minimize the loss function, analogous to Eq.~\ref{eq:pca_rec_error}, 
\begin{equation}
    \min_{\mathbf{w}} \mathbb{E}[\| x - AE(x, \mathbf{w}) \|^2] \, ,
    \label{eq:ae_loss}
\end{equation}
where $w$ are the trainable parameters of the $AE$ neural network, and $x$ are the input data.
The $AE$ learns how to project the data into a lower dimensional space, the latent space $z$ with dimension $z_{dim}$ and then reconstructs the inputs back to their original form.
While the minimization of the error function Eq.~(\ref{eq:pca_rec_error}) for the PCA finds the optimal orthogonal rotation of the observable basis, the loss function Eq.~(\ref{eq:ae_loss}) finds the optimal non-linear map implicit in the $AE$. The reconstruction quality is quantified by the coefficient of determination given in Eq.~(\ref{eq:r2}), with $\hat x = AE(x, \mathbf{w})$.

\begin{figure}[!htbp]
    \begin{subfigure}[b]{0.49\textwidth}
        \centering
        \includegraphics[width=\textwidth]{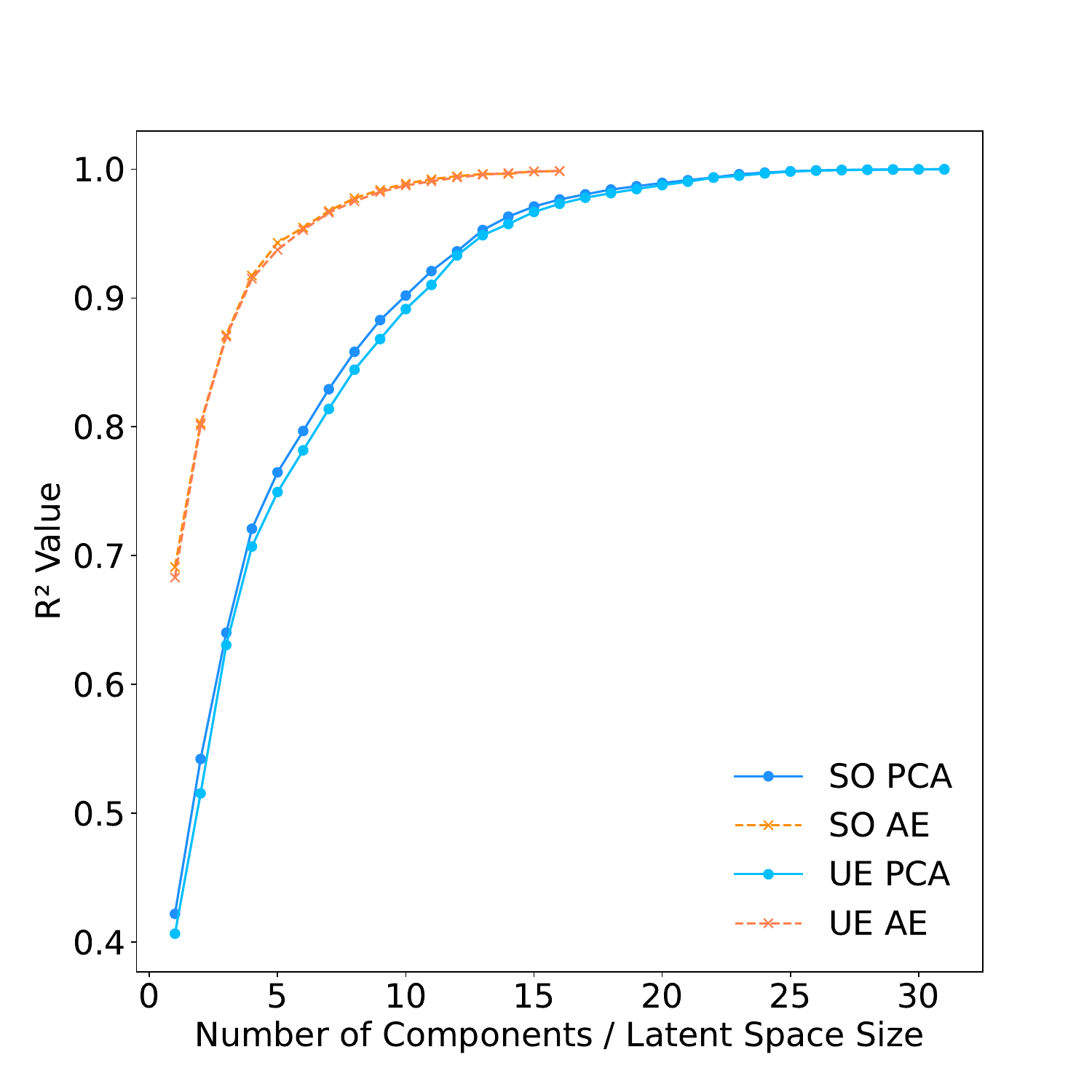}
        \caption{}
        \label{sfig:r2}
    \end{subfigure}  
    \hfill
    \begin{subfigure}[b]{0.49\textwidth}
        \centering
        \includegraphics[width=\textwidth]{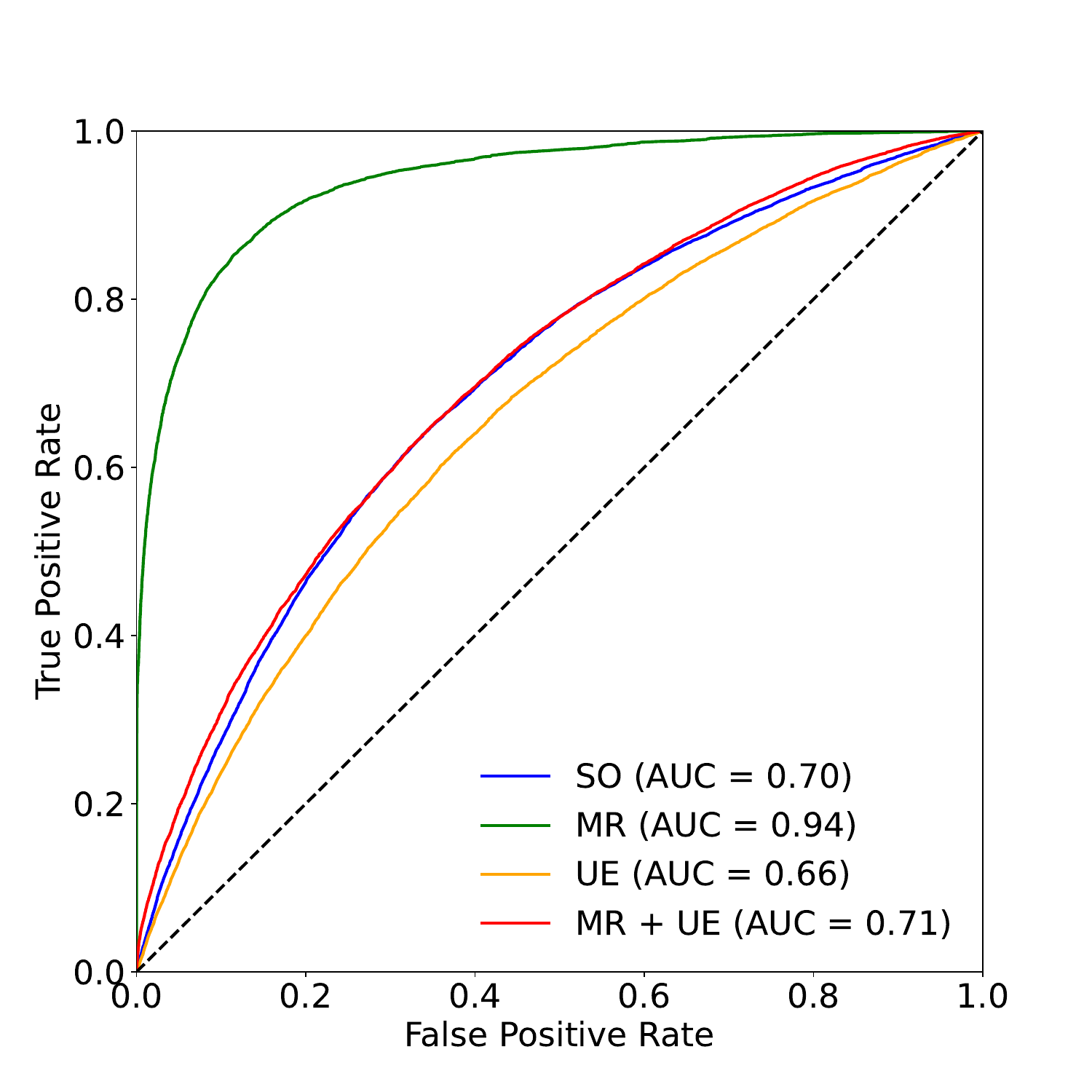}
        \caption{}
        \label{sfig:r2roc}
    \end{subfigure}    
    \caption{Performance metrics for both ML analysis, performed across all four cases. (a) \(R^2\) as a function of the number of included effective degrees of freedom (the number of principal components in the PCA analysis or the dimension of the latent space \(z_{dim}\) in the AE), for pp samples without UE (SO) and including it (UE); and (b) Receiver Operating Characteristic (ROC) curve for BDT analysis (right), where we denote the \textbf{pp} vs \textbf{PbPb + MR} as MR and the most realistic case as MR + UE.}
    \label{fig:mlperf}
\end{figure}

To assess the effect of UE contamination on correlations between observables we compare the results, for both the PCA and AE analyses, for pp samples without UE (\textbf{pp}) as considered in \cite{CrispimRomao:2023ssj} and including UE contamination (\textbf{pp+UE}). 
This comparison is shown in Fig.~\ref{sfig:r2}. 
As previously noted in \cite{CrispimRomao:2023ssj}, the ability of the AE to capture non-linear relations between observables leads to a significant reduction of the number of effective degrees of freedom (here the dimension $z_{dim}$ of the latent space) relative to the PCA case (number of principal components) needed to describe the full data set with a given quality.
The only effect of UE contamination is to (very) slightly reduce the reconstruction quality in the PCA case for an intermediate number of principal components. 
Overall, UE contamination does not seem to modify the underlying correlations between these observables.

\subsection{Discrimination of quenched and unquenched jets -- BDT}

We now discuss the effect of medium response and UE contamination, both separately and jointly, on the discrimination power of a BDT trained on the full set of observables. 
Fig.~\ref{sfig:r2roc} shows the receiver operating characteristic (ROC), which relates the true positive rate (fraction of signal jets that are identified as such) with the false positive rate (background jets erroneously identified as signal), for discrimination between pp and PbPb samples in different cases. The area under each of these curves (AUC), shown in Table~\ref{tab:roc}, quantifies how well the classifier performs in each case. A perfect classifier would have ROC AUC equal to 1, while a random classifier would have a ROC AUC of 0.5.

The case addressed in \cite{CrispimRomao:2023ssj}, in which the discrimination is done using samples without MR and UE for both pp and PbPb, is denoted Signal Only (SO). The ROC AUC of $0.697$ in that case serves as a baseline for the remaining discrimination cases.

\begin{figure}[!htbp]
    \centering
    \begin{subfigure}[b]{0.49\textwidth}
        \centering
        \includegraphics[width=\textwidth]{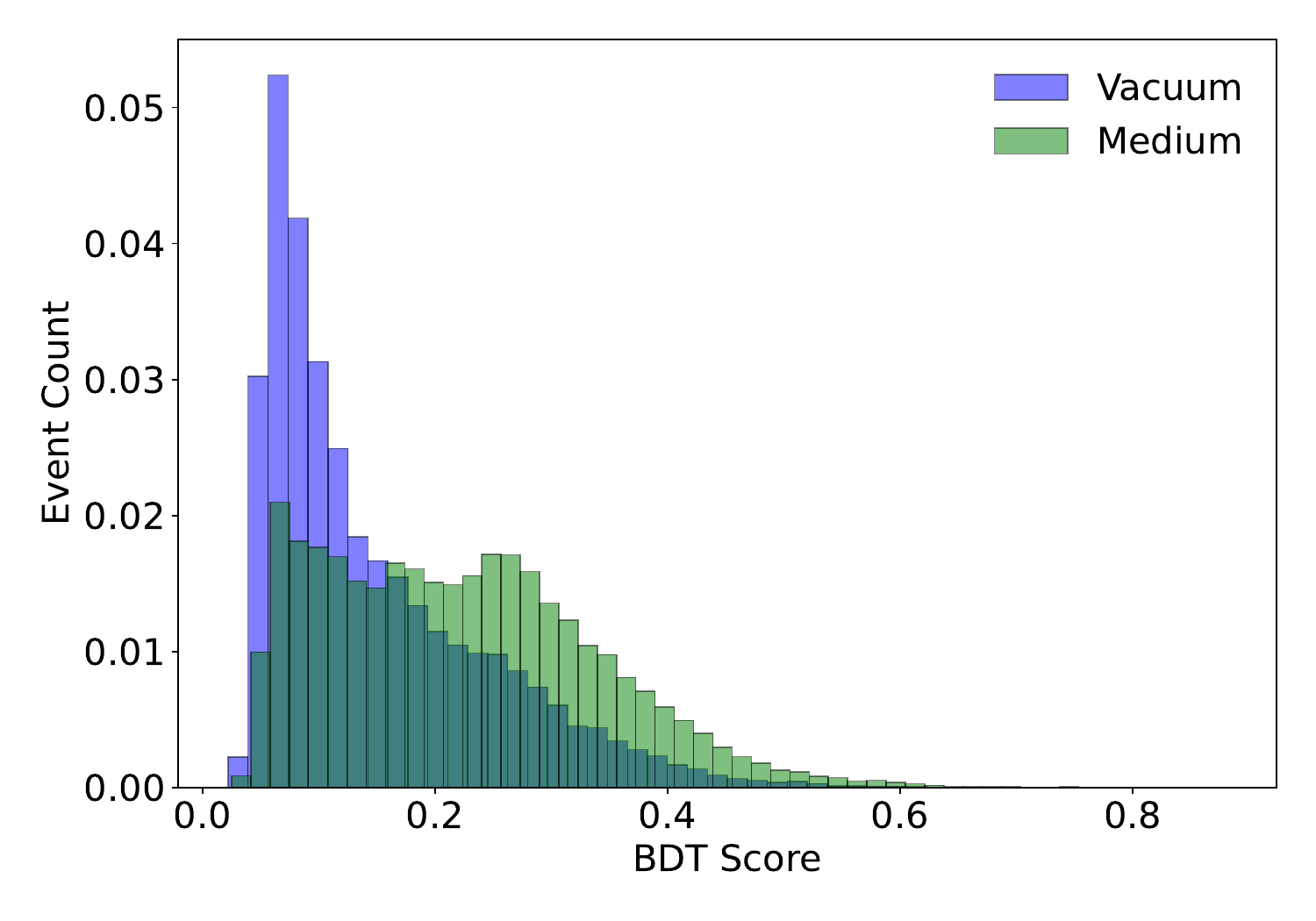}
         \caption{}
         \label{sfig:bdt_out_so}
    \end{subfigure}
    \hfill
    \begin{subfigure}[b]{0.49\textwidth}
        \centering
        \includegraphics[width=\textwidth]{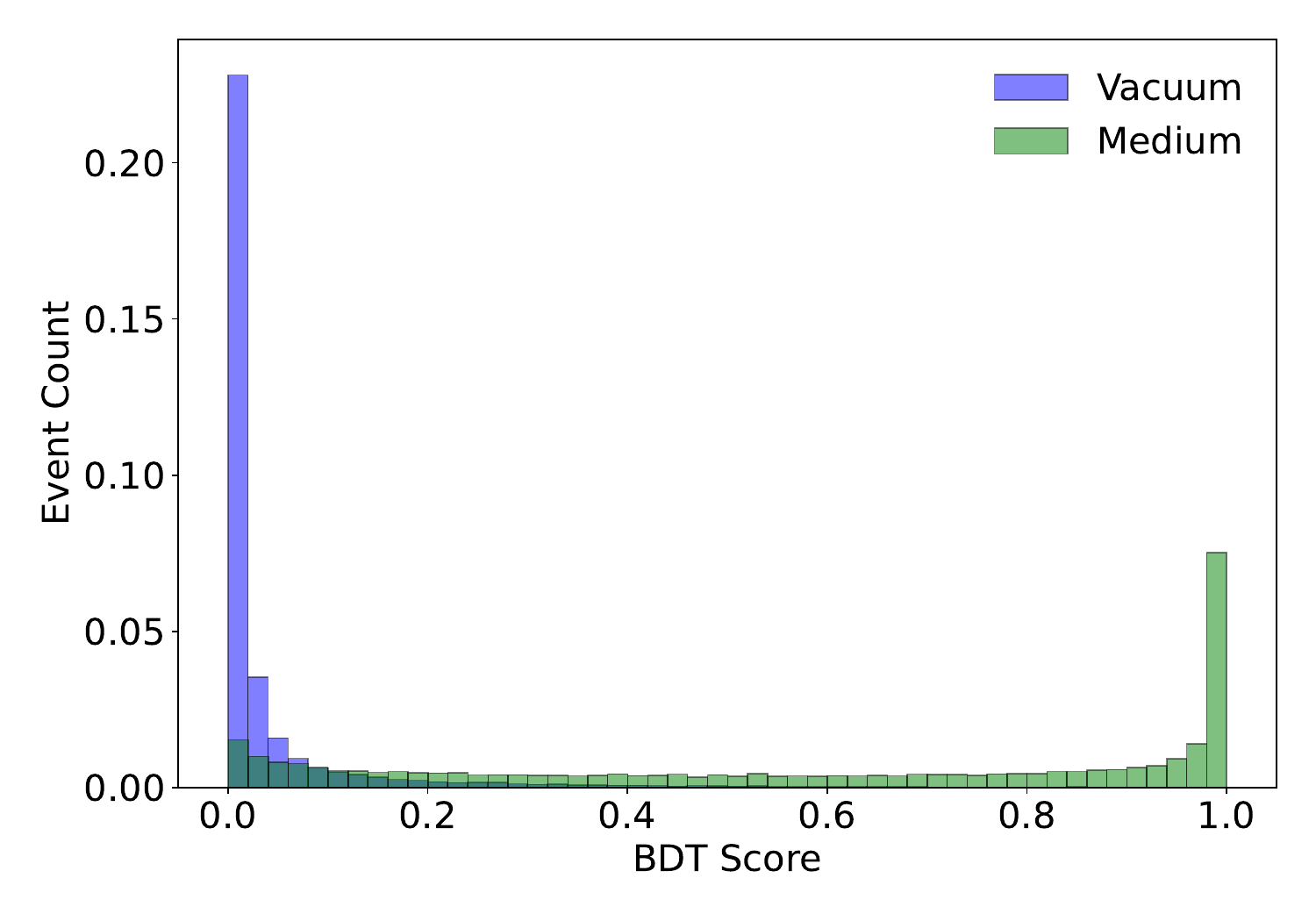}
        \caption{}
         \label{sfig:bdt_out_mr}
    \end{subfigure}
    \\
    \begin{subfigure}[b]{0.49\textwidth}
        \centering
        \includegraphics[width=\textwidth]{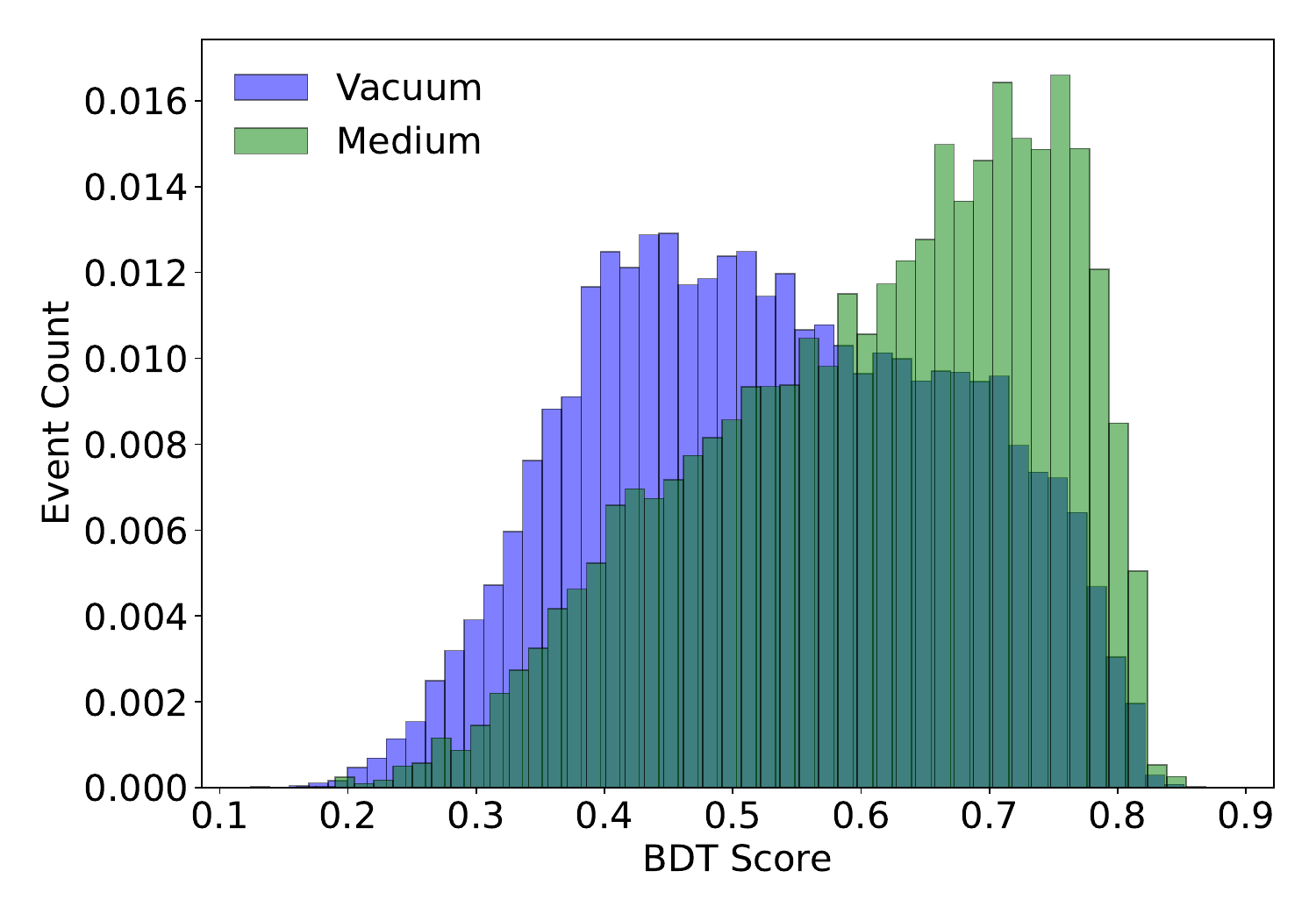}
        \caption{}
         \label{sfig:bdt_out_ue}
    \end{subfigure}
    \hfill
    \begin{subfigure}[b]{0.49\textwidth}
        \centering
        \includegraphics[width=\textwidth]{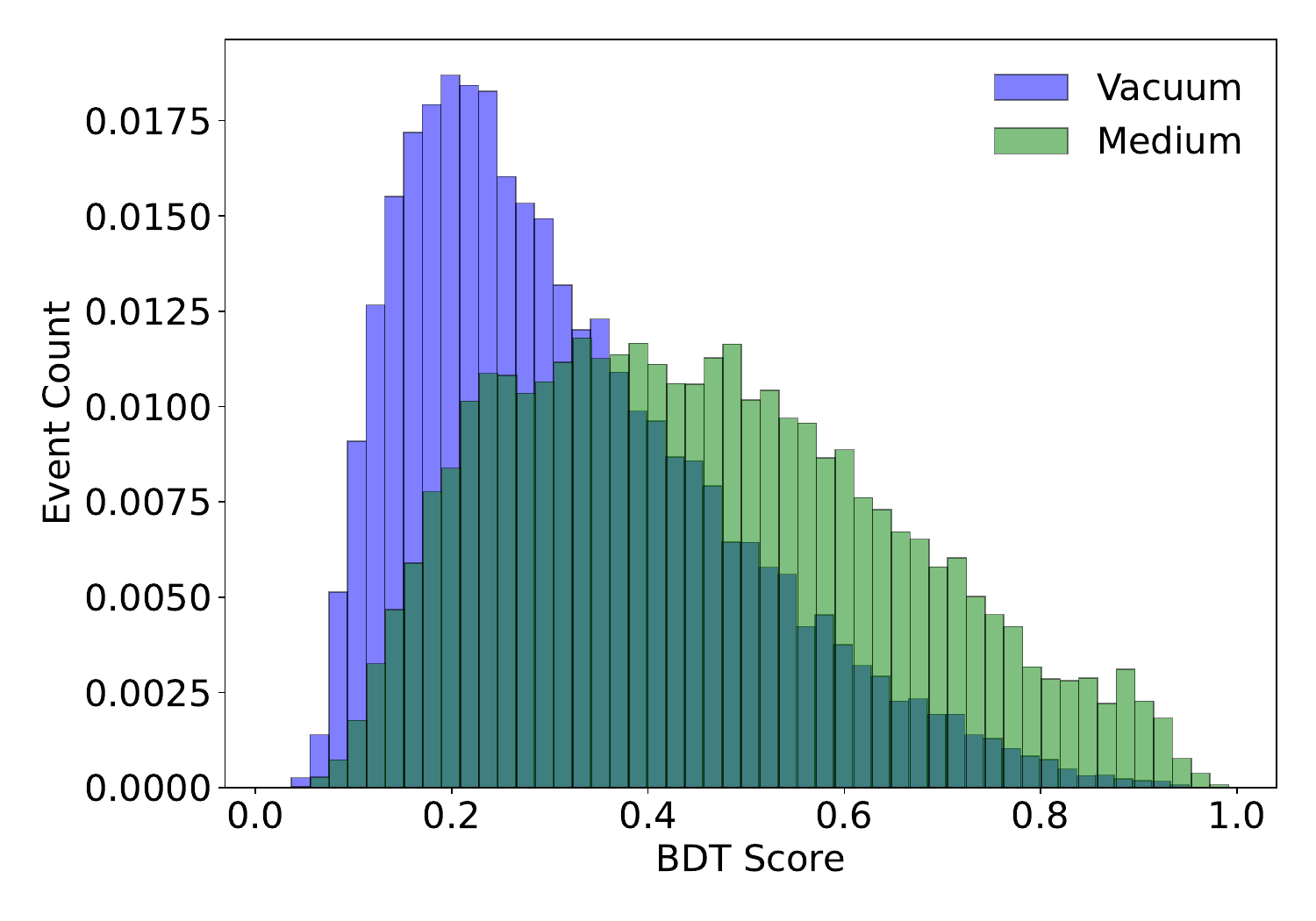}
        \caption{}
         \label{sfig:bdt_out_mrue}
    \end{subfigure}
\caption{BDT model output for vacuum (indigo) and medium (green) samples for: (a) Signal Only samples (\textbf{pp} vs \textbf{PbPb}); (b) including Medium Response (\textbf{pp} vs \textbf{PbPb + MR}); (c) including Underlying Event contamination (\textbf{pp + UE} vs \textbf{PbPb + UE}); (d) including both Medium Response and Underlying Event contamination (\textbf{pp + UE} vs \textbf{PbPb + UE + MR}).}
    \label{fig:modelout}
\end{figure}

When MR is included, that is when MR is accounted for in PbPb (\textbf{PbPb+MR}), the discrimination against the reference pp sample (\textbf{pp}) is almost perfect with a ROC AUC of $0.941$. In this case, see Fig.~\ref{sfig:bdt_out_mr}, the BDT output scores show the hallmark (two fully separated peaks at the extrema of output values) of the identification of a clear distinguishing feature in the data. The BDT has simply identified the presence of MR in one sample and its absence in the other, classifying the jets accordingly. Naively, this could be interpreted as a success if one did not recall that the case at hand is not portable to a realistic analysis where UE contamination is also necessarily present. 

Before addressing our most realistic case, the one where the discrimination is done between a PbPb sample with both MR and UE (\textbf{PbPb+MR+UE}) and a pp sample contaminated with PbPb UE (\textbf{pp+UE}), we discuss the effect of UE contamination alone. When samples including UE contamination are used the performance of the classifier is reduced to a ROC AUC of $0.660$ signaling that UE makes jets in pp and PbPb appear more alike. This is clearly seen by comparing Fig.~\ref{sfig:bdt_out_ue} where UE is included with Fig.~\ref{sfig:bdt_out_so}, where UE is not included.

The realistic discrimination power of the BDT classifier will then be the balance between the existing MR as a clear distinguishing feature and the smearing effect due to UE contamination. That the ROC AUC for the realistic case, \textbf{PbPb+MR+UE} vs. \textbf{pp+UE}, is larger ($0.707$) than the baseline case (SO), and smaller than the case where only MR (MR) is present, shows that some of the discriminative power of MR survives the confounding effect of UE contamination. The BDT output in this case, Fig.~\ref{sfig:bdt_out_mrue}, shows that for high score values a high purity sample of quenched jets could be potentially isolated.

The survival of MR has a distinguishing feature requires that the BDT has to be able to, at least to some extent, distinguish between MR and UE. We carried out such a cross-check resulting in a ROC AUC of $0.804$, as shown in Table~\ref{tab:roc}.

\begin{table}[!htbp]
\centering
\begin{tabular}{@{}lcccc@{}}
\toprule
 & PbPb & PbPb + MR & PbPb + UE & PbPb + MR + UE \\ 
\midrule
pp            & 0.697 & 0.941 & -.--- & -.--- \\
pp + UE       & -.--- & -.--- & 0.660 & 0.707 \\
PbPb + UE     & -.--- & 0.804 & -.--- & -.--- \\
\bottomrule
\end{tabular}
\caption{ROC AUCs for models trained with samples including different effects.}
\label{tab:roc}
\end{table}

\subsection{A cross-check}
\label{subsec:crosscheck}

An important check that must be carried out as part of any ML-based study is to assess whether the discrimination performed by the classifier is indeed between quenched and unquenched jets or whether the discrimination is based on some spurious feature unrelated to quenching.

\begin{figure}[!htbp]
    \centering
    \begin{subfigure}[b]{0.49\textwidth}
        \centering
        \includegraphics[width=\textwidth]{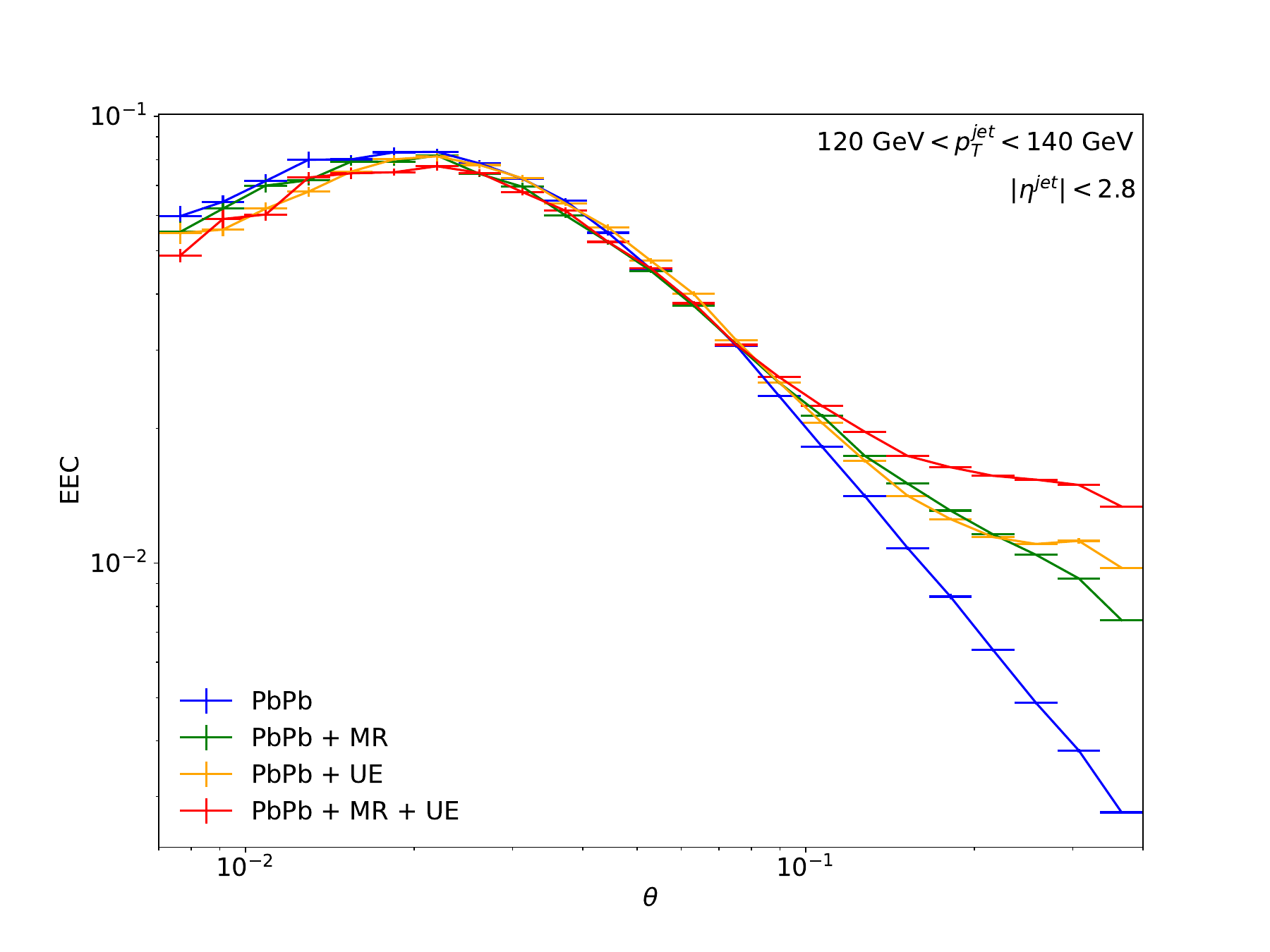}
        \caption{}
        \label{sfig:eec_med}
    \end{subfigure}
    \hfill
    \begin{subfigure}[b]{0.49\textwidth}
        \centering
        \includegraphics[width=\textwidth]{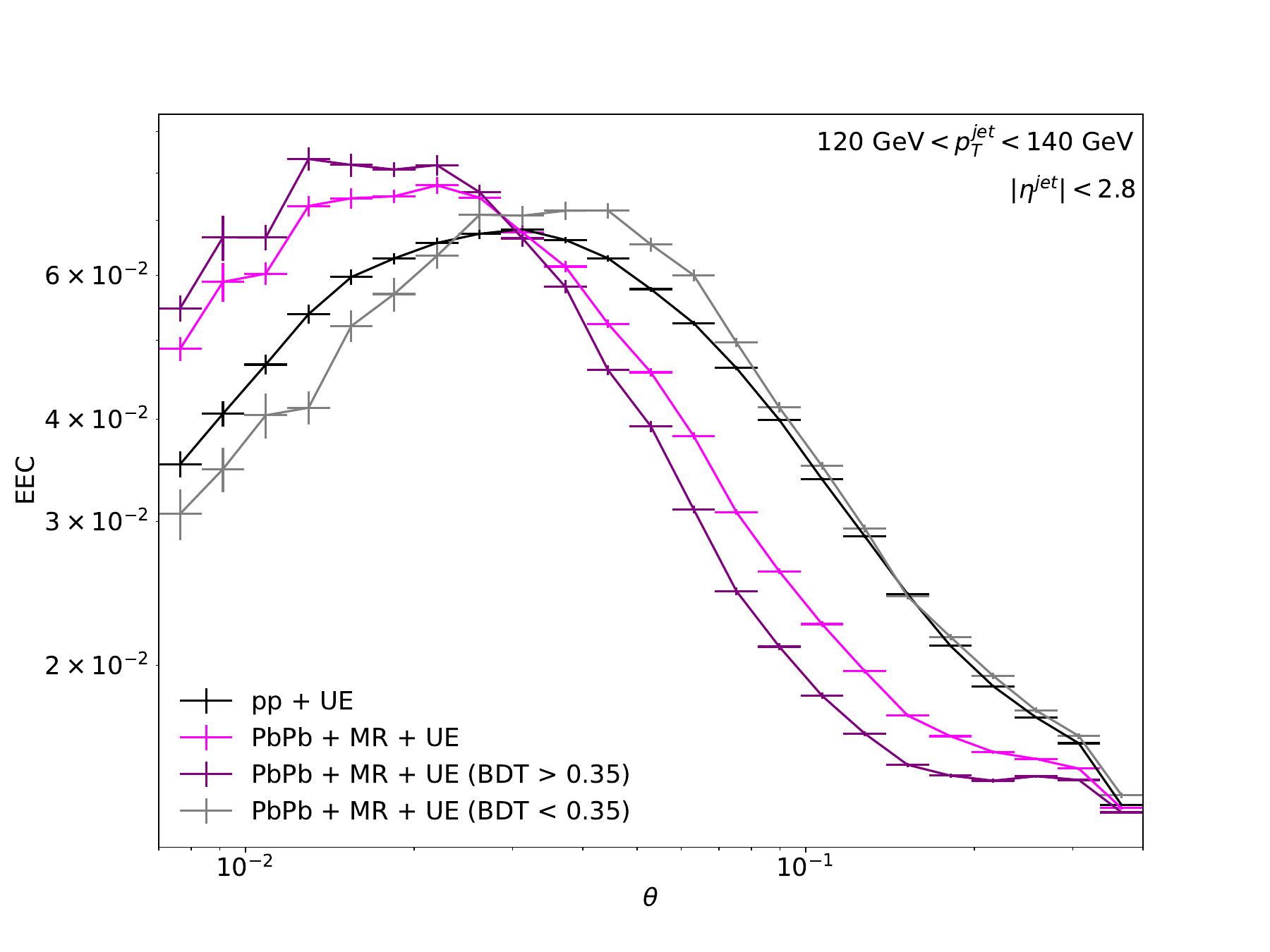}
        \caption{}
        \label{sfig:eec_main_cut}
    \end{subfigure}
    \caption{Normalized EECs over test split for jets in: (a) PbPb samples without MR and UE, \textbf{PbPb} (blue), including MR, \textbf{PbPb + MR} (green), including UE, \textbf{PbPb + UE} (yellow), and including both MR and UE, \textbf{PbPb + MR + UE} (red); and (b) pp sample including UE, \textbf{pp + UE} (black), PbPb including both MR and UE, \textbf{PbPb + MR + UE} (violet), PbPb including both MR and UE for BDT output below 0.35 (gray), and PbPb including both MR and UE for BDT output over 0.35 (purple).}
    \label{fig:eecs}
\end{figure}

\begin{figure}[!htbp]
    \centering
    \begin{subfigure}[b]{0.49\textwidth}
        \centering
        \includegraphics[width=\textwidth]{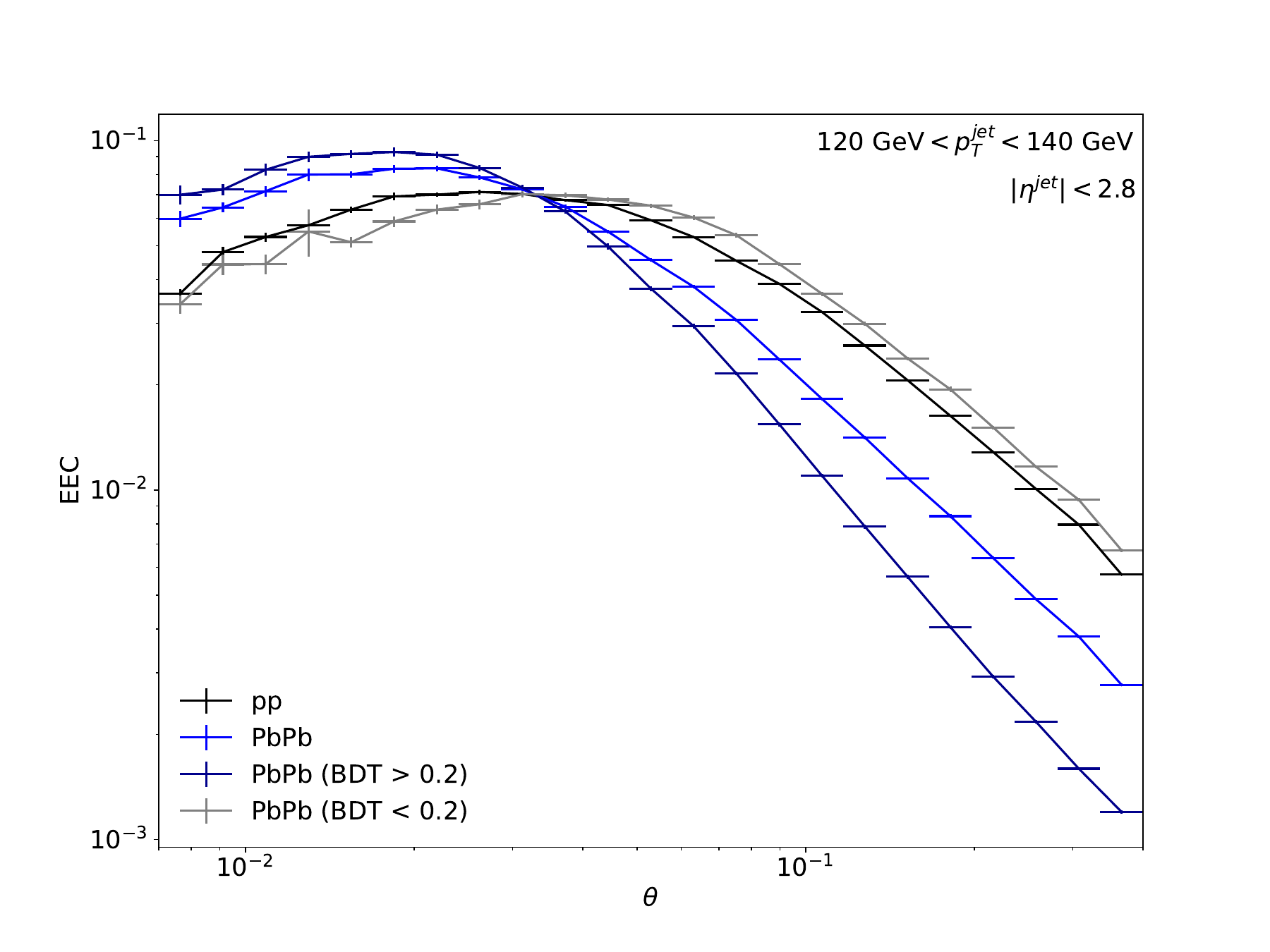}
        \caption{}
        \label{sfig:eec_so}
    \end{subfigure}
    \\
    \centering
    \begin{subfigure}[b]{0.49\textwidth}
        \centering
        \includegraphics[width=\textwidth]{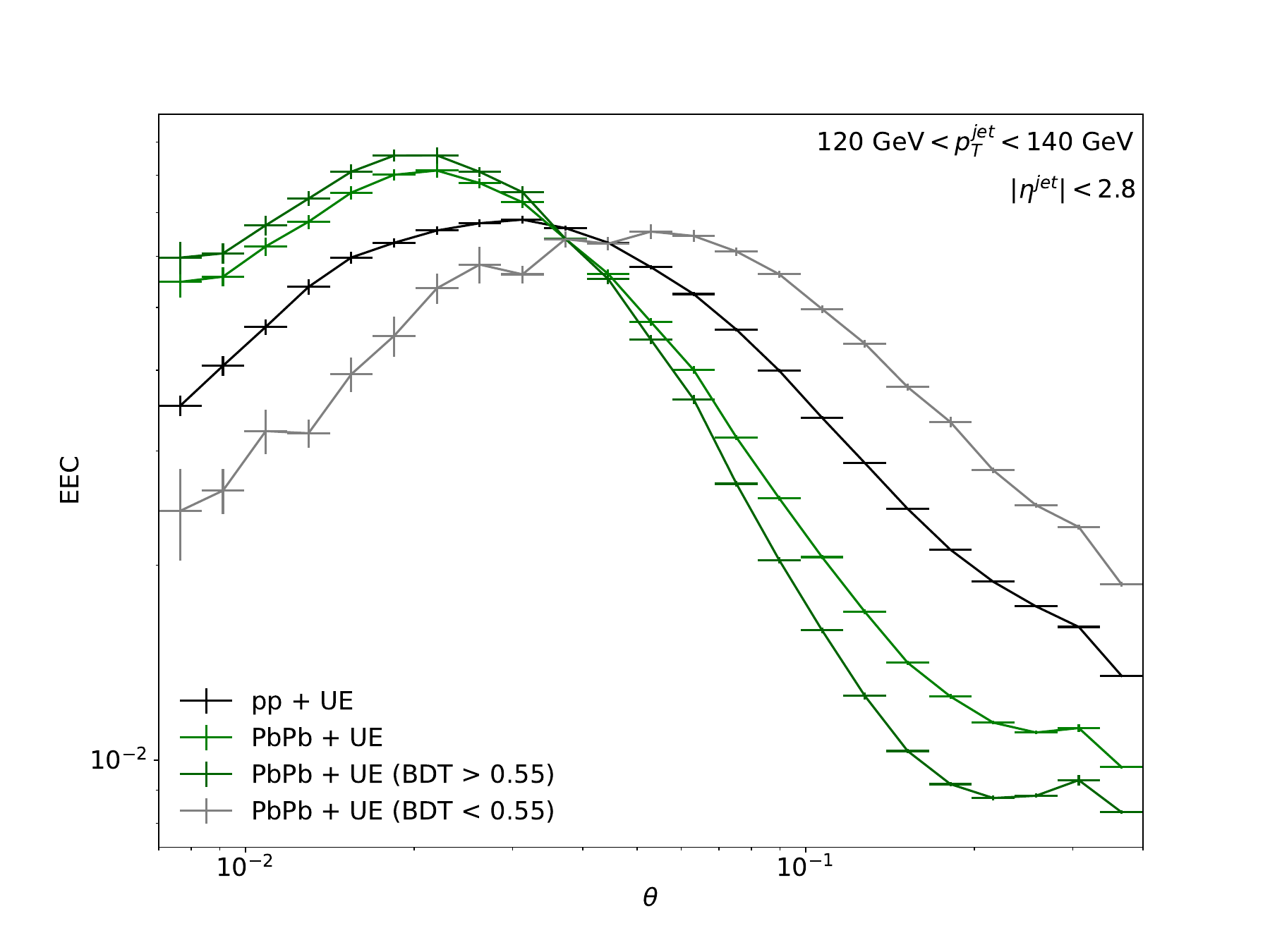}
        \caption{}
        \label{sfig:eec_ue}
    \end{subfigure}
    \hfill
    \begin{subfigure}[b]{0.49\textwidth}
        \centering
        \includegraphics[width=\textwidth]{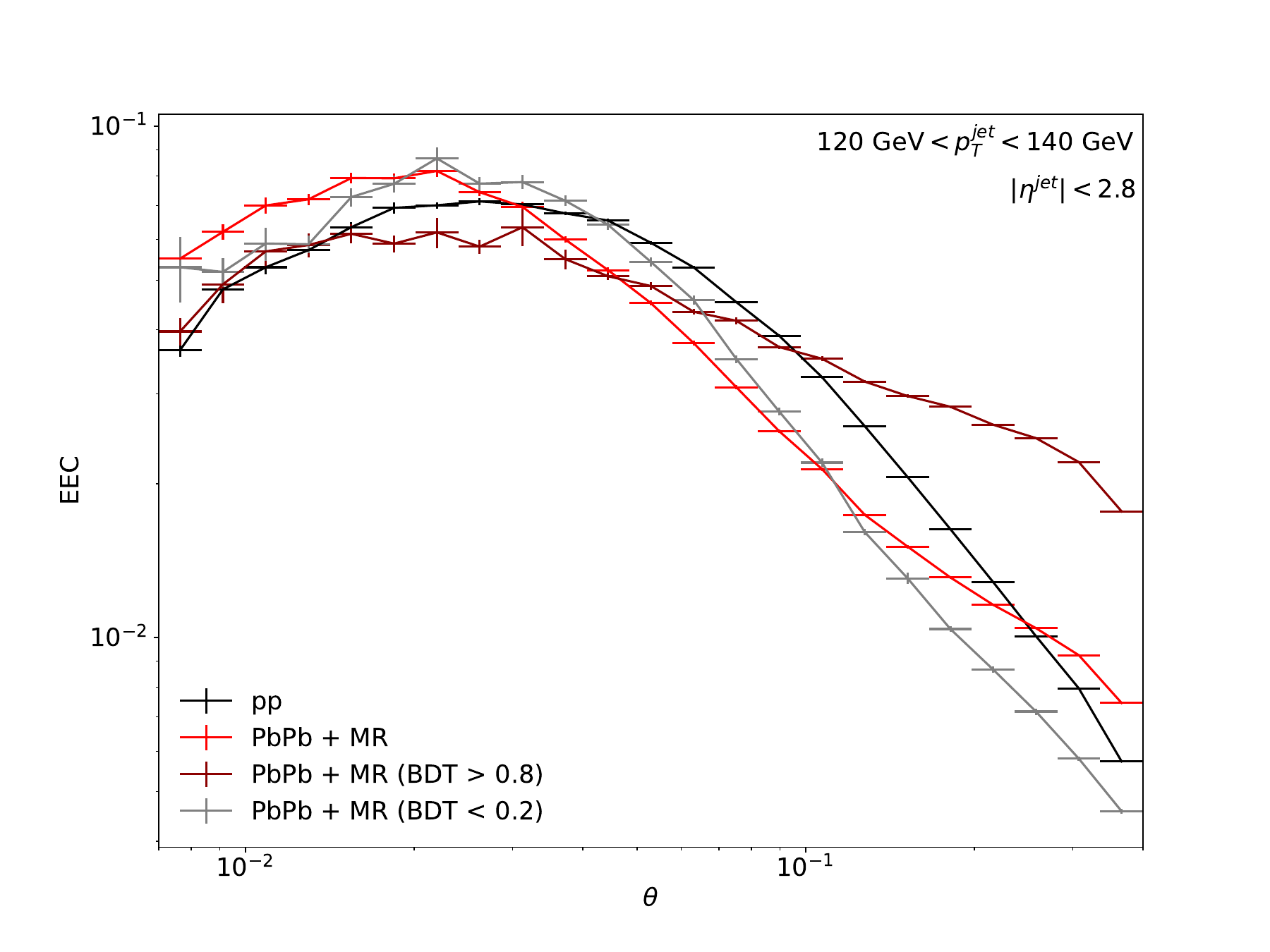}
        \caption{}
        \label{sfig:eec_mr}
    \end{subfigure}
    \caption{Normalized EECs over test split for medium samples within given BDT cuts and corresponding references for: (a) , \textbf{pp} (black), \textbf{PbPb} (blue), PbPb for BDT output below 0.2 (gray), and PbPb for BDT output over 0.2 (dark blue); (b) pp including UE, \textbf{pp + UE} (black), PbPb including UE, \textbf{PbPb + UE} (green), PbPb including UE for BDT output below 0.55 (gray), and PbPb including UE for BDT output over 0.55 (dark green); and (c) \textbf{pp} (black), PbPb including MR, \textbf{PbPb + MR} (red), PbPb including MR for BDT output below 0.55 (gray), and PbPb including MR for BDT output over 0.55 (dark red).}
    \label{fig:eec2}
\end{figure}

The EEC, introduced in  Sec.~\ref{subsubsection:eec}, is a particularly suitable observable to conduct this cross-check. As shown in Figs.~\ref{fig:eecs} and~\ref{fig:eec2}, when plotted for the relevant ranges of the BDT output (the ones we expect to correspond to quenched and unquenched jets), the EEC displays clear features identifying what jet properties drove the classification.

In Fig.~\ref{sfig:eec_med} we show the effect of MR and UE, both separately and jointly, on the EEC. Both MR and UE enhance the large angle part of the EEC which, when no medium is present, is fully described perturbatively \cite{Komiske:2022enw}.
With the UE subtraction we considered in this work, the contribution of UE contamination has a similar magnitude to that of MR being larger for the larger angles. It has been shown \cite{CMS-PAS-HIN-23-004} that the contribution of UE contamination to the EEC can be 
suppressed very substantially. For our purposes, the possibly overestimated contribution from UE contamination simply makes the distinction of unquenched and quenched jets more complicated. As such, the conclusions we draw regarding that separability should be taken as rather conservative. 

To assess the extent to which the BDT is able to classify jets according to how modified by the QGP they are, we consider separately jets for which the BDT output is below (above) 0.35. This value corresponds, see Fig.~\ref{sfig:bdt_out_mrue}, to the transition point from dominance of pp (embedded in UE) jets to dominance of jets from \textbf{PbPb + MR + UE}. 

The results shown in Fig.~\ref{sfig:eec_main_cut} confirm that the BDT classifies jets in the PbPb sample according to the extent of their quenching. The EEC for the subsample with output lower than 0.35 is rather similar to that of pp jets while PbPb jets for which the output is larger than 0.35 have an EEC that is further separated from the pp EEC than that of the inclusive PbPb jet sample.

We have also assessed the nature of the discrimination in the cases where no MR nor UE are accounted for (SO), when only MR is included, and when only UE is included.

In the SO case, Fig.~\ref{sfig:eec_so}, we find that the BDT discriminates in a form analogous to that described for the realistic case where both MR and UE were included (Fig.~\ref{sfig:eec_main_cut}). Jets for which the BDT output is lower than the transition value, here 0.2 (see Fig.~\ref{sfig:bdt_out_so}), have an EEC that is more modified with respect to pp than that of the inclusive PbPb sample and jets with BDT output below the transition value have an EEC very similar to that of pp jets.

The same is true to some extent when only UE contamination (in both pp and PbPb) is included. However, in this case (see Fig.~\ref{sfig:eec_ue}) the PbPb EEC for jets with BDT output lower than the transition value is significantly shifted towards 
large angles. This shift indicates that the BDT is classifying with a low output, jets that have lower average transverse momentum than those in pp. 

When only MR is included, see Fig.~\ref{sfig:eec_mr}, our earlier conclusion on the artificiality of the discrimination is confirmed. Jets for which the BDT output is above the transition value show a very prominent enhancement characteristic of MR and those with low BDT output do not appear to have an EEC alike that of pp jets. In this case the BDT simply identifies the presence of MR with no discrimination related to any further modification of jet properties.

\section{Conclusions}
\label{sec:conclusions}

In this paper we set out to identify jet quenching effects in the presence of UE contamination. We found that many observables are affected by UE contamination in a form akin to true quenching effects arising from the response of the QGP to the traversing partons. This makes the interpretation of the observed modifications of those observables as quenching ill-defined.

We established that true quenching effects can be defined as those observed in realistic PbPb samples where both MR and UE have been accounted for but that are not observed in a reference pp sample that has been embedded in a PbPb UE for which the same UE subtraction procedure as used for PbPb is employed. 
This reference pp sample corresponds to a putative situation where no interaction between the developing jet and QGP takes place, such that observed modifications arise solely from UE contamination.

Within this setup we characterized observables according to their robustness to UE contamination. In broad terms, we found that observables that do not depend on intra-jet angular information appear to be robust and those that do, appear not to be.

The extent to which this sensitivity to UE contamination affects ML studies that aim to discriminate between quenched and unquenched jets is the central result of our work. We found the discrimination power of a simple BDT architecture between a realistic PbPb sample (including both MR and UE) and a pp sample (including UE) to be comparable, in fact slightly larger, than that obtained in proof-of-principle studies that neglect both UE and MR. We take this as strong indication that ML based jet quenching discriminators can be used with experimental data. Importantly, the BDT used as input jet observables including many that are sensitive to underlying event contamination. Dedicated work on further ML architectures whose input is less sensitive to UE is ongoing.

Further, we showed the EEC to be an extremely useful validation tool for ML studies, giving highly interpretable information regarding the physical features identified as relevant by the ML classifier.

While our results have been obtained for a specific event generator (\jewel) and for a specific implementation of UE (and its subtraction), we expect the main conclusion of our work -- that ML classification of quanched jets is robust to UE contamination -- to hold more generally. 

\jewel\ has been stress tested against most of the available jet quenching experimental data and providing an overall very good description. Any alternative event generator that also describes the available data will necessarily have very similar predictions to those of \jewel\ and thus not have a significant effect on our analysis.
Similarly, our implementation of UE is based on experimental measurements and any other approach than is consistent with data will yield a very similar UE to that used here.
Since our UE subtraction procedure not optimized, our UE contamination is a worst case scenario. Improved UE subtraction strategies, including observable specific mitigation, will reduce UE and make the classification task easier.

The results of this work consolidate the relevance of ML classification of quenched jets as a viable strategy to explore experimental data and pave the way for future work using state-of-the-art ML architectures from which we expect a significant increase in discrimination power.

\acknowledgments

This work is supported by European Research Council project ERC-2018-ADG-835105 YoctoLHC. JAG acknowledges support by FCT under contract PRT/BD/151554/2021. JGM thanks the hospitality of the Theory Department at CERN where part of this work was done. 
The authors would like to thank Miguel Crispim Romão and Liliana Apolin\' ario for several helpful discussions and Marco Leitão for kindly providing starting code to produce the EECs.

\appendix

\section{Thermal Momenta Subtraction in \jewel}
\label{app:jewsub}

\jewel\ does not simulate complete heavy ion events. Thermal partons from the QGP that are not involved in interactions with the developing parton shower are never included in the event record. 
Thermal partons that interact with shower partons can be included in the event record. 
For these scattered thermal partons, referred to as \textit{recoils}, to provide a description of medium response, their original (prior to scattering) four-momenta have to be subtracted. In other words, medium response is the difference between the effect of the recoils and what would have been the final state if the scattered thermal partons had not been involved in any interactions.
The limitations of the first implementation of such a subtraction \cite{KunnawalkamElayavalli:2017hxo} were addressed and resolved in \cite{Milhano:2022kzx}.

The subtraction algorithm introduced in \cite{Milhano:2022kzx}, which we use here, is based on the constituent subtraction method \cite{Berta:2014eza} with ghosts replaced by the thermal momenta. For completeness we reproduce it here.

All four-momenta (final state particles and thermal momenta) are represented by their transverse momentum $p_\perp$, mass $m_\delta = \sqrt{m^2 + p_\perp^2} - p_\perp$, rapidity $y$ and azimuthal angle $\phi$:
\begin{equation}
 p^\mu = \left( (m_\delta + p_\perp) \cosh(y),\ p_\perp \cos(\phi),\ p_\perp \sin(\phi),\ (m_\delta + p_\perp) \sinh(y) \right)\, .
\end{equation}

A list of all possible pairs consisting of a final state particle $i$ and a thermal momentum $k$ is sorted by distance $\Delta R_{ik} = \sqrt{(y_i-y_k)^2 + (\phi_i-\phi_k)^2} $.
The subtraction proceeds by going through the list (starting from the smallest distance) and in each pair subtracting the smaller $p_\perp$ from the larger and the smaller $m_\delta$ from the larger:
\begin{equation}
\begin{array}{lcl}
\text{if} \quad p_{\perp}^{(i)} \ge p_{\perp}^{(k)} & \quad : \qquad & p_{\perp}^{(i)} \to p_{\perp}^{(i)} - p_{\perp}^{(k)} \\
                                 &  & p_{\perp}^{(k)} \to 0\\
\text{if} \quad p_{\perp}^{(i)} < p_{\perp}^{(k)} & \quad : \qquad & p_{\perp}^{(i)} \to 0\\
                                 &  & p_{\perp}^{(k)} \to p_{\perp}^{(k)} - p_{\perp}^{(i)}\, ,
\end{array}
\end{equation}
and
\begin{equation}
\begin{array}{lcl}
\text{if} \quad m_\delta^{(i)} \ge m_\delta^{(k)}& \quad : \qquad & m_\delta^{(i)} \to m_\delta^{(i)} - m_\delta^{(k)} \\
                                 &  & m_\delta^{(k)} \to 0\\
\text{if} \quad m_\delta^{(i)} < m_\delta^{(k)}& \quad : \qquad & m_\delta^{(i)} \to 0\\
                                 &  & m_\delta^{(k)} \to m_\delta^{(k)} - m_\delta^{(i)}.
\end{array}
\end{equation}

To avoid subtractions from particles too far away from each other, we impose a cut-off at 0.5 in $\Delta R_{ik}$. After all subtractions are done, all momenta with $p_\perp = 0$ are removed. The remaining momenta constitute the subtracted ensemble. We perform the subtraction on the entire event prior to jet reconstruction. 
Inside jets the final state particles carry much more momentum than the thermal momenta. This constituent subtraction procedure thus leads to all thermal momenta being used up and disappearing from the ensemble.

\section{PbPb underlying event}
\label{app:UE}

\begin{figure}[!htbp]
    \centering
    \includegraphics[scale=.37]{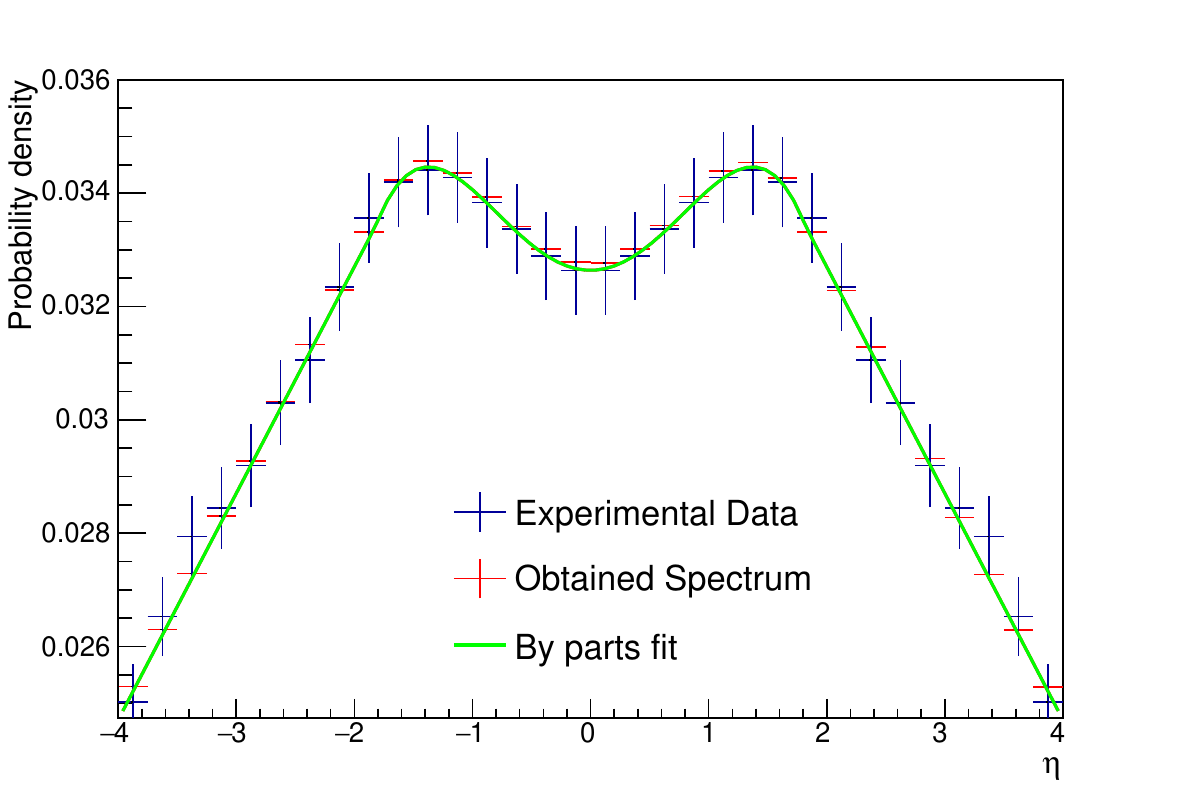}
    \includegraphics[scale=.37]{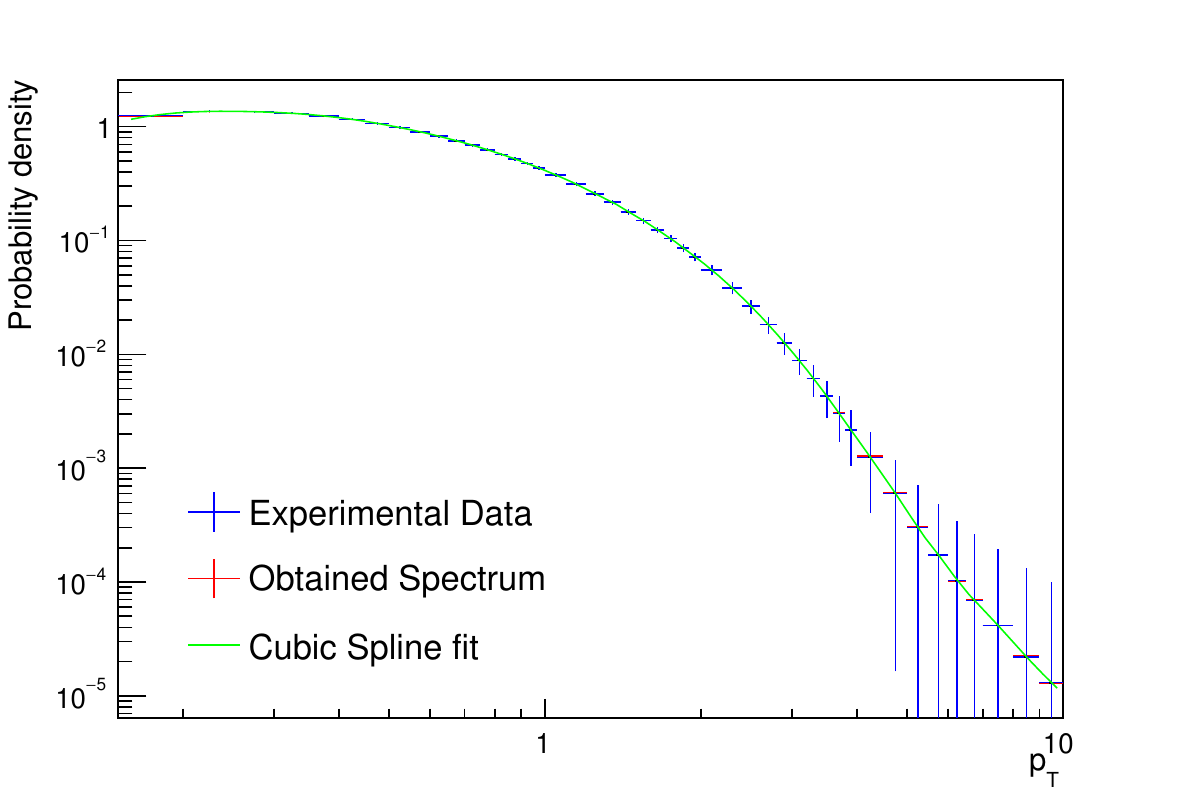}
    \caption{(left) Pseudorapidity charged particle distribution \cite{ALICE:2016fbt} compared to our fit, and an average of sampling over \(10000\) events; (right) Transverse momenta spectrum \cite{ALICE:2018vuu} compared to our fit, and the spectrum obtained from averaging over \(10000\) sampled events.}
    \label{fig:fits}
\end{figure}

To account for fluctuations on an event-by-event basis we have generated one underlying event (UE) for each hard scattering.
Each UE is generated, for pseudorapidity $|\eta| < 4$, as follows:

\begin{itemize}
    \item The total number of particles is sampled from a Gaussian distribution with mean $20178$ and standard deviation of $142$ (the square root of the mean). The mean total number of particles was obtained by integrating, over $|\eta| < 4$, the pseudorapidity distribution of charged particles for the $0-10\%$ centrality class measured by the ALICE collaboration \cite{ALICE:2016fbt}, and scaling it by a factor of $1.5$ to roughly account for neutral particles.
    
    \item The pseudorapidity of each particle is obtained by sampling the same pseudorapidity distribution \cite{ALICE:2016fbt} which was fitted piece-wise with a second and fourth order polynomial fit for mid-rapidity, and linearly in the backward/forward regions. Fig.~\ref{fig:fits} (left) shows a comparison of the measured distribution, our fit, and the distribution obtained from averaging over $10000$ sampled events.
    
    \item The transverse momentum of each particle is obtained by sampling the spectrum in \cite{ALICE:2018vuu} fitted with a cubic spline. 
    Fig.~\ref{fig:fits} (right) shows a comparison of the measured spectrum, our fit, and the spectrum obtained from averaging over $10000$ sampled events.
    
    \item The azimuthal position of each particle is sampled from a uniform distribution.
    
    \item All particles are assumed to be pions (with a $1/3$ probability each), and masses are assigned accordingly.
\end{itemize}

\begin{figure}[!htbp]
    \centering
     \includegraphics[width=\textwidth]{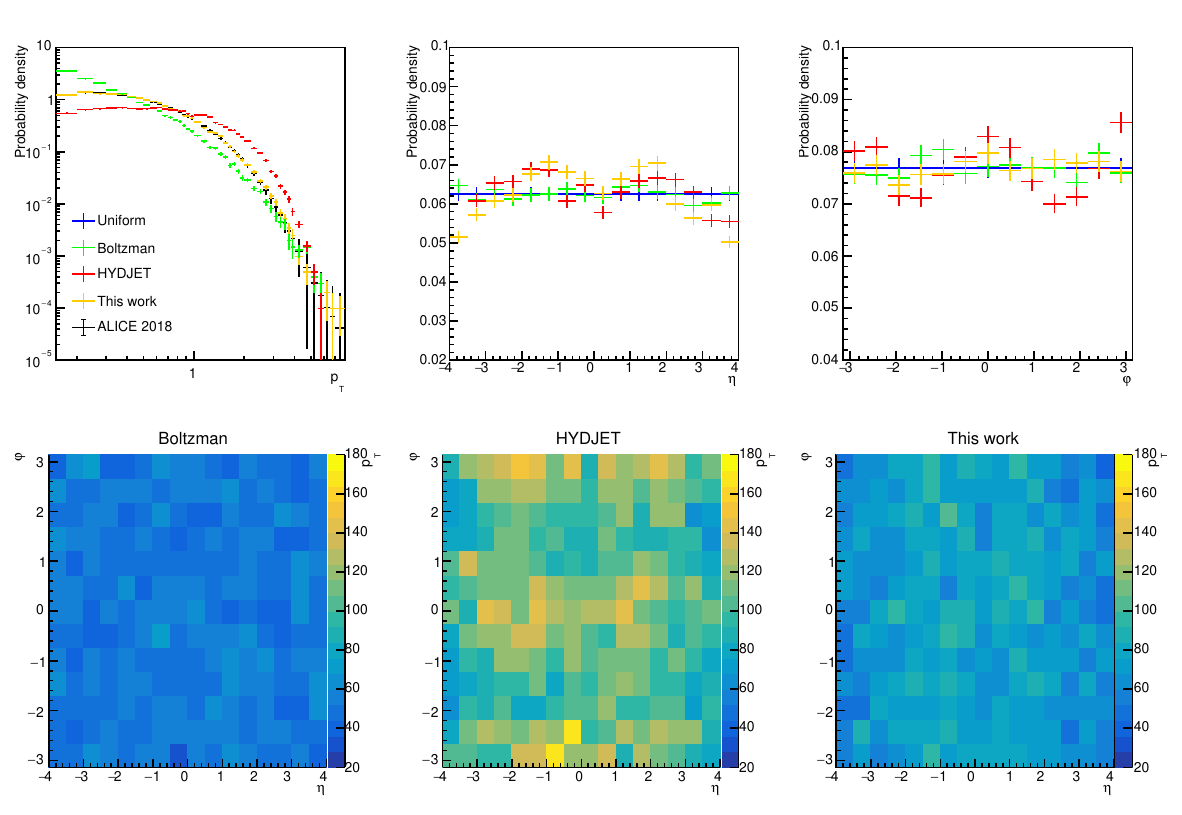}
    \caption{Comparison of UEs generated through different methods: Boltzmann based (green), HYDJET based (red) and our procedure (yellow). Experimental data from \cite{ALICE:2016fbt} and \cite{ALICE:2018vuu} is shown in black for comparison. (top) \(p_T\), \(\eta\) and \(\phi\) distribution for all three methods; (bottom) transverse momenta distribution across the full \(\eta\), \(\phi\) plane for the three methods (each case corresponds to a single generated event, not an average over events).}
    \label{fig:backval}
\end{figure}

In Fig.~\ref{fig:backval} we show a comparison of the UE generated as described above, with UE generated by sampling a Boltzmann(-like) distribution for the transverse momentum spectrum (as in \cite{Liu:2022hzd}) and assuming uniform distributions in both azimuth and pseudorapidity, and UE obtained from randomly sampling HYDJET \cite{Lokhtin:2008xi} events generated with no hard-scattering.

\section{UE subtraction} 
\label{app:uesub}

UE subtraction is carried out using event-wide Iterative Constituent Subtraction (ICS) \cite{Berta:2019hnj} with two iterations and parameters \(\Delta R^{max}_1 = 0.2\), \(\Delta R^{max}_2 = 0.1\), \(\alpha =1\) and \(A_g=0.0025\) as recommended for \(R=0.4\) anti-\(k_T\) jets.
ICS is an extension of the Constituent Subtraction (CS) \cite{Berta:2014eza} method where CS subtraction is applied iteratively. After each iteration, the remaining unsubtracted background is redistributed uniformly before performing the next subtraction. This approach provides improved performance in terms of bias reduction and resolution, particularly in jet kinematics and substructure observables.

CS uses very soft particles, referred to as ghosts, evenly distributed in the $y-\phi$ plane, each occupying a fixed area $A_g$, to perform the subtraction. The background transverse momentum ($p_T$) density, $\rho$, is estimated as a function of other variables (typically rapidity, $y$). In our case, we estimate $\rho$ using the grid median background estimator without $y$ modulation, as described in \cite{Berta:2014eza, Berta:2019hnj}.

The subtraction proceeds as follows: for each particle, the distance between the particle $i$ and ghost $k$ in the $y-\phi$ plane, $\Delta R_{ik}$, is calculated and scaled by a power $\alpha$ of the particle's $p_T$. Distances are sorted from smallest to largest, and subtraction is performed iteratively. If a particle's momentum is larger, the ghost’s momentum is subtracted; otherwise, the particle's momentum is reduced by the ghost’s momentum. This process continues until $\Delta R_{ik}$ exceeds a threshold value $\Delta R_{max}$.

\section{Additional jet observables}
\label{app:jetobs}

\subsection{Jet fragmentation functions}
\label{app:fragfunc}

\begin{figure}[!htbp]
     \centering
     \begin{subfigure}[b]{0.49\textwidth}
         \centering
         \includegraphics[width=\textwidth]{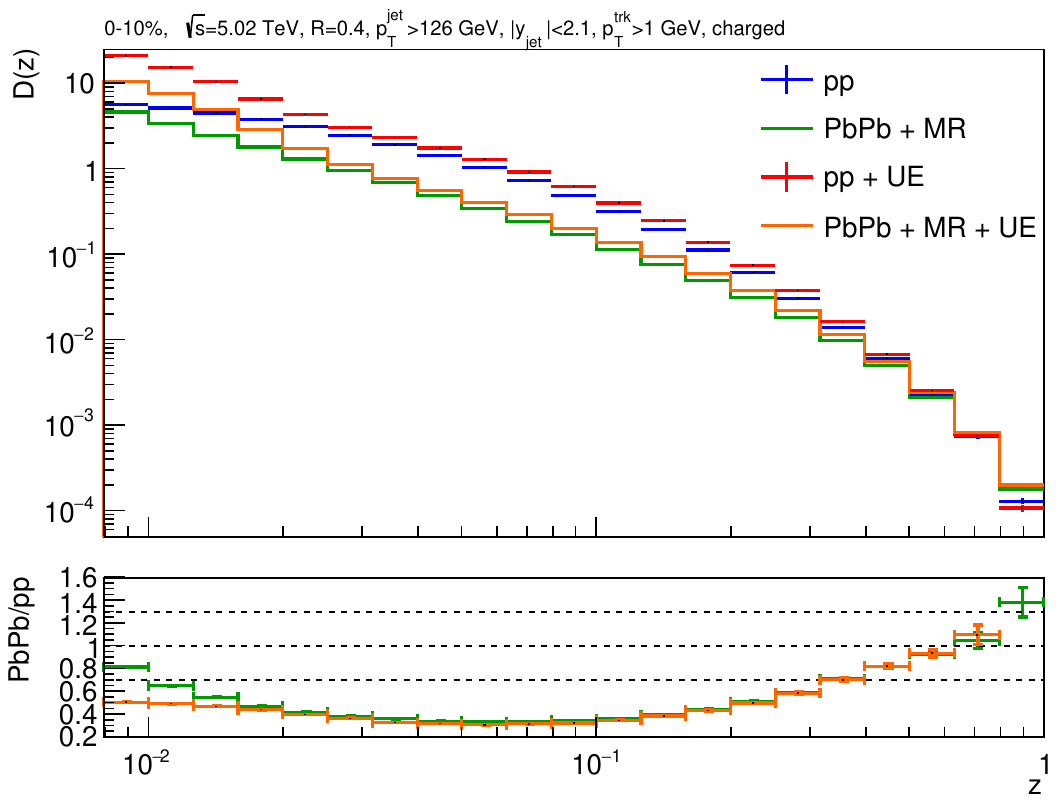}
         \caption{}
     \end{subfigure}
     \hfill
     \begin{subfigure}[b]{0.49\textwidth}
         \centering
         \includegraphics[width=\textwidth]{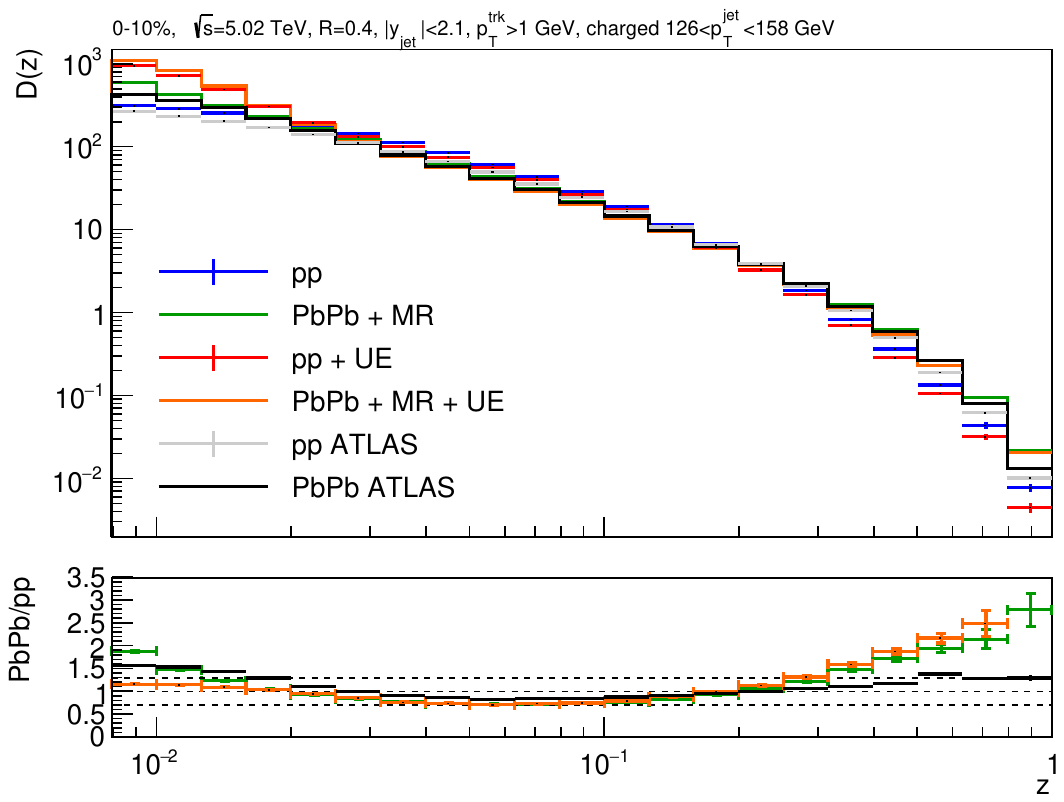}
        \caption{}
     \end{subfigure}
     \\
     \begin{subfigure}[b]{0.49\textwidth}
         \centering
         \includegraphics[width=\textwidth]{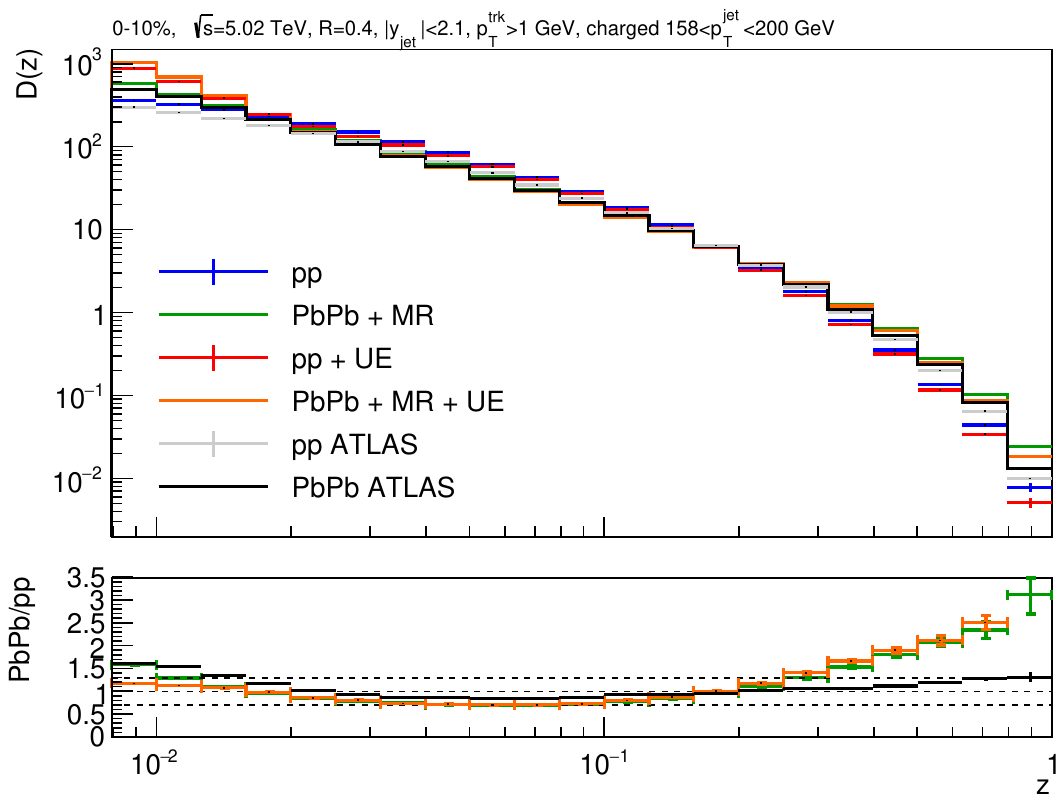}
          \caption{}
     \end{subfigure}
     \hfill
     \begin{subfigure}[b]{0.49\textwidth}
         \centering
         \includegraphics[width=\textwidth]{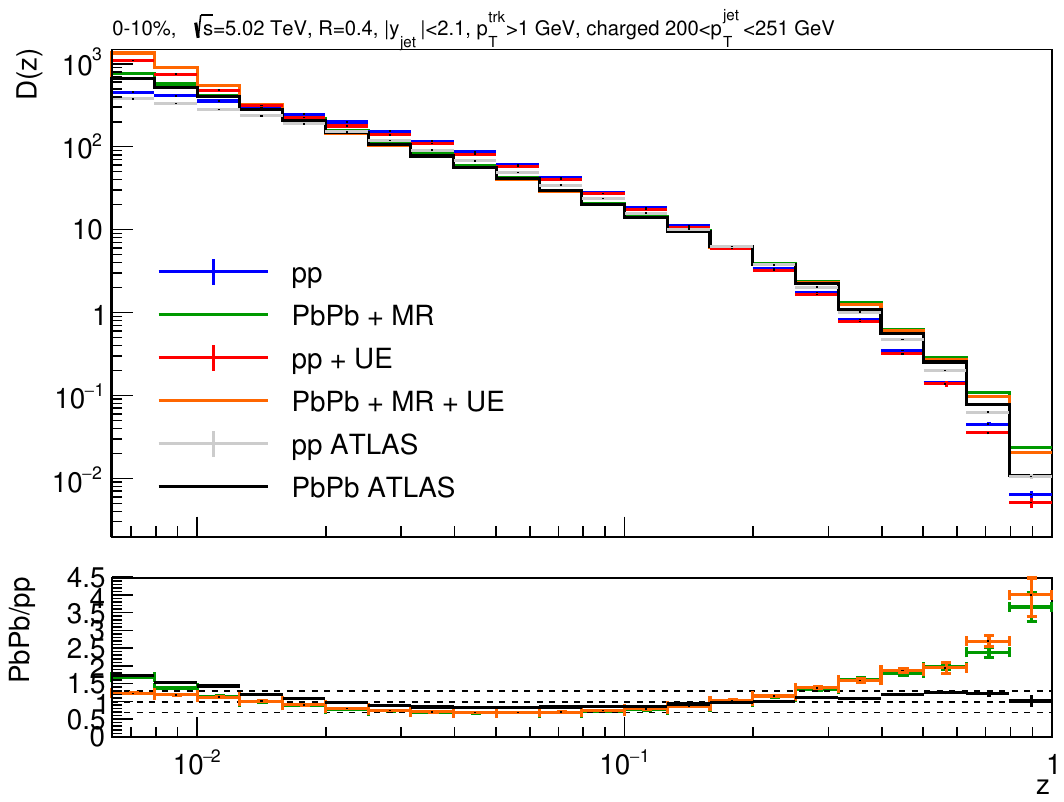}
          \caption{}
     \end{subfigure}
      \\
     \begin{subfigure}[b]{0.49\textwidth}
         \centering
         \includegraphics[width=\textwidth]{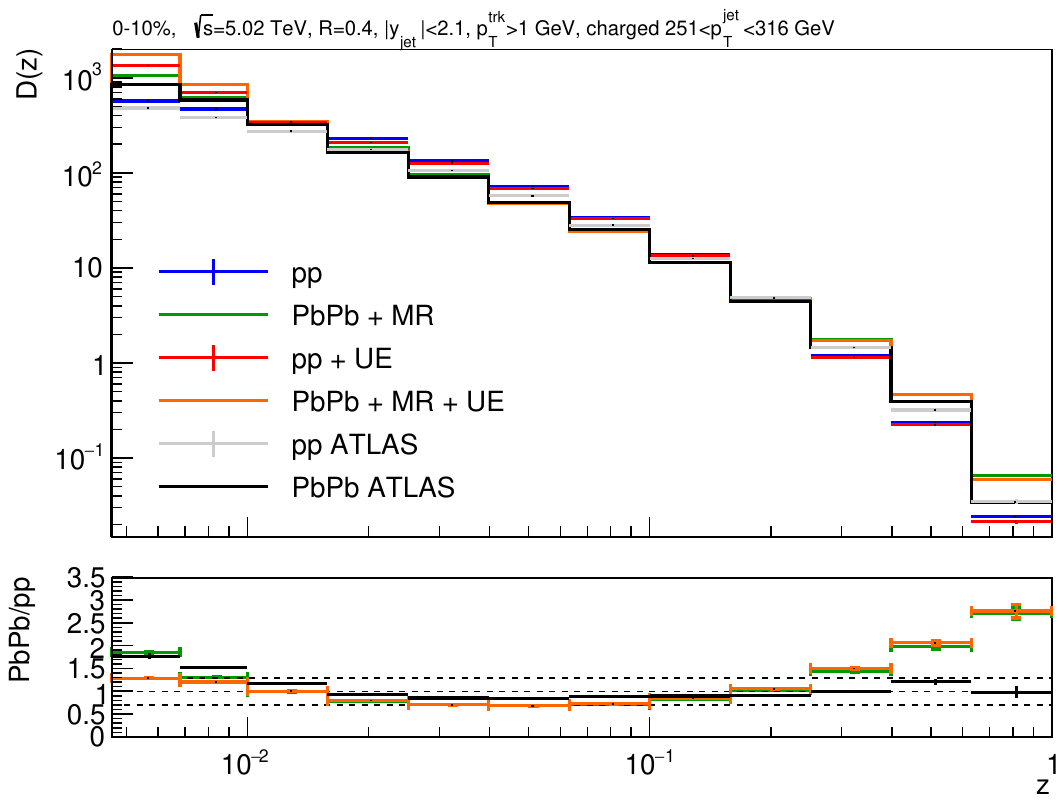}
          \caption{}
     \end{subfigure}
     \hfill
     \begin{subfigure}[b]{0.49\textwidth}
         \centering
         \includegraphics[width=\textwidth]{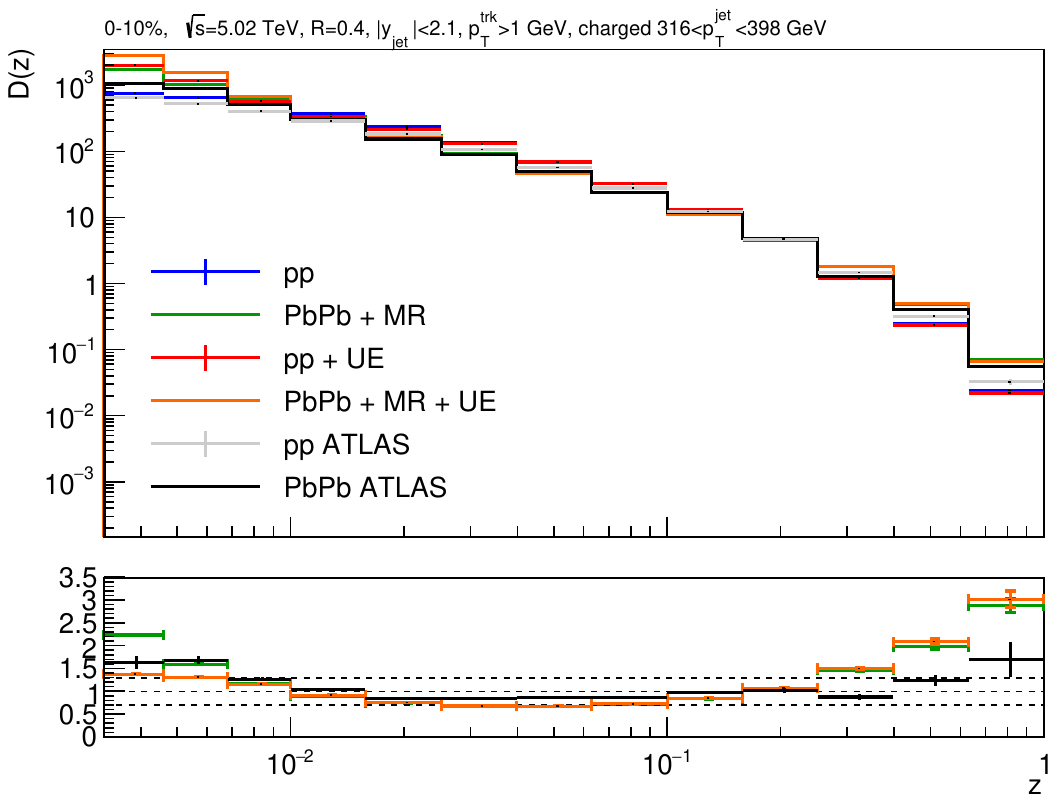}
          \caption{}
     \end{subfigure}
     \caption{Jet fragmentation functions for \textbf{pp} and \textbf{PbPb + MR} with and without UE contamination, with PbPb to pp ratios in the bottom panels and experimental data from \cite{ATLAS:2018bvp}, for transverse momentum ranges: (a) \(p_T^\mathsf{jet} > 126\,\)GeV; (b) \(126 > p_T^\mathsf{jet} > 158\,\)GeV; (c) \(158 > p_T^\mathsf{jet} > 200\,\)GeV; (d) \(200 > p_T^\mathsf{jet} > 251\,\)GeV; (e) \(251 > p_T^\mathsf{jet} > 316\,\)GeV; and (f) \(316 > p_T^\mathsf{jet} > 398\,\)GeV.}
        \label{fig:ff}
\end{figure}

The jet fragmentation function, shown in Fig.~\ref{fig:ff}, is given by:
\begin{equation}
D(z) = \frac{1}{N_{\text{jet}}} \frac{dN_{\text{ch}}}{dz}\, ,
\end{equation}
where $z = \frac{p_T^{\text{ch}} \cdot \cos(\Delta R)}{p_T^{\text{jet}}}$ is the charged-particle longitudinal momentum fraction relative to the jet, with $\Delta R$ the distance of the particle to the jet axis. Results are shown for several transverse momentum ranges, according to experimental results in \cite{ATLAS:2018bvp}, as well as for the full transverse momentum range considered.

This observable shows significant robustness across its range, except at low \(z\). In this region both pp and PbPb samples present an excess when including UE contamination. This seems to be explained when considering that UE contamination tends to come in fluctuations of low transverse momentum fraction relative to the jet, particularly when employing constituent subtraction based methods.

\subsection{Jet mass}

\begin{figure}[!htbp]
     \centering
     \begin{subfigure}[b]{0.49\textwidth}
         \centering
         \includegraphics[width=\textwidth]{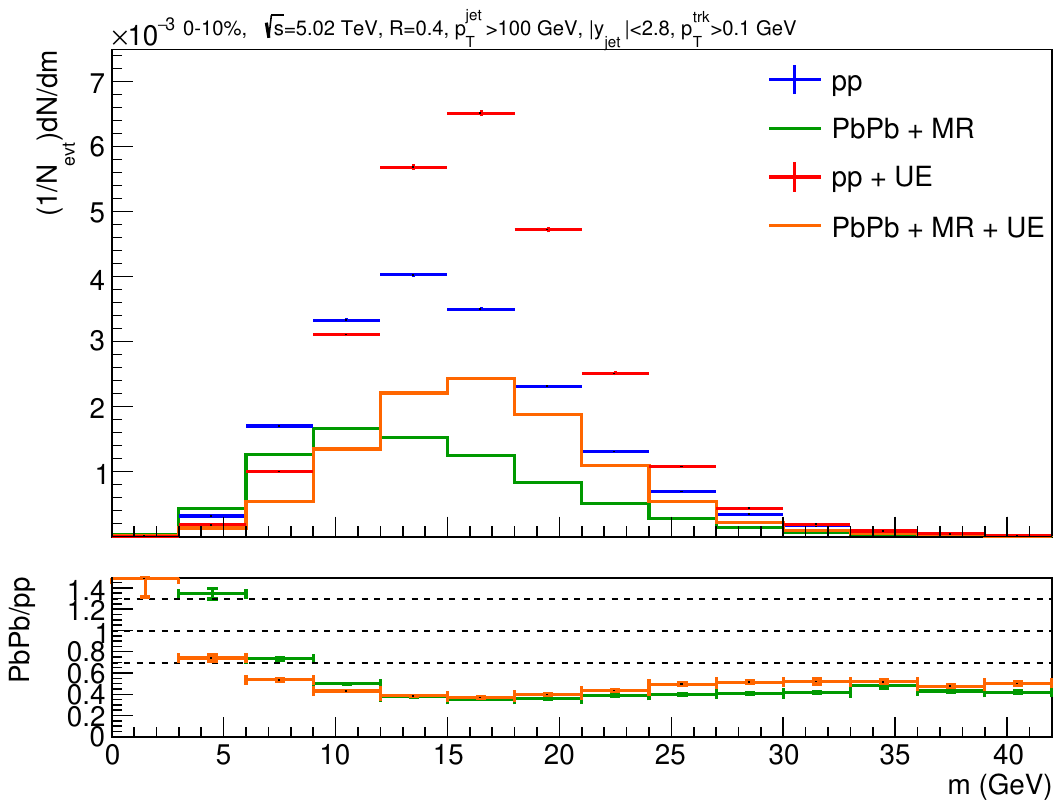}
         \caption{}
     \end{subfigure}
     \hfill
     \begin{subfigure}[b]{0.49\textwidth}
         \centering
         \includegraphics[width=\textwidth]{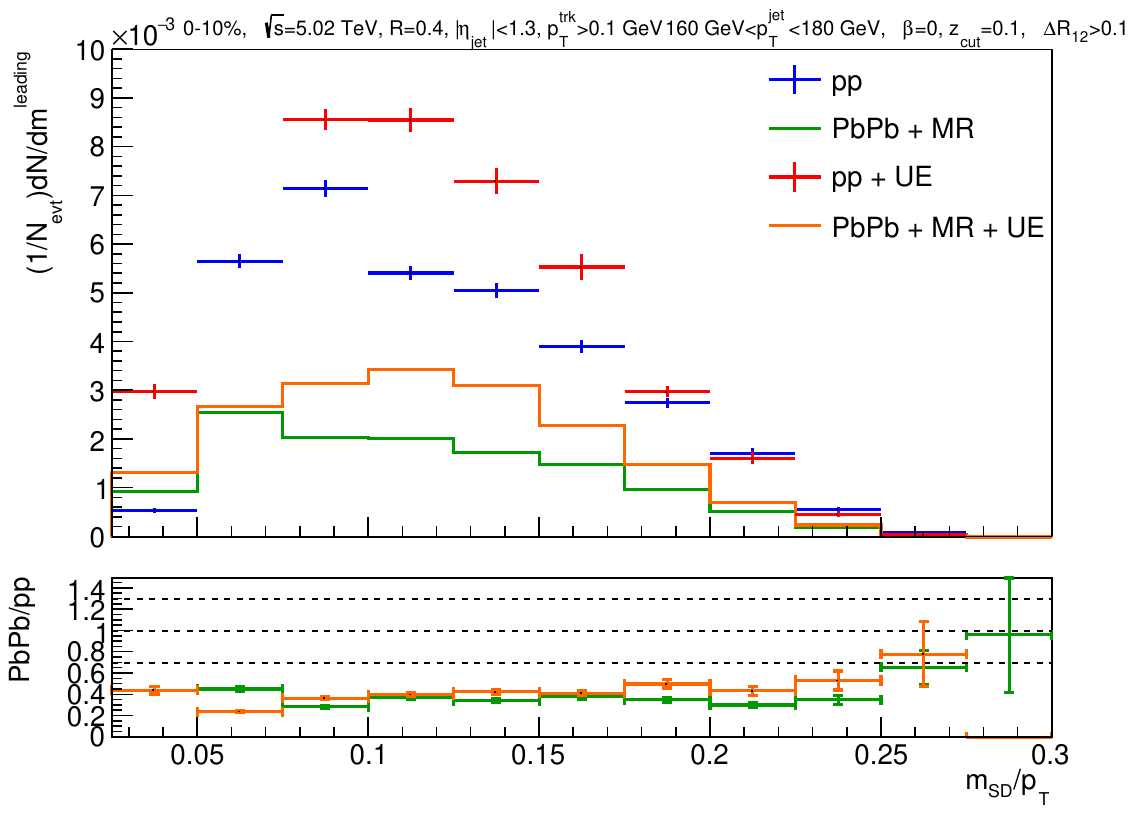}
          \caption{}
     \end{subfigure}
     \\
     \begin{subfigure}[b]{0.49\textwidth}
         \centering
         \includegraphics[width=\textwidth]{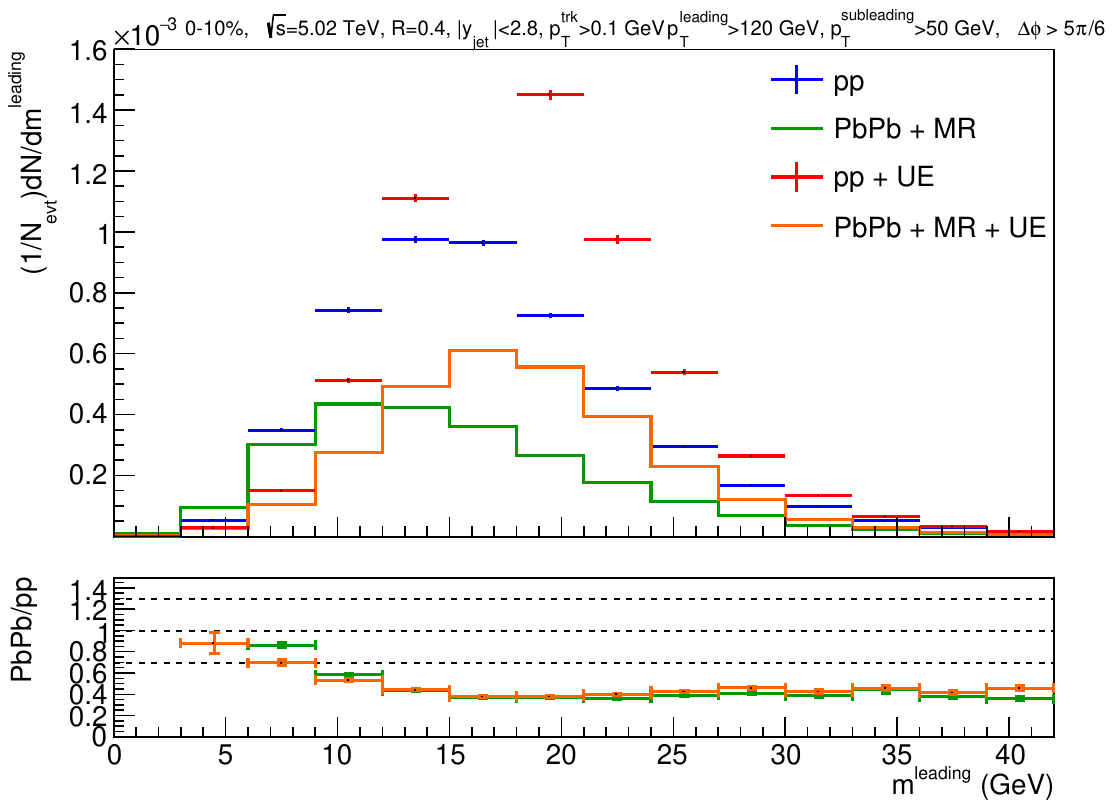}
          \caption{}
     \end{subfigure}
     \hfill
     \begin{subfigure}[b]{0.49\textwidth}
         \centering
         \includegraphics[width=\textwidth]{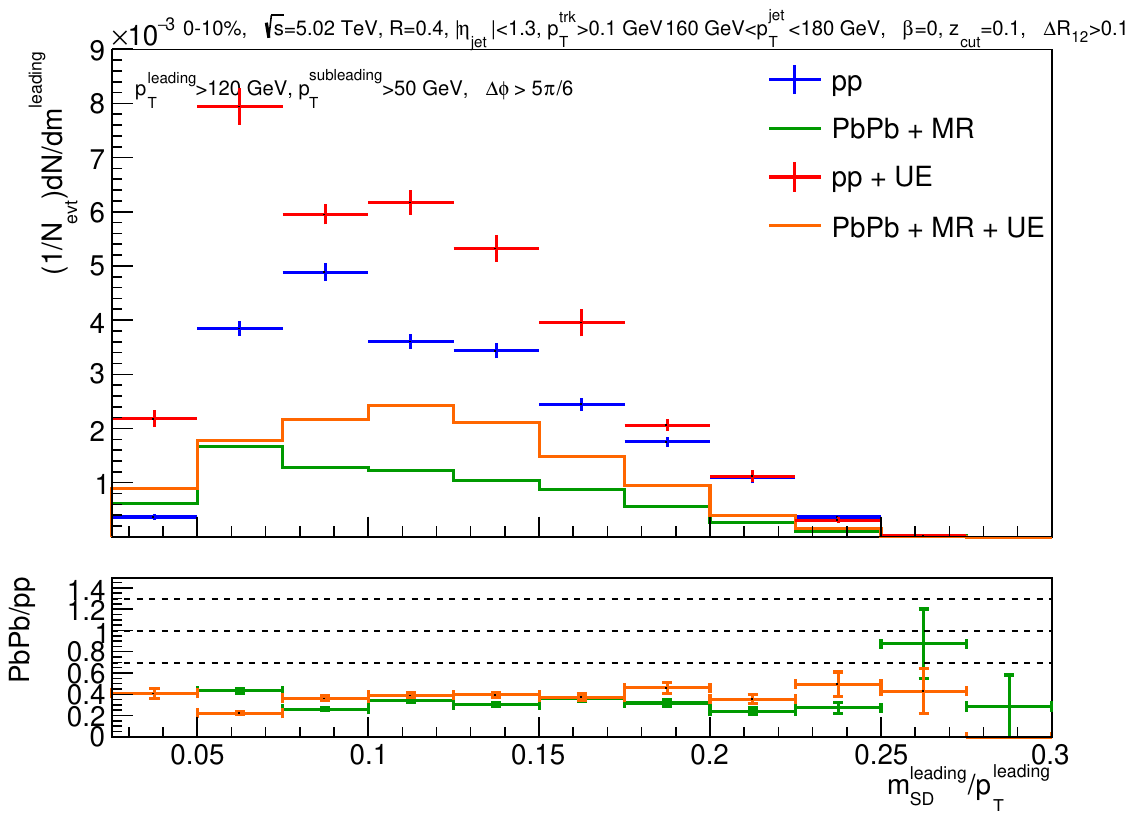}
          \caption{}
     \end{subfigure}
      \\
     \begin{subfigure}[b]{0.49\textwidth}
         \centering
         \includegraphics[width=\textwidth]{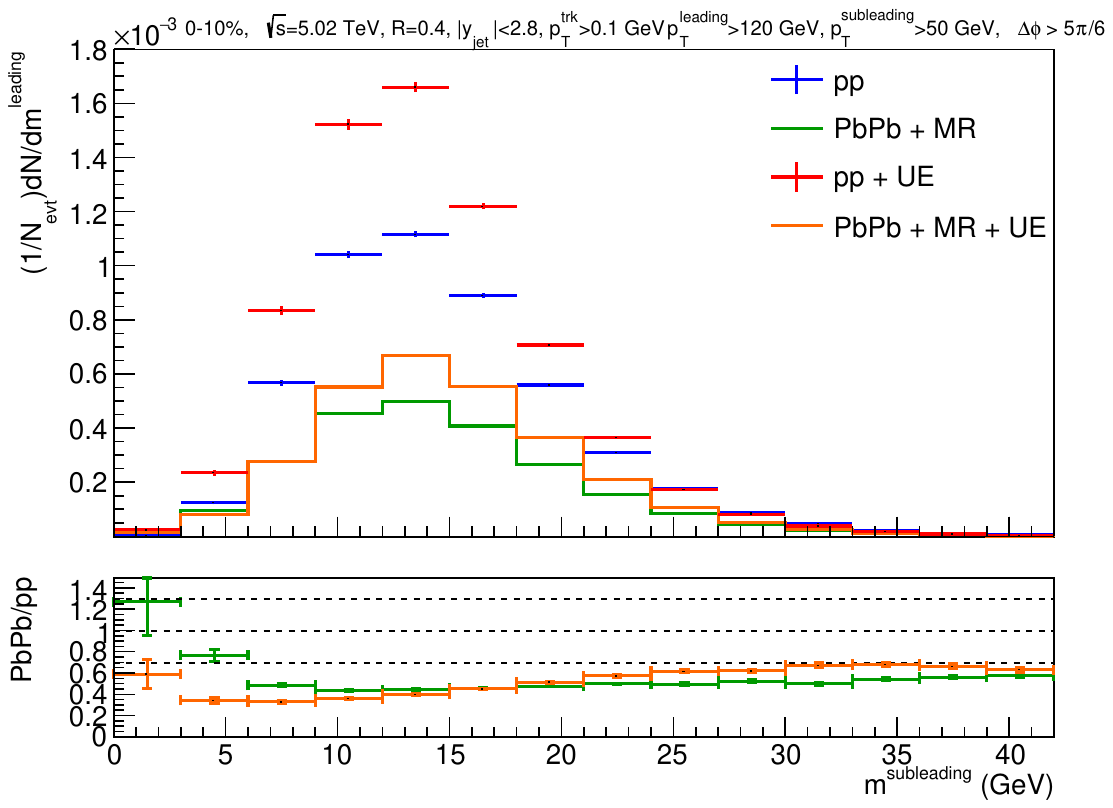}
          \caption{}
     \end{subfigure}
     \hfill
     \begin{subfigure}[b]{0.49\textwidth}
         \centering
         \includegraphics[width=\textwidth]{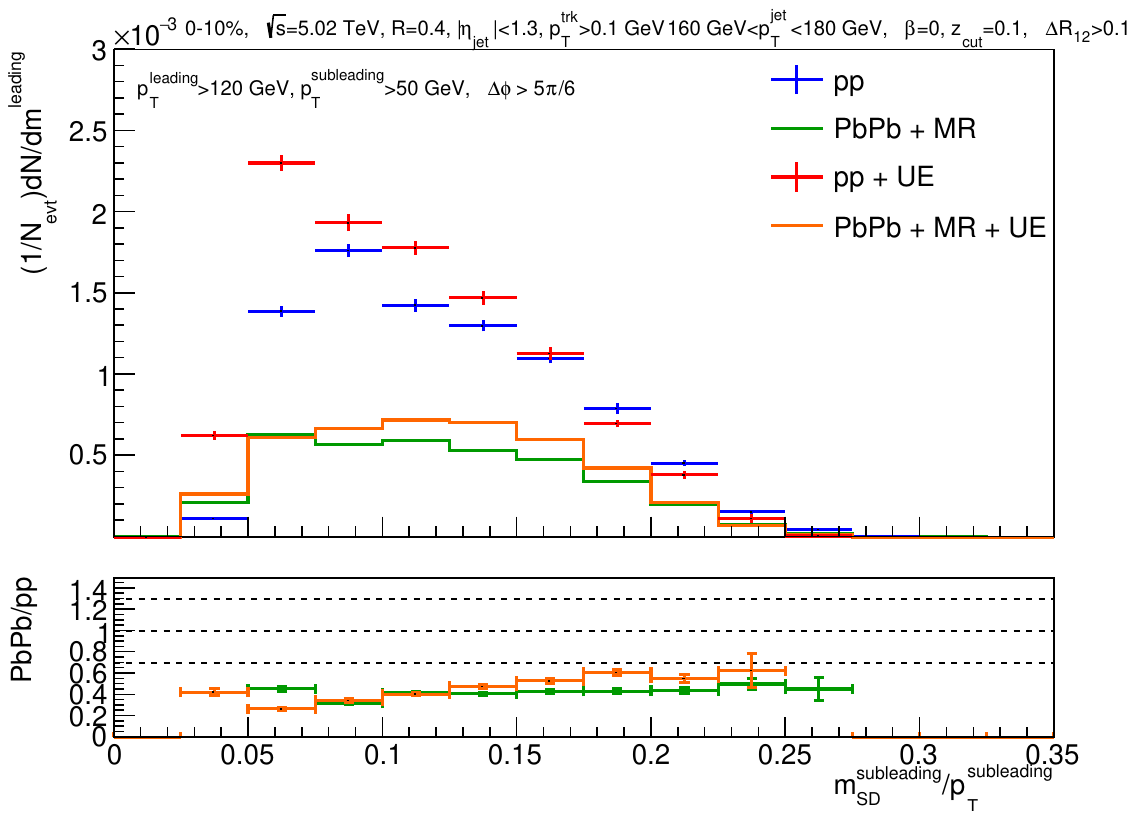}
          \caption{}
     \end{subfigure}
     \caption{Jet mass for \textbf{pp} and \textbf{PbPb + MR} with and without UE contamination, with PbPb to pp ratios in the lower panels, with SD (right) and without SD (left), for: the inclusive samples (top); samples only including the leading jet of the dijet pair (middle) and samples including only the subleading jet of the dijet pair (bottom).}
        \label{fig:jetmass}
\end{figure}

Jet mass, shown in Fig.~\ref{fig:jetmass}, is given by:
\begin{equation}
    m_{jet} = \sqrt{E_{jet}^2 - |\vec{p}_{jet}|^2}\, ,
\end{equation}
where \(E_{jet}\) is the energy of the jet and \(\vec{p}_{jet}\) is the 3-momentum of the jet. Results are shown for both ungroomed and Soft Drop groomed jets in an inclusive jet sample, and separately for leading and sub-leading jets in a dijet pair.

In this case, we do seem to have some level of robustness to the procedure but only at the level of the PbPb to pp ratios. Overall UE contamination seems to create a positive shift in the average value of the mass across all cases, albeit softer for the subleading jet of the dijet pair. The fact that the ratios are robust despite this, tells us that, for both pp and PbPb samples, the shift must be between very similar and exactly the same.

\subsection{Jet girth}

\begin{figure}[!htbp]
    \centering
    \includegraphics[scale=.59]{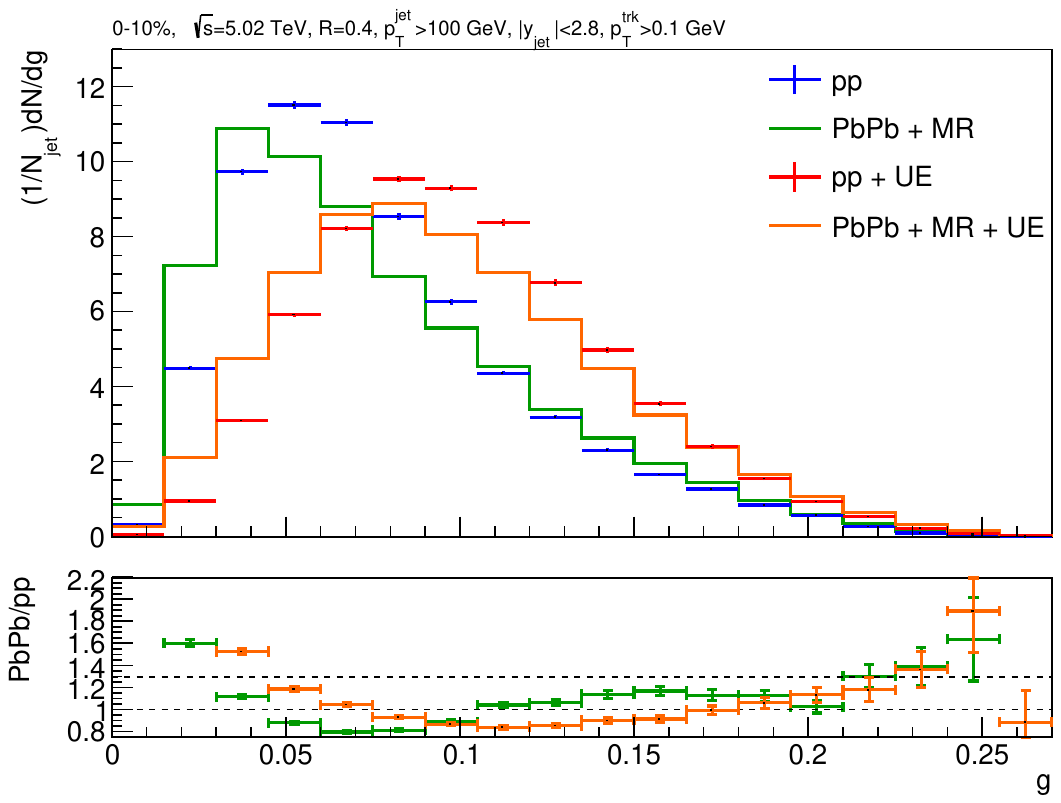}
    \caption{Jet girth, \(g\), for \textbf{pp} and \textbf{PbPb + MR} collisions with and without UE contamination. The ratios with and without UE contamination are presented in the bottom panel.}
    \label{fig:l11}
\end{figure}

Jet girth, shown in Fig.~\ref{fig:l11}, is given by:
\begin{equation}
    g  = \sum_{i\in jet} z_i \Delta R_{i,jet}
\end{equation}
where \(z_i\) is the transverse momentum fraction of constituent \(i\) relative to the jet and \(\Delta R_{i,jet}\) the distance of the particle to the jet axis.

For this observable, neither the distributions nor the PbPb to pp ratios show strong robustness to UE contamination across the whole range for this observable. At the level of the distributions UE contamination increases the average value of both the pp and PbPb distributions.

\subsection{N-Subjetiness}

\begin{figure}[!htbp]
     \centering
     \begin{subfigure}[b]{0.49\textwidth}
         \centering
         \includegraphics[width=\textwidth]{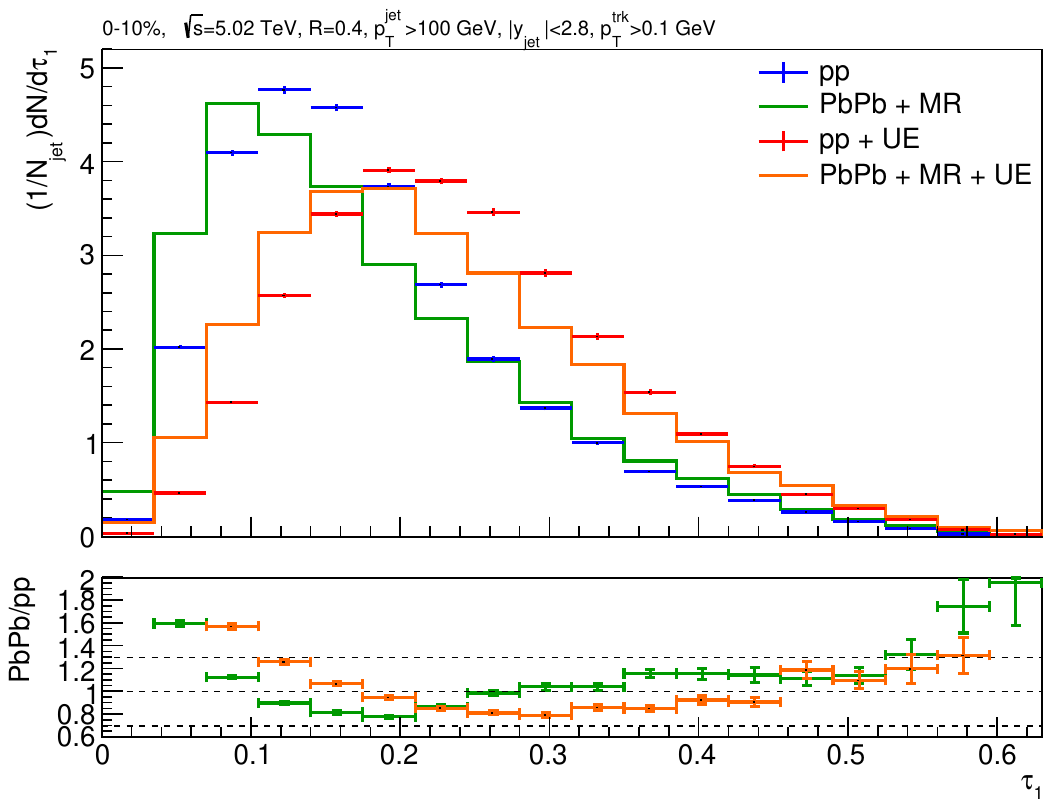}
         \caption{}
     \end{subfigure}
     \\
     \begin{subfigure}[b]{0.49\textwidth}
         \centering
         \includegraphics[width=\textwidth]{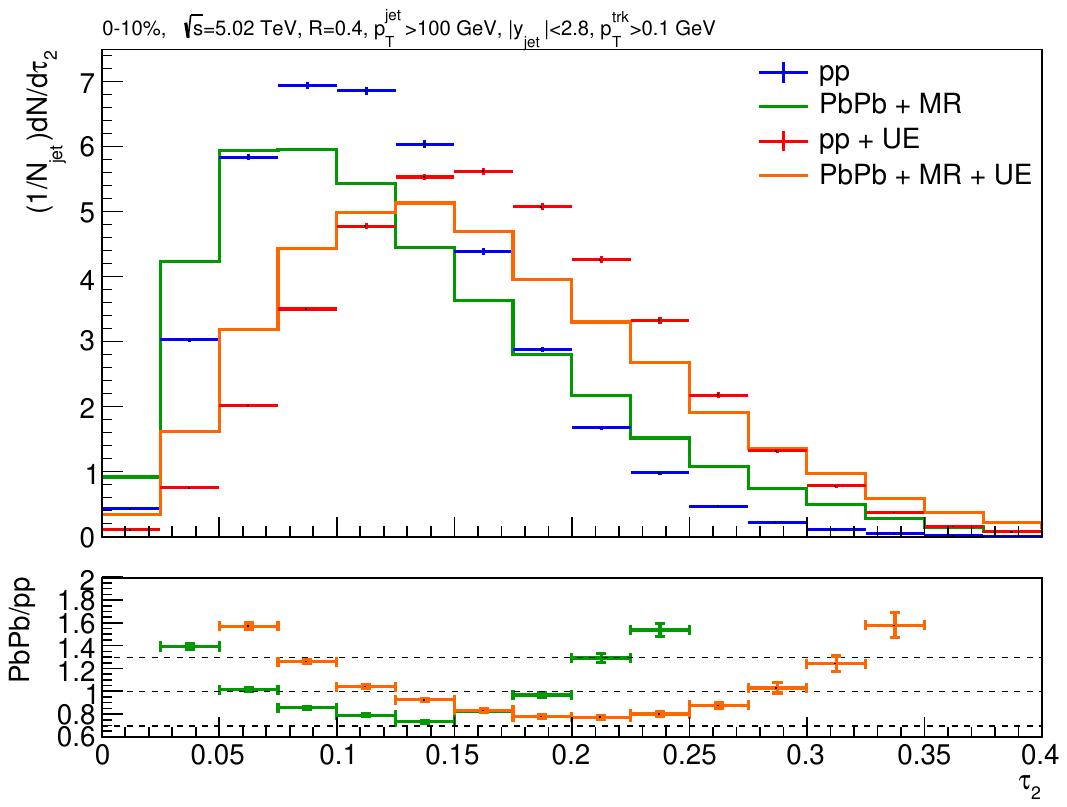}
            \caption{}
     \end{subfigure}
     \hfill
     \begin{subfigure}[b]{0.49\textwidth}
         \centering
         \includegraphics[width=\textwidth]{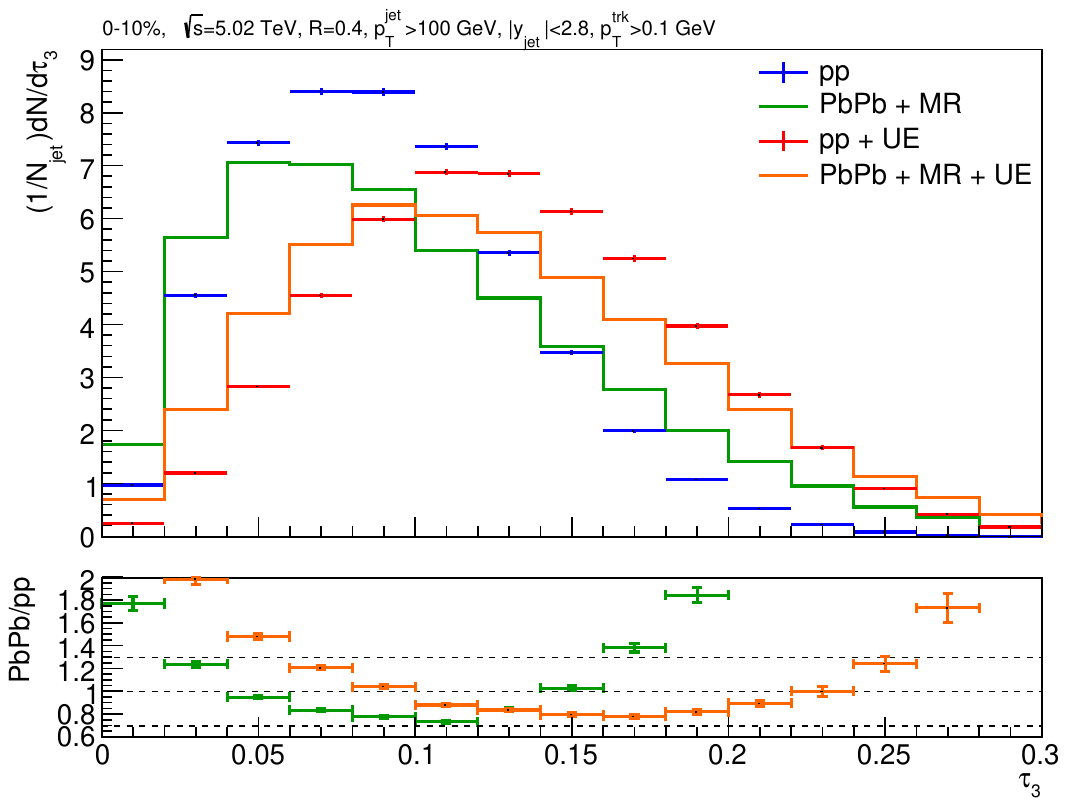}
            \caption{}
     \end{subfigure}
     \caption{N-subjetiness for \textbf{pp} and \textbf{PbPb + MR} collisions with and without UE contamination, with the PbPb to pp ratios presented in the bottom panels for: (a) \(N=1\); (b) \(N=2\); (c) \(N=3\).}
        \label{fig:nsub}
\end{figure}

N-Subjetiness, for \(N = 1, 2, 3\) , shown in Fig.~\ref{fig:nsub}, is given by:
\begin{align}
    \tau_1 &= \frac{1}{p_T^{jet} R} \sum_{i} p_{T,i} \min \left( \Delta R_{1,i} \right), \\
    \tau_2 &= \frac{1}{p_T^{jet} R} \sum_{i} p_{T,i} \min \left( \Delta R_{1,i}, \Delta R_{2,i} \right), \\
    \tau_3 &= \frac{1}{p_T^{jet} R} \sum_{i} p_{T,i} \min \left( \Delta R_{1,i}, \Delta R_{2,i}, \Delta R_{3,i} \right),
\end{align}
where $p_T^{jet}$ represents the transverse momentum of the jet, $R = 0.4$ refers to the jet parameter and each constituent has a transverse momentum $p_{T,i}$. The term $\Delta R_{k,i}$ describes the angular distance in the $\eta$-$\phi$ plane between the $i$-th constituent and the axis of the $k$-th subjet.

For these observables, both the distributions and the ratios do not seem to be robust to UE contamination. For all three cases we have a shift of the average of the distribution to the right, with a similar shift on the ratio. Furthermore, the parabolic shape of the ratios seems to be widened by the procedure.

\begin{figure}[!htbp]
     \begin{subfigure}[b]{0.49\textwidth}
         \centering
         \includegraphics[width=\textwidth]{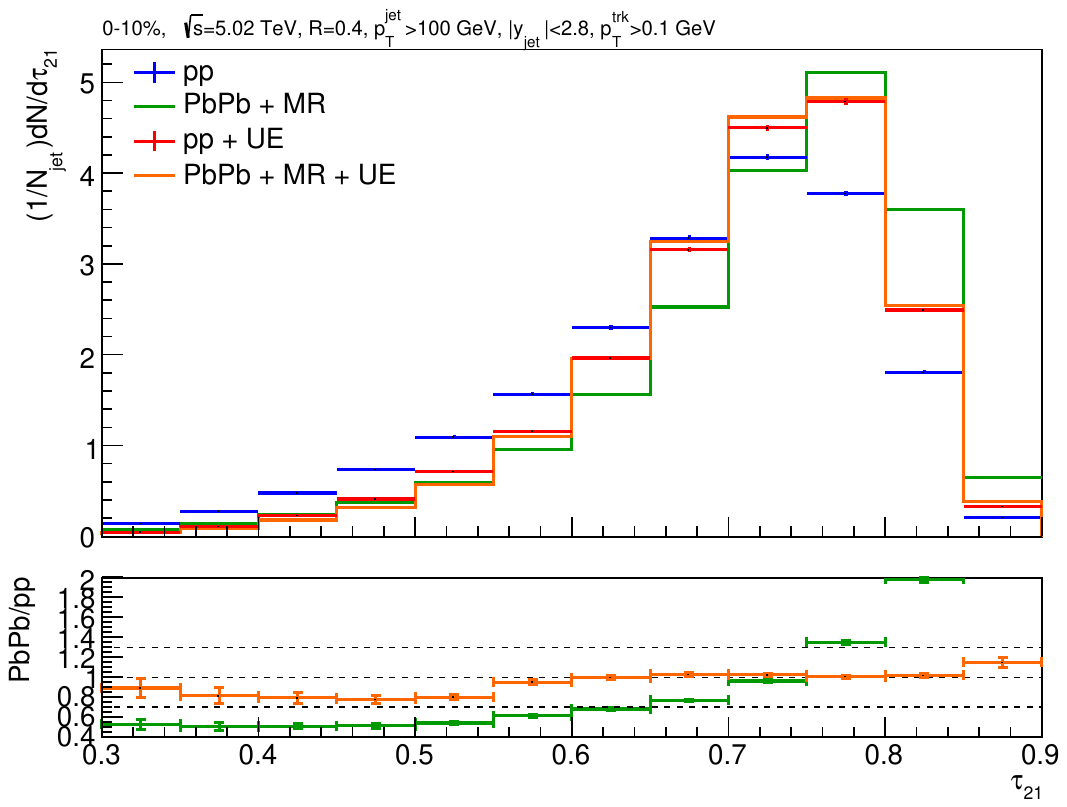}
            \caption{}
     \end{subfigure}
     \hfill
     \begin{subfigure}[b]{0.49\textwidth}
         \centering
         \includegraphics[width=\textwidth]{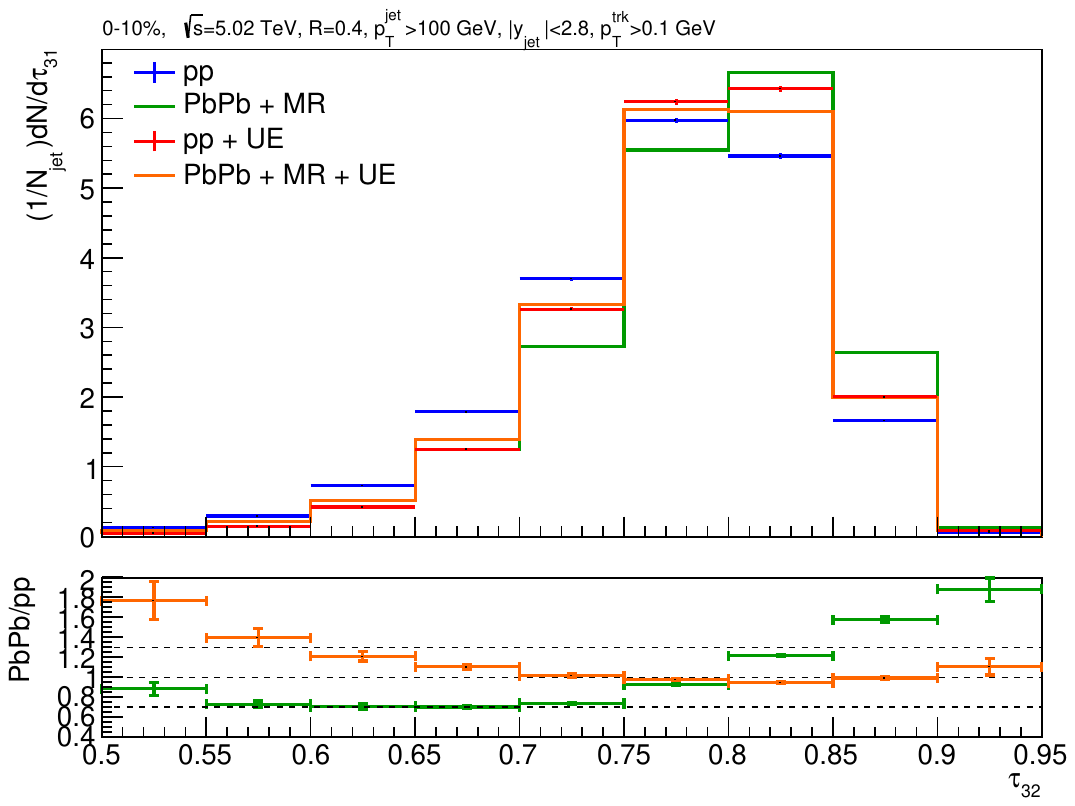}
            \caption{}
     \end{subfigure}
     \caption{Ratio of N-subjetiness to (N-1)-subjetiness for \textbf{pp} and \textbf{PbPb + MR} collisions with and without UE contamination, with the PbPb to pp ratios presented in the bottom panel for: (a) \(N=2\); (b) \(N=3\).}
        \label{fig:nsubrat}
\end{figure}

N-Subjetiness ratios, shown in Fig.~\ref{fig:nsubrat}, seem to no longer present such a pronounced shift to the right at the level of the distributions, but still show strong modifications on the PbPb to pp ratios. 

\subsection{Soft Drop Specific Observables}

\begin{figure}[!htbp]
     \centering
     \begin{subfigure}[b]{0.49\textwidth}
         \centering
         \includegraphics[width=\textwidth]{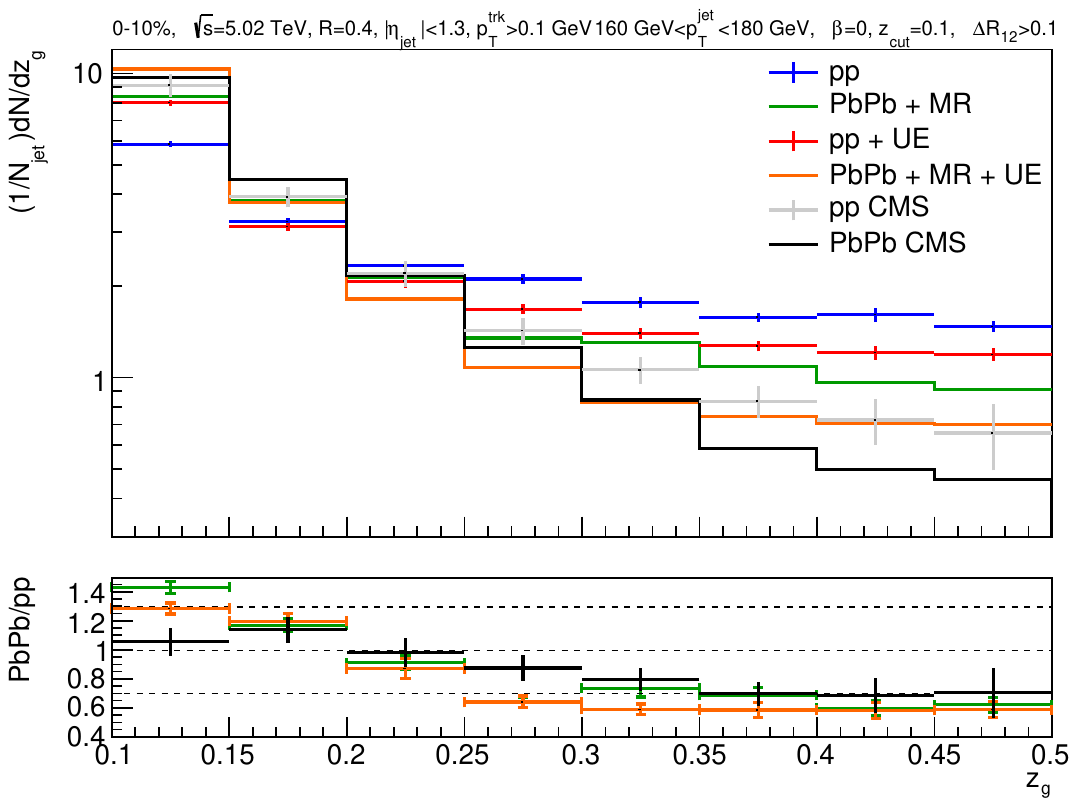}
         \caption{}
     \end{subfigure}
     \hfill
     \begin{subfigure}[b]{0.49\textwidth}
         \centering
         \includegraphics[width=\textwidth]{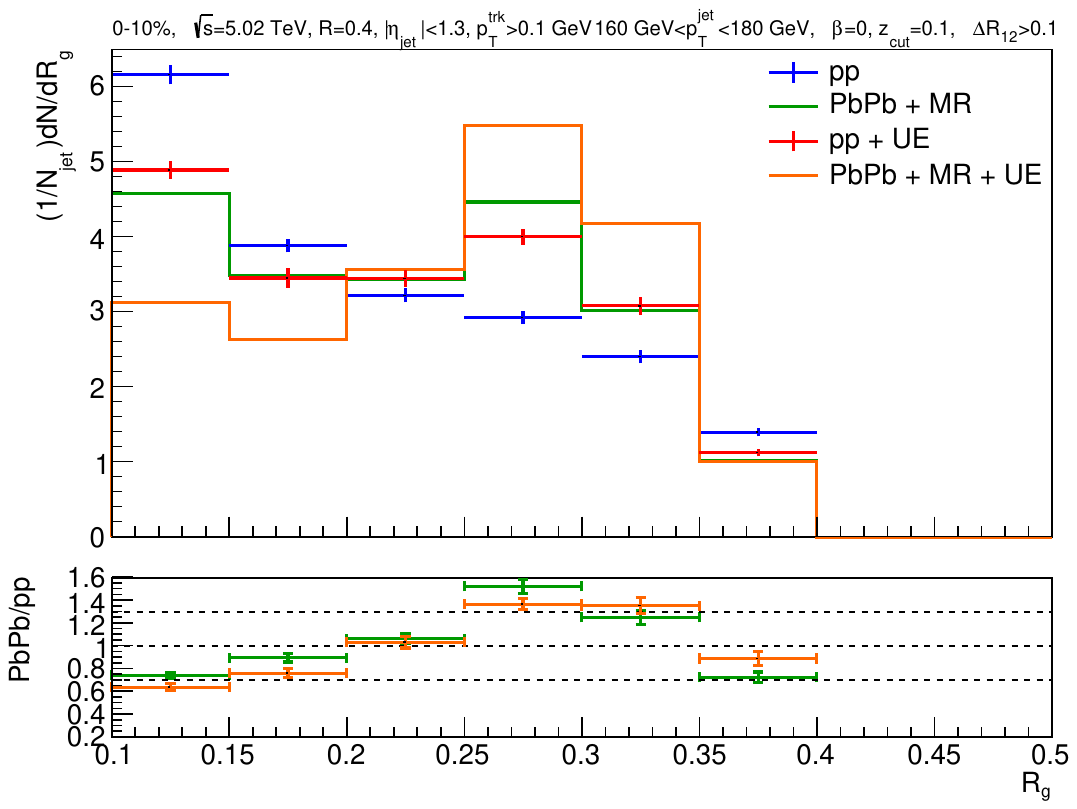}
          \caption{}
     \end{subfigure}
     \\
     \begin{subfigure}[b]{0.49\textwidth}
         \centering
         \includegraphics[width=\textwidth]{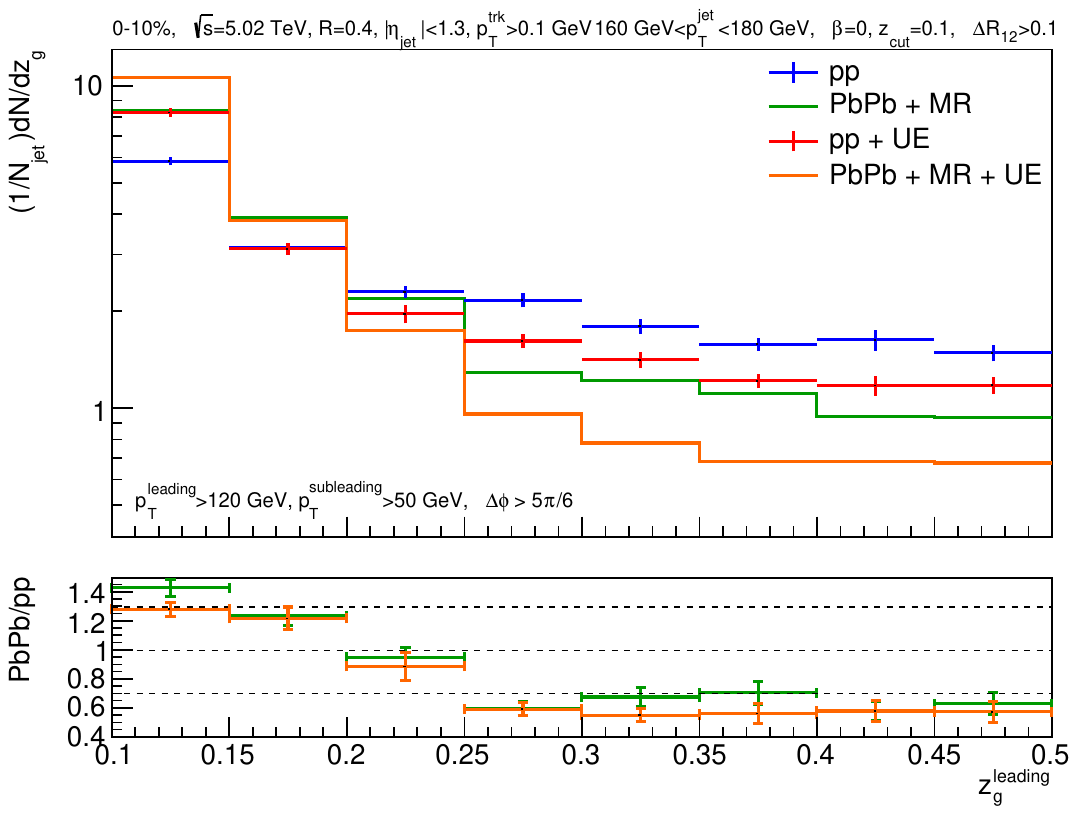}
          \caption{}
     \end{subfigure}
     \hfill
     \begin{subfigure}[b]{0.49\textwidth}
         \centering
         \includegraphics[width=\textwidth]{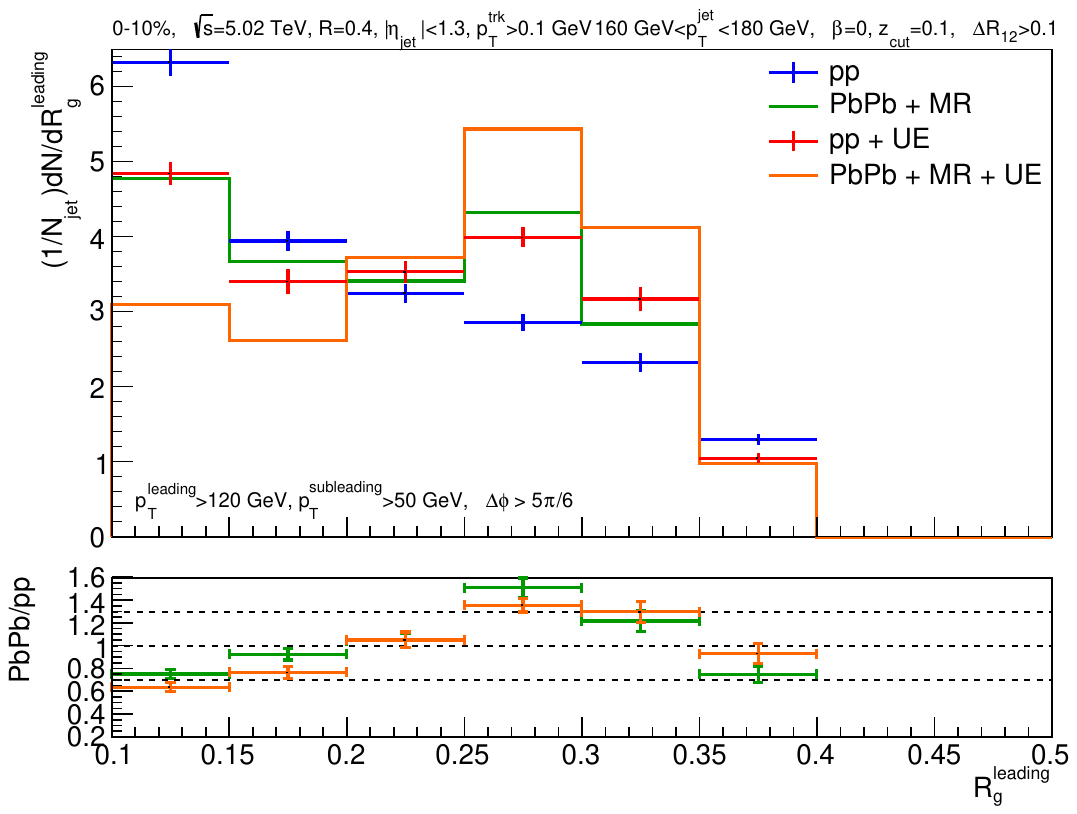}
          \caption{}
     \end{subfigure}
      \\
     \begin{subfigure}[b]{0.49\textwidth}
         \centering
         \includegraphics[width=\textwidth]{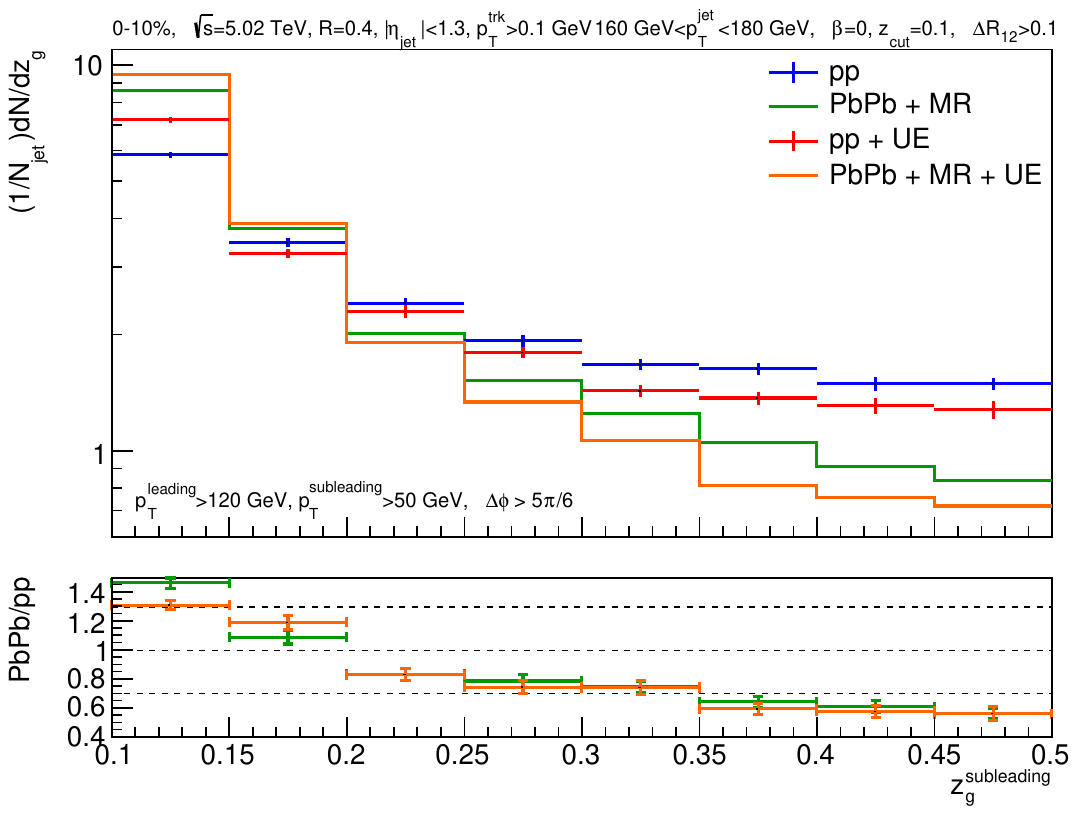}
          \caption{}
     \end{subfigure}
     \hfill
     \begin{subfigure}[b]{0.49\textwidth}
         \centering
         \includegraphics[width=\textwidth]{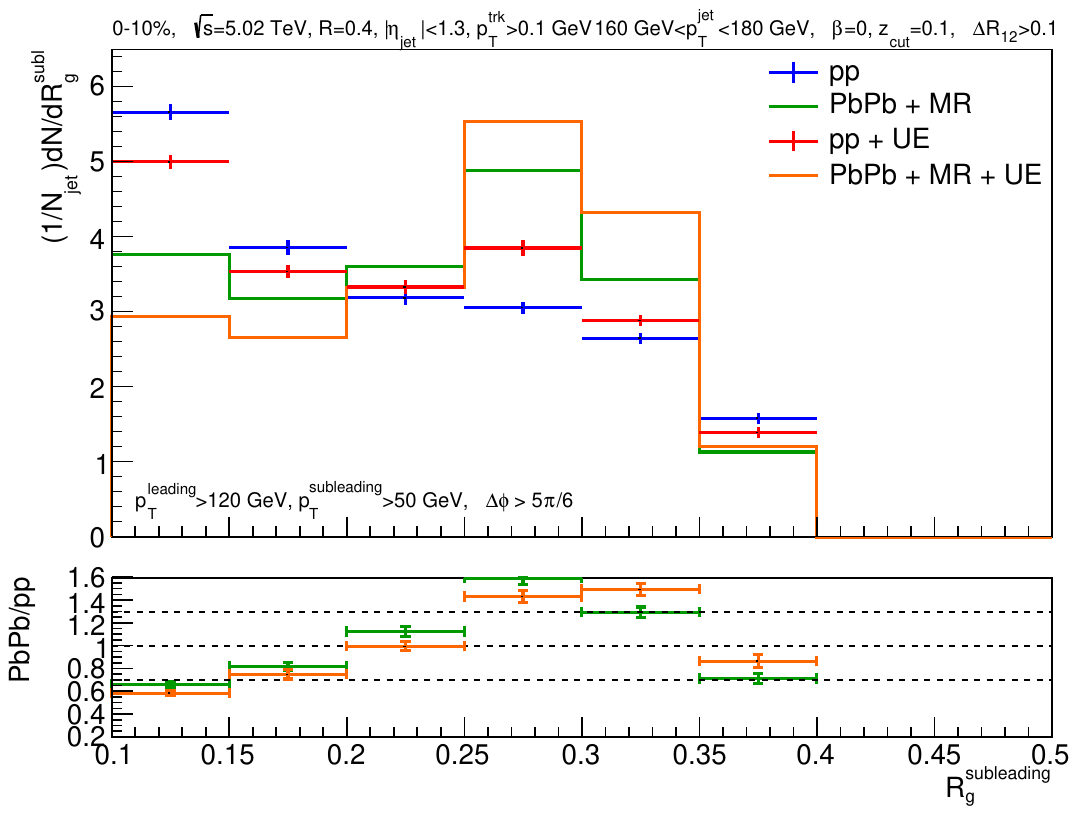}
          \caption{}
     \end{subfigure}
     \caption{Momentum sharing in SD declustering (left) and splitting opening (right) for \textbf{pp} and \textbf{PbPb + MR} with and without UE contamination, with PbPb to pp ratios in the bottom panels and with experimental data from \cite{CMS:2017qlm} for: the inclusive samples (top); samples only including the leading jet of the dijet pair (middle) and samples including only the subleading jet of the dijet pair (bottom).}
        \label{fig:zgrg}
\end{figure}

The transverse momentum sharing fraction for the first declustering that satisfies the SoftDrop condition, shown on the left of Fig.~\ref{fig:zgrg}, is given by:
\begin{equation}
z_g = \frac{\min(p_{T,i}, p_{T,j})}{p_{T,i} + p_{T,j}} > z_{\text{cut}} \left( \frac{\Delta R_{ij}}{R_0} \right)^{\beta}\, ,
\end{equation}
where \(z_{\text{cut}}\) and \(\beta\) are parameters of the SD procedure. Results are presented for both an inclusive jet sample and separately for leading and subleading jets in a dijet pair.

For this observable, the distributions seem to be enhanced at lower values and suppressed at higher values within the considered range. This contamination however, seems to be the same for pp and PbPb, yielding a robust ratio for this observable 

The opening angle of the declustering that meets the Soft Drop condition, shown on the right of Fig.~\ref{fig:zgrg}, is given by:
\begin{equation}
R_g = \Delta R_{ij}: z_g = \frac{\min(p_{T,i}, p_{T,j})}{p_{T,i} + p_{T,j}} > z_{\text{cut}} \left( \frac{\Delta R_{ij}}{R_0} \right)^{\beta}.
\end{equation}
Results are presented for both an inclusive jet sample and separately for leading and sub-leading jets in a dijet pair.

In this case, we find an analogous situation to the previous case. Both pp and PbPb distributions seem to be modified in similar ways, yielding robust ratios for all cases. The difference is simply that the contamination leads to a suppression of lower values and an enhancement of mid to high values instead.

\begin{figure}[!htbp]
     \centering
     \begin{subfigure}[b]{0.49\textwidth}
         \centering
         \includegraphics[width=\textwidth]{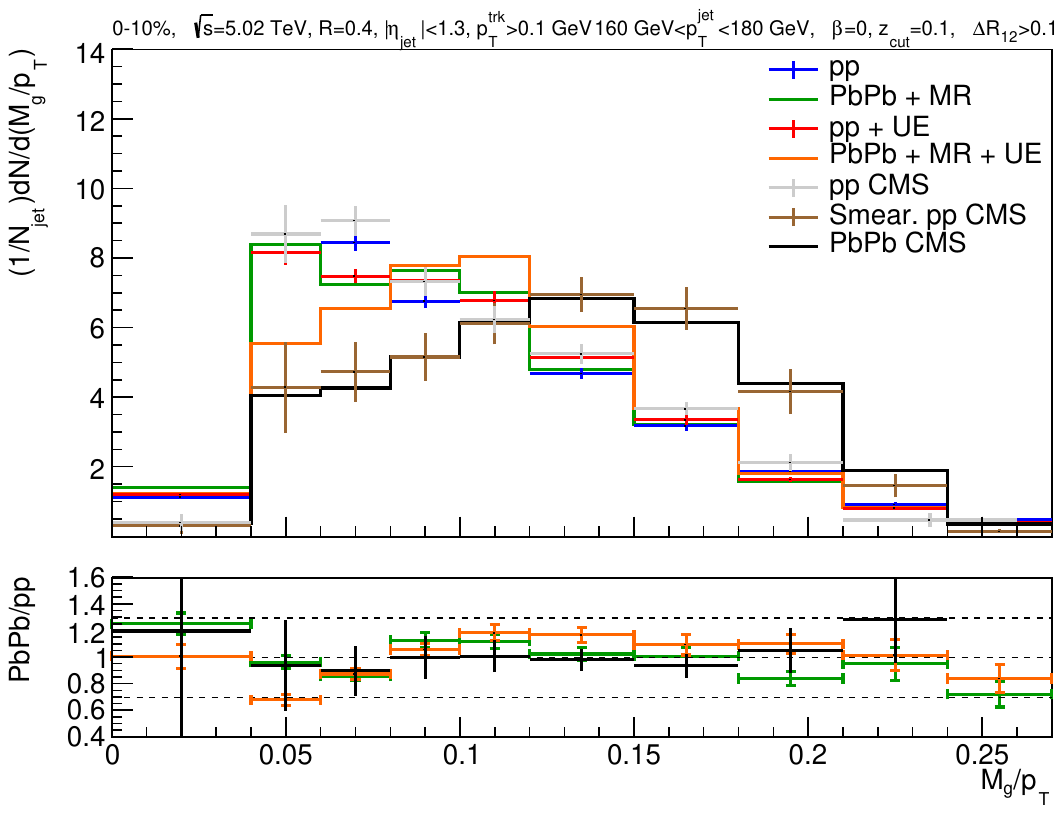}
         \caption{}
     \end{subfigure}
     \\
     \begin{subfigure}[b]{0.49\textwidth}
         \centering
         \includegraphics[width=\textwidth]{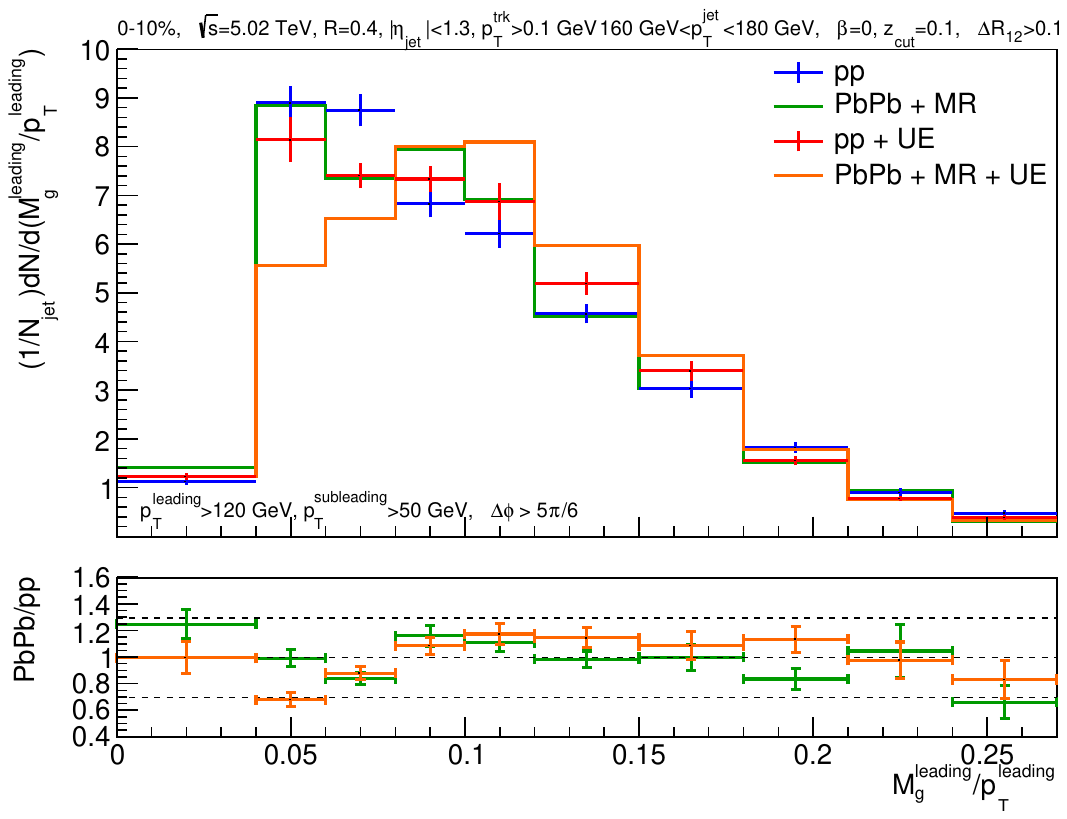}
            \caption{}
     \end{subfigure}
     \hfill
     \begin{subfigure}[b]{0.49\textwidth}
         \centering
         \includegraphics[width=\textwidth]{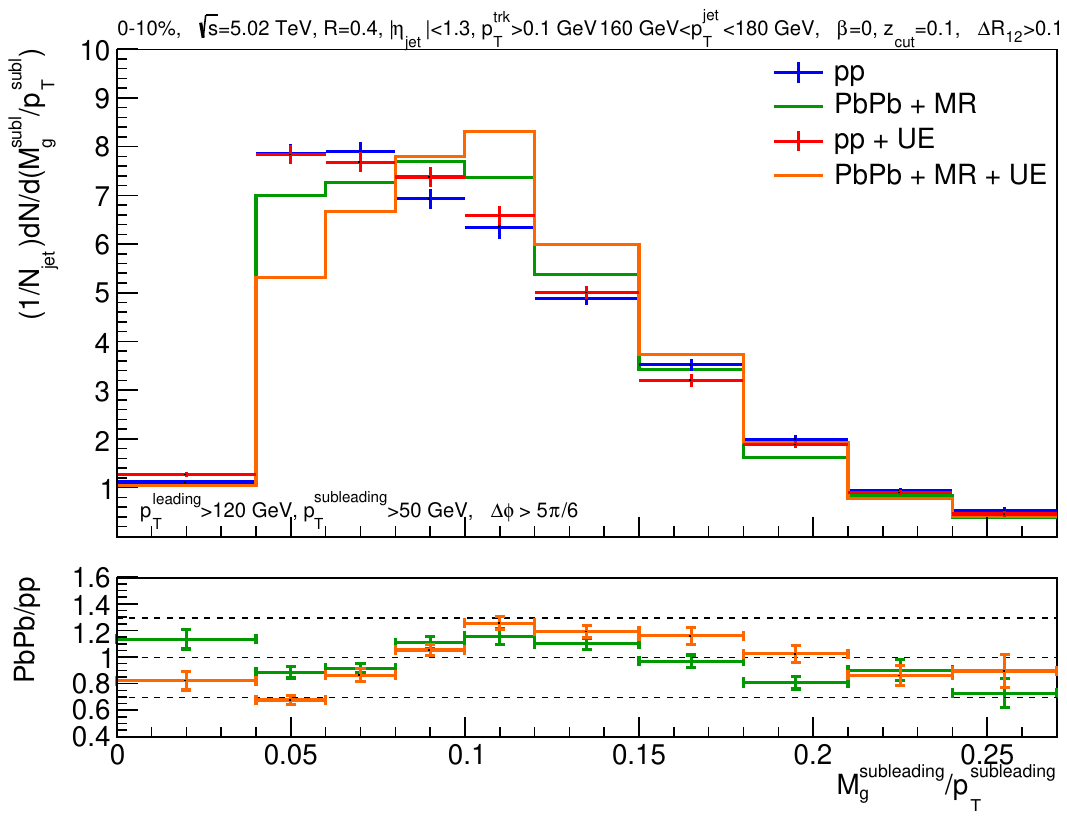}
            \caption{}
     \end{subfigure}
     \caption{Groomed jet mass for \textbf{pp} and \textbf{PbPb + MR} with and without UE contamination, with PbPb to pp ratios in the bottom panels and with experimental data from \cite{CMS:2018fof} for: (a) the inclusive samples; (b) samples only including the leading jet of the dijet pair; and (c) and samples including only the subleading jet of the dijet pair.}
        \label{fig:mg}
\end{figure}

The groomed jet mass, shown in Fig.~\ref{fig:mg}, is given by:
\begin{equation}
M_g = \sqrt{2E_1 E_2 (1 - \cos \theta_{12})}\, ,
\end{equation}
where \(E_i\) is the energy of each of the declustered subjets and \(\theta_{12}\) the angle between them. Results are presented for both an inclusive jet sample and separately for leading and sub-leading jets in a dijet pair.
Again, the PbPb to pp ratios are fairly robust to UE contamination.

\subsection{Lund Planes}

\begin{figure}[!h]
    \centering
    \centering
    \begin{subfigure}[b]{0.48\textwidth}
        \centering
        \includegraphics[width=\textwidth]{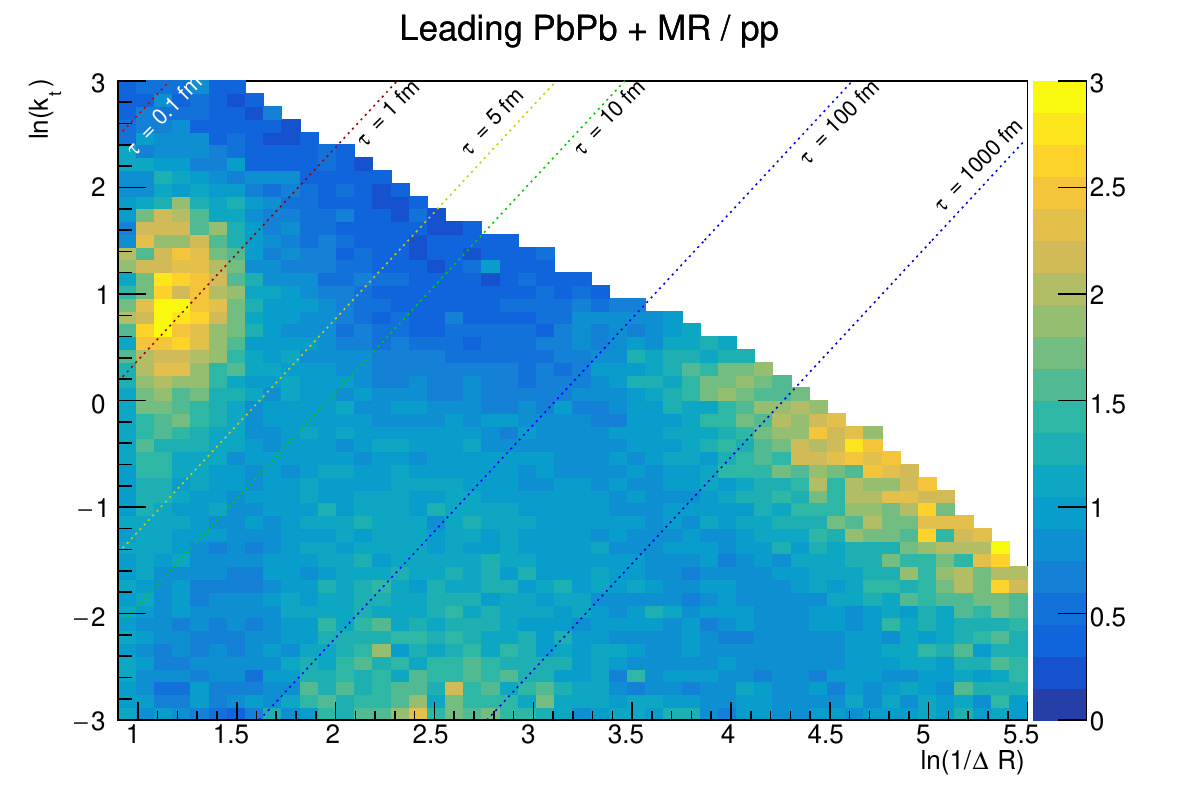}
         \caption{}
        \label{sfig:lundl_gen}
    \end{subfigure}
     \hfill
    \begin{subfigure}[b]{0.48\textwidth}
        \centering
        \includegraphics[width=\textwidth]{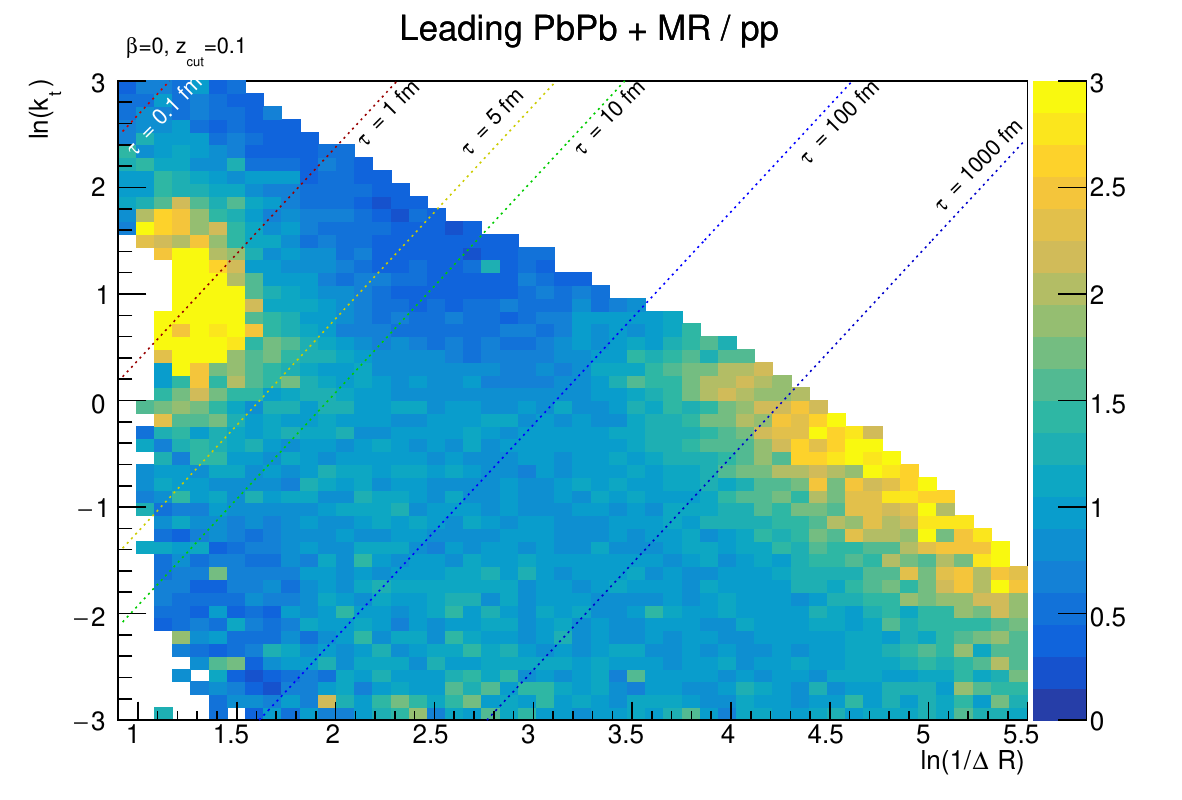}
         \caption{}
        \label{sfig:lundl_gen_sd}
    \end{subfigure}
    \\
    \begin{subfigure}[b]{0.48\textwidth}<
        \centering
        \includegraphics[width=\textwidth]{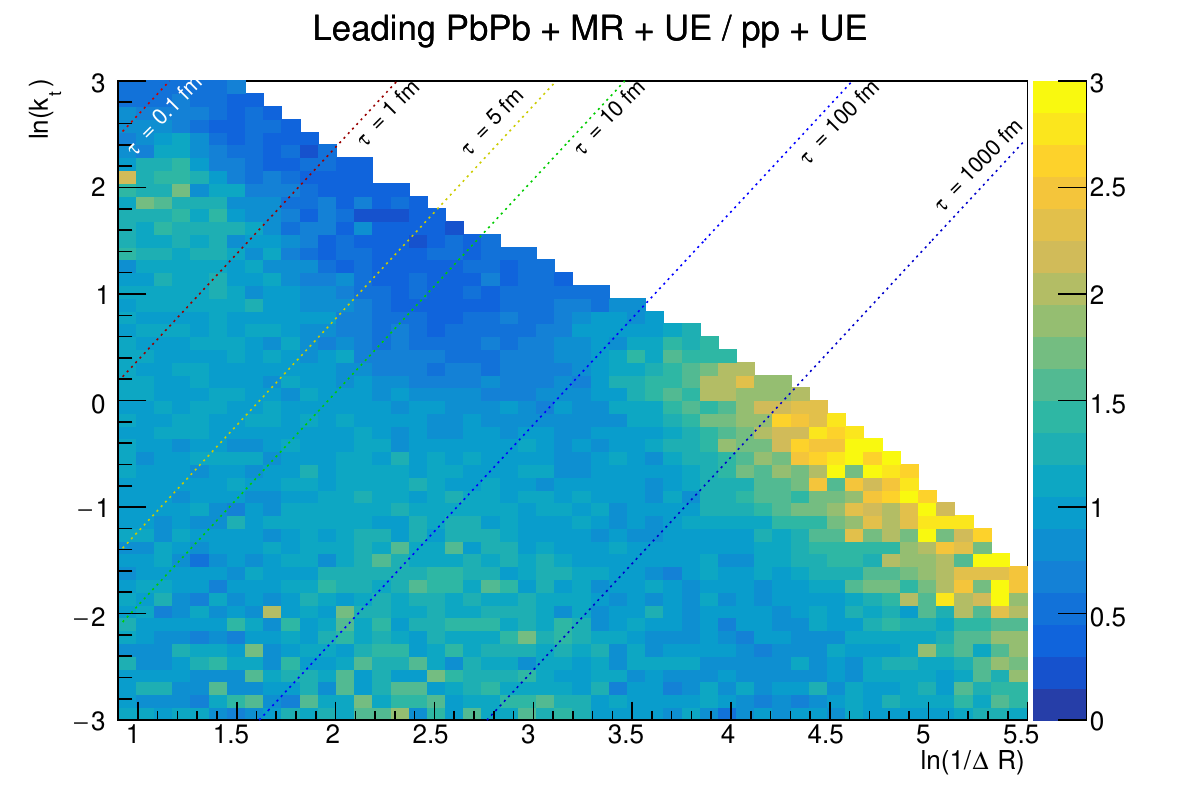}
         \caption{}
        \label{sfig:lundl_sub}
    \end{subfigure}
     \hfill
    \begin{subfigure}[b]{0.48\textwidth}
        \centering
        \includegraphics[width=\textwidth]{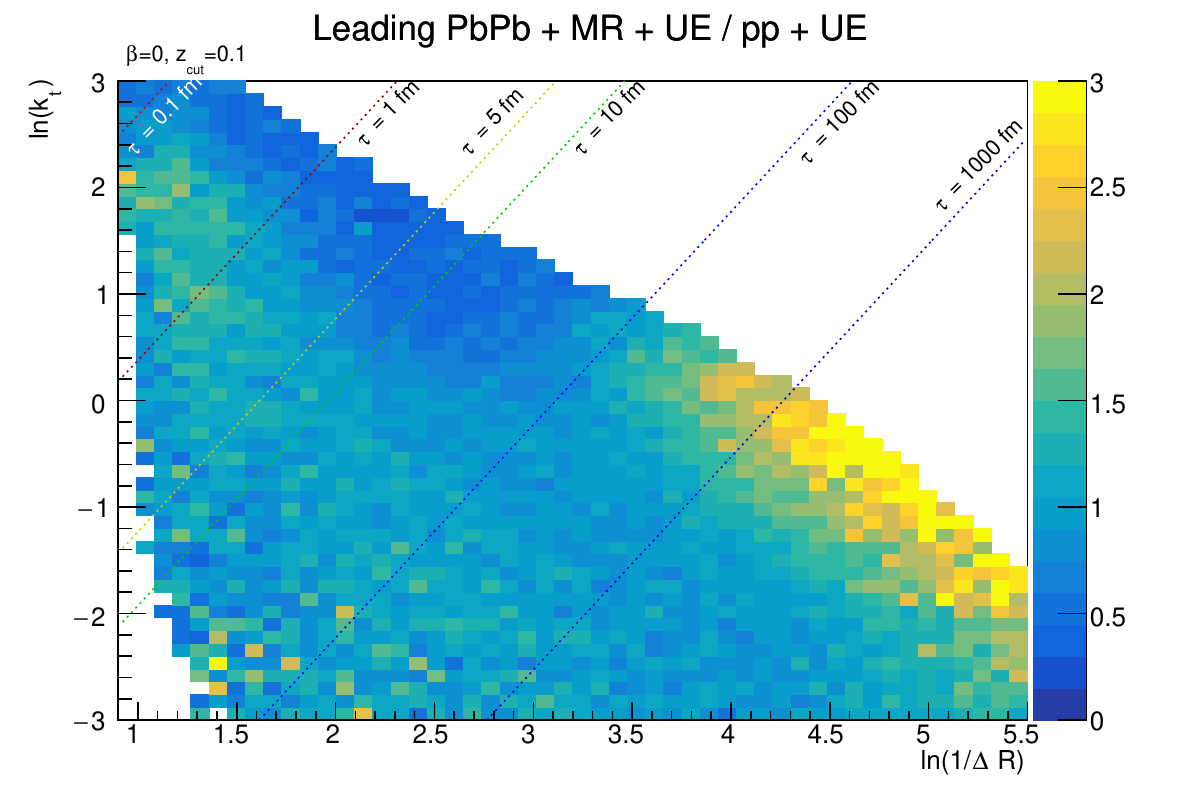}
         \caption{}
        \label{sfig:lundl_sub_sd}
    \end{subfigure}
    \caption{Ratios of \textbf{PbPb + MR} to \textbf{pp} Jet Lund Planes for the leading jet, including (bottom) and not including (top) UE contamination, with (right) and without (left) SD applied.}
    \label{fig:lundRl}
\end{figure}

\begin{figure}[!h]
    \centering
    \centering
    \begin{subfigure}[b]{0.48\textwidth}
        \centering
        \includegraphics[width=\textwidth]{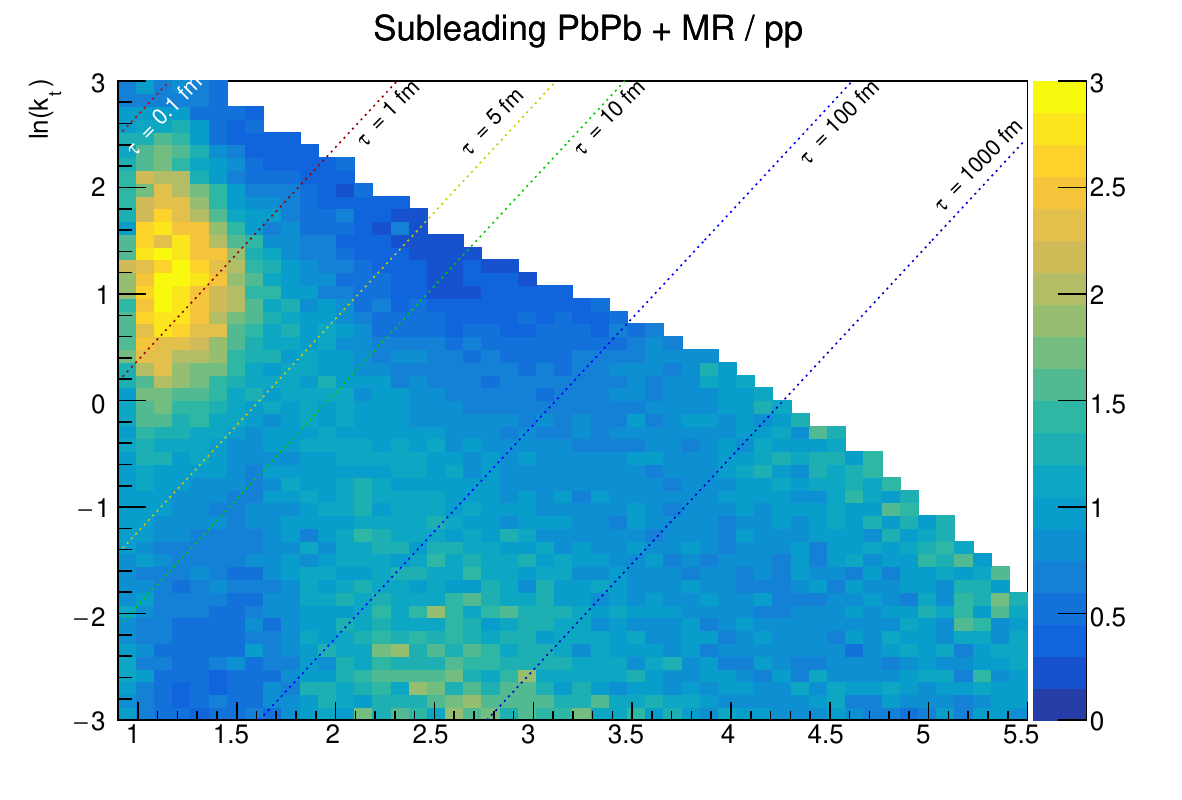}
         \caption{}
        \label{sfig:lunds_gen}
    \end{subfigure}
     \hfill
    \begin{subfigure}[b]{0.48\textwidth}
        \centering
        \includegraphics[width=\textwidth]{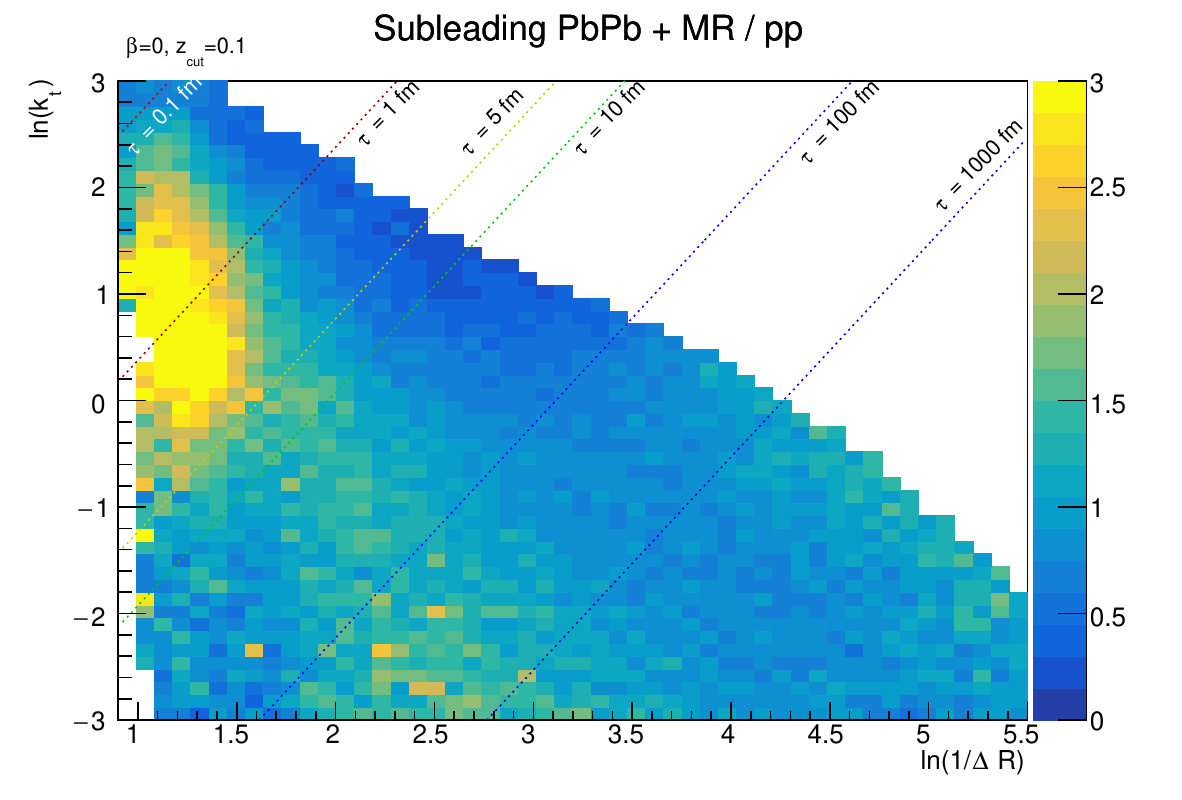}
         \caption{}
        \label{sfig:lunds_gen_sd}
    \end{subfigure}
    \\
    \begin{subfigure}[b]{0.48\textwidth}
        \centering
        \includegraphics[width=\textwidth]{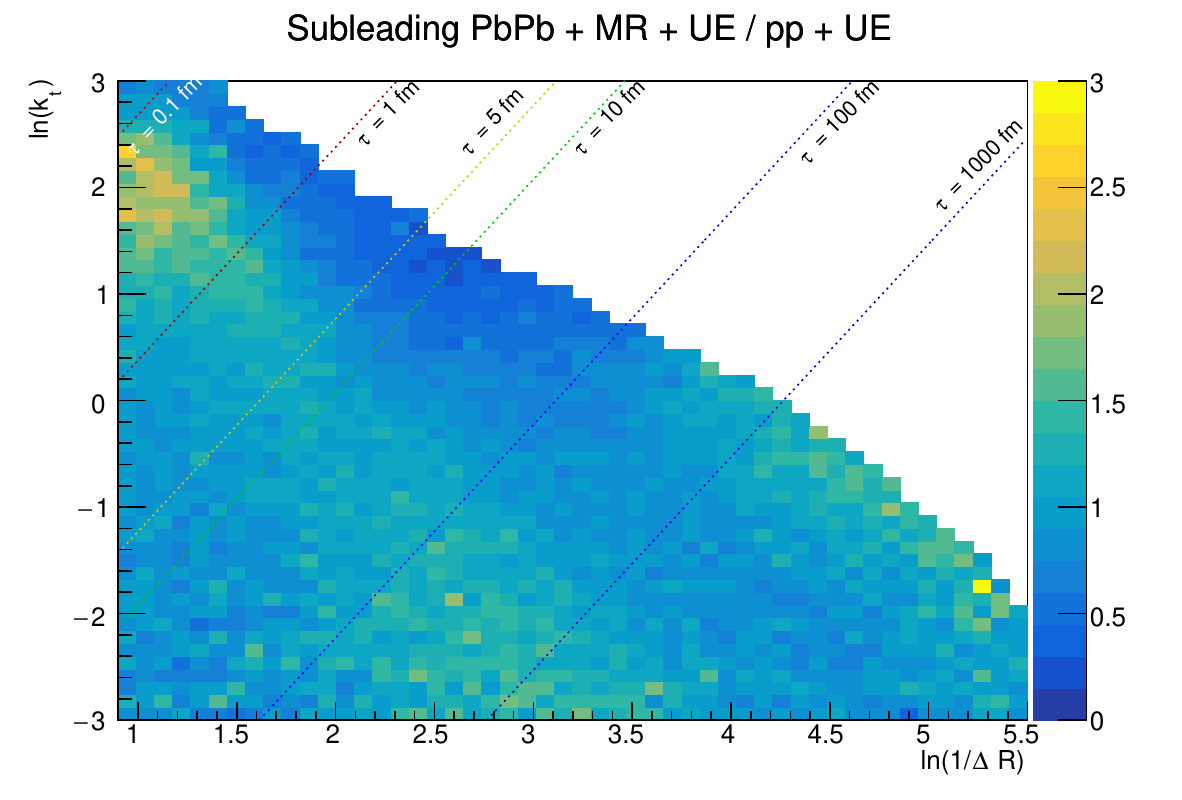}
         \caption{}
        \label{sfig:lunds_sub}
    \end{subfigure}
     \hfill
    \begin{subfigure}[b]{0.48\textwidth}
        \centering
        \includegraphics[width=\textwidth]{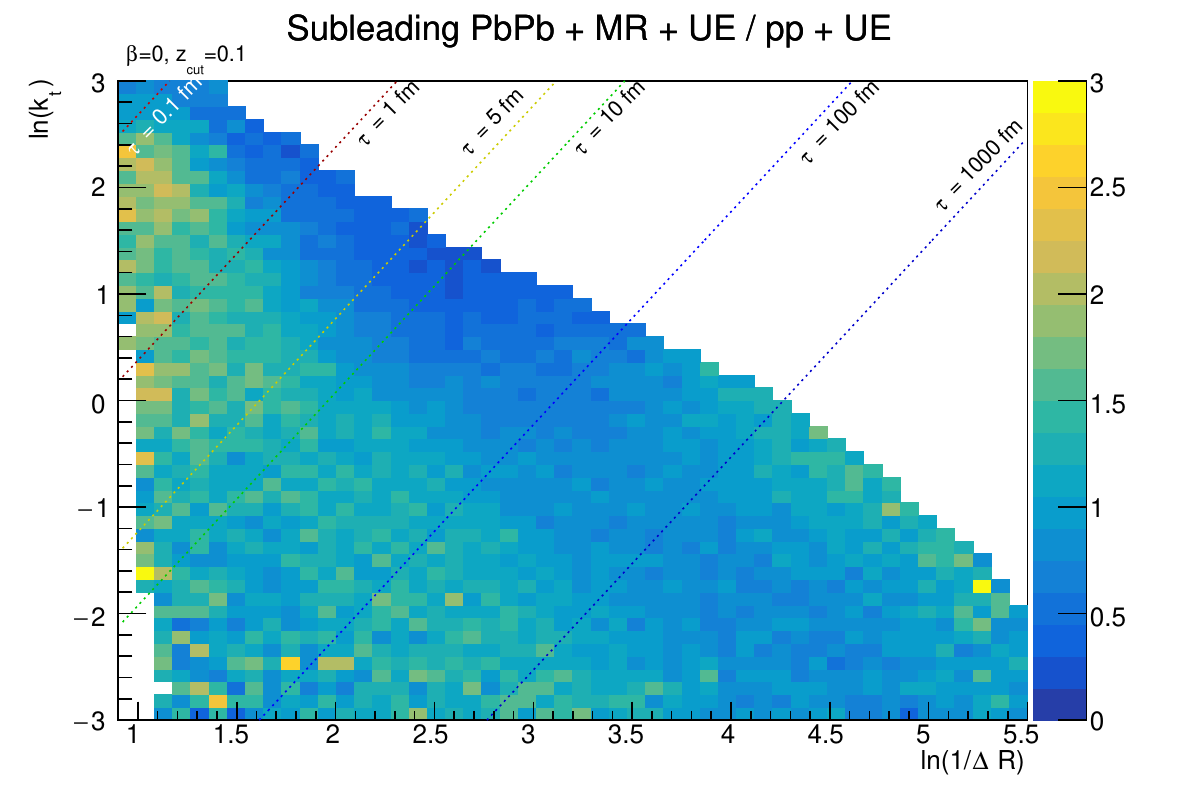}
         \caption{}
        \label{sfig:lunds_sub_sd}
    \end{subfigure}
    \caption{Ratios of \textbf{PbPb + MR} to \textbf{pp} Jet Lund Planes for the subleading jet, including (bottom) and not including (top) UE contamination, with (right) and without (left) SD applied.}
    \label{fig:lundRs}
\end{figure}

Lund Planes and how to construct them is described in subsection \ref{subsubsec:lund}. In Fig.~\ref{fig:lundRl} and Fig.~\ref{fig:lundRs} the same PbPb to pp Lund planes ratios as in Fig.~\ref{fig:lundR} are shown but for the leading and subleading jets of the dijet pair respectively. The MR attributed enhancement seen in Fig.~\ref{fig:lundR}, is yet again present to a slightly more significant degree for the leading jet of the dijet pair (Fig.~\ref{fig:lundRl}) and significantly more for the subleading jet of the dijet pair (Fig.~\ref{fig:lundRs}). In both cases, and analogously to the inclusive case of Fig.~\ref{fig:lundR}, the feature is washed out once UE is accounted for.

\clearpage

\bibliographystyle{jhep}

\bibliography{apples}
 
\end{document}